\documentclass[aps,prx,superscriptaddress,twocolumn,eprint,amsmath,amssymb,longbibliography,nofootinbib]{revtex4-1}
\usepackage{amsmath,amsthm,amsfonts,amssymb,amscd}
\usepackage{fullpage}
\usepackage{lastpage}
\usepackage{graphicx}
\usepackage{hyperref}
\usepackage{wrapfig}
\usepackage{enumerate}
\usepackage{fancyhdr}
\usepackage{mathrsfs}
\usepackage{mathtools}
\usepackage{braket}
\usepackage{slashed}
\usepackage{bm}
\usepackage[dvipsnames]{xcolor}
\usepackage{float}
\usepackage[margin=0.55in]{geometry}
\usepackage[english]{babel}
\usepackage{geometry}                
\usepackage{graphicx}
\usepackage{amssymb}
\usepackage{epstopdf}
\newcommand{\mcm}[1]{\textcolor{black}{#1}}

\newcommand{\SR}[1]{\textcolor{black}{#1}}

\newcommand{\mjb}[1]{\textcolor{black}{#1}}
\newcommand{\nf}[1]{\textcolor{black}{#1}}
\begin{document}

\title{Symmetry, Thermodynamics and Topology in Active Matter}
\author{Mark J. Bowick}
\affiliation{Kavli Institute for Theoretical Physics, University of California Santa Barbara, Santa Barbara, CA 93106, USA}
\author{Nikta Fakhri}
\affiliation{Department of Physics, Massachusetts Institute of Technology, Cambridge, MA 02139, USA}
\author{M. Cristina Marchetti}
\affiliation{Department of Physics, University of California Santa Barbara, Santa Barbara, CA 93106, USA}
\author{Sriram Ramaswamy}
\affiliation{Centre for Condensed Matter Theory, Department of Physics,
Indian Institute of Science, Bangalore 560 012, India}

\begin{abstract}
This article 
\mcm{outlines a selection of  current and emerging directions in active matter research. }
The topics highlighted include: the ubiquitous occurrence of spontaneous flows and active turbulence and the theoretical and experimental challenges associated with controlling and harnessing such flows; the role of motile topological defects in ordered states of active matter and their possible biological relevance; the emergence of nonreciprocal effective interactions and the role of chirality in active systems, \mcm{with} intriguing connections to  non-Hermitian quantum mechanics; the progress towards a  formulation of the thermodynamics of active systems; 
the impact of the active matter framework on our understanding of the emergent mechanics of biological tissue. These diverse phenomena all stem from the defining property of active matter as an assembly of components that individually and dissipatively break time-reversal symmetry. 
\end{abstract}

\maketitle

\section{Active Matters!}
\label{sec:intro}

\paragraph{\mcm{A bit of history.}} In 1995 the Hungarian physicist Tam\'{a}s Vicsek proposed a minimal model of bird flocking inspired by the physics of magnetism~\cite{vicsek1995novel}. He showed that a collection of flying spins - self-propelled point particles traveling with fixed speed in a direction updated by alignment with neighbors in a noisy environment - can undergo a phase transition from a disordered state where the \mcm{spins} fly randomly in all directions to an ordered state of collective motion. Just a few months later, Vicsek presented his work in a seminar at IBM Yorktown Heights. John Toner and Yuhai Tu realized that they could turn Vicsek's agent-based model into a field theory and formulated what are now known as the Toner-Tu equations of flocking~\cite{toner1995long}.

Truth be told, what physicists now know as the Vicsek model had effectively been previously formulated by Craig Reynolds, a computer scientist working for the animation industry, who in 1986 created \emph{Boids}~\cite{reynolds1987flocks} -- an agent-based simulation of collective motion that he employed, for instance, to generate the animation of flying bats in the 1992 feature \emph{Batman Returns}. Indeed the model appears even earlier in the literature, in theories of fish schools by Aoki \cite{Aoki1982simulation} and Partridge \cite{partridge1982structure}. While these contributions remained unnoticed by the physics community for some years, Reynolds, a leader in the development of three-dimensional animation, was awarded a Scientific and Technical Award by the Academy of Motion Picture Arts and Sciences in 1998. In the intervening years, the notion introduced by Vicsek that the collective dynamics of self-driven entities can be described as a nonequilibrium phase transition gained enormous popularity among physicists and was shown to provide a powerful framework for describing spontaneous organization on many scales. For their key contribution to the creation of the field of active matter, Vicsek, Toner and Tu  received the Lars Onsager prize of the American Physical Society in 2020.

The first published use of the term Active Matter appears to be in Ref.~\cite{ramaswamy2006mechanics}. Active Membranes appear a little earlier in the physics literature~\cite{prost1996shape,ramaswamy2000nonequilibrium}. The term ``active'' in reference to fuel-driven transport across a membrane, against a concentration gradient, is standard in biology~\cite{darnell1990molecular}.
Active stresses in a fluid medium suffused with sustained energy conversion make their first appearance in Ref~\cite{finlayson1969convective}.

\paragraph{\mcm{What is active matter?}} Today the name active matter refers to any collection of entities that individually \mcm{use} free energy to generate their own motion and forces~\cite{ramaswamy2010mechanics,marchetti2013hydrodynamics}. Through interactions, these active particles  spontaneously organize in emergent large-scale structures with a rich range of materials properties.
\mcm{The defining property of an active system is that the 
\SR{energy input} that maintains the system out of equilibrium, whether truly internal or created by contact with a proximate surface, acts individually and independenly on each ``active particle''.   Hence, once the chemo-mechanical processes that convert fuel into motion are integrated out, the dynamics of such active entities breaks time reversal symmetry (TRS) in a local and sustained matter. This should be contrasted with more conventional nonequilibrium systems that are displaced from equilibrium  globally by an external force, 
\SR{as in sedimentation under gravity}, or are forced at the boundaries, such as through an imposed mechanical shear.}  Due to the breaking of TRS at the microscale, active systems do not obey detailed balance and can generate self-sustained flows and cyclical currents. Thus, steady-state movies of active dynamics run forwards and backwards do not look the same, as they would in Newtonian mechanics. \mcm{Active matter therefore poses a fundamental problem in nonequilibrium statistical mechanics: what laws govern order, phase transitions and fluctuations in systems in which the very particles constituting the system break detailed balance.}

\paragraph{\mcm{Why is active matter important?}} Examples of active matter abound in the living world~\cite{vicsek2012collective}, from the coordinated motion of cells in wound healing~\cite{Poujade15988} to bird flocks~\cite{Ballerini1232} and human crowds~\cite{Helbing2000,Bain2019}. Over the last twenty years, a number of synthetic analogues have also been engineered. These systems exploit energy-conversion avenues, ranging from the interplay of vibration and static friction~\cite{narayan2007long,deseigne2010collective,kumar2014flocking} to \mcm{ a variety of autophoretic processes~\footnote{\mcm{Autophoresis denotes particle drifts induced by  concentration gradients created by the particles themselves.}}} \cite{golestanian2019phoretic}, to endow particles from colloidal to macroscopic scales with self-propulsion, often tunable through suitably applied light or electric fields~\cite{Bechinger-RMP2018}. 
\mcm{Active matter spans an enormous range of scales, with realizations found inside the cell nucleus and  in flocks of hundreds of thousands of starlings. It  embodies a  unique
property of  living systems, which is the ability to convert energy
injected at the molecular scale into organized motion and function at the macroscopic scale. 
Research in active matter aims at understanding this defining property and at transfering this fundamental understanding to other disclipines. The importance of the field stems therefore from both the deep intellectual challenge of developing a predictive theory of the hierarchical organization ubiquitous in living systems and the potential for applying this understanding to the design of new bioinspired materials.}

\paragraph{\mcm{Active20.}} \mcm{Over the past two decades, continuum \SR{and particle-scale} theories, simulations and experiments have uncovered striking properties of active particles that require conceptually new approaches and have brought key questions into sharper focus. Yet the field remains at the forefront of current research in soft matter, biological physics and statistical physics, with high levels of activity and excitement.} This motivated the plan to run in Spring 2020 a nearly three-month-long program on active matter physics at the Kavli Institute for Theoretical Physics (KITP) at UC Santa Barbara. The program was due to focus on recent developments and especially future directions and connections with other fields, from stochastic thermodynamics to topological phenomena analogous to those familiar in quantum systems. The Covid-19 pandemic was well upon us in the Spring and it was decided to run the program virtually just a week before the planned start of March 16. Indeed the state of California issued a mandatory stay-at-home order on March 19. The KITP and the program coordinators quickly regrouped and on March 30 they started the first virtual KITP program which ran for nine weeks, with exceptionally high participation of scientists from all over the world, in spite of the challenges of time zones.  In this article we describe some of the outcomes of that program and highlight new directions of inquiry that emerged during those nine weeks.  The following presents
the four authors' personal view, informed by contributions from all program participants, of some of the emerging directions and open questions in active matter research.

\paragraph{\mcm{Outline.}} The rest of the article is organized as follows. In Section~ \ref{sec:turbulence} we describe recent developments and future directions in understanding the self-sustained dynamics of orientationally-ordered active fluids. We focus on systems with nematic order and on the ubiquitous phenomenon of active turbulence. 
In Section~\ref{sec:defects} we highlight the dynamics of topological defects \mcm{in the nematic texture} which spontaneously proliferate in the turbulent state. An important open direction here is investigating the possible biological relevance of defects in organ and organism development and collective cell migration. Understanding the dynamics of defects may also provide a framework for controlling and directing active flows, as described in Section~\ref{sec:harness}. 
A closely related challenge, highlighted in Section~\ref{sec:directing}, is developing predictive theories of active assembly, with the ultimate goal of mimicking nature to construct machines made of smaller machines and capable of performing specified functions. To achieve this, we need to understand collective behavior of entities with effective interactions that are nonreciprocal, i.e., do not obey Newton's third law. This raises important and deep questions discussed in Section~\ref{sec:NR}, intimately related to the role of chirality (Section~\ref{sec:chiral}), and with surprising connections to non-Hermitian quantum mechanics. Section~\ref{sec:thermo} focuses on the rapidly evolving use of ideas from stochastic thermodynamics to quantify deviations from equilibrium and formulate the nonequilibrium thermodynamics of active matter. In Section~\ref{sec:cells} we highlight the role of  active matter physics in developing a framework for describing the mechanical properties of cell collectives and biological tissue and conclude in \mcm{Section~\ref{sec:outlook} highlighting the breadth and connectivity of the field.}

\section{From spontaneous flow to active turbulence}
\label{sec:turbulence}

\mcm{Liquids can be made to flow by applying external forces at their boundaries.} Collections of self-driven  entities form active fluids that flow \emph{spontaneously} with no externally applied forces or pressure gradients~\cite{Simha2002,voituriez2005spontaneous}.  Since individual active units are often elongated or polarized along an axis,  such self-driven fluids can exhibit liquid crystalline order~\cite{marchetti2013hydrodynamics}. \mcm{Order in turn couples to spontaneous flows, resulting in 
unusual rheological properties such as ``superfluid''-like behavior, excitability and chaotic spatio-temporal dynamics known as bacterial or active turbulence. An important goal is to understand and classify dynamical phase transitions in these active fluids and the properties of various phases in terms of the symmetry of orientational order  and the mechanisms that control energy dissipation. In particular, the ubiquitous observation of self-sustained spatiotemporal chaotic dynamics in both engineered and living systems begs the question:  what are the universal properties of active turbulence? Can we identify and classify various scenarios that drive the transition from spontaneous laminar flows to chaotic ones? Are the metrics used to quantify high-Reynolds number turbulence useful in the active context? }

\paragraph{\mcm{Polar and nematic active fluids.}} \mcm{Active particles with a head and a tail, such as bacteria~\cite{wensink2012meso}, birds~\cite{toner1995long,Ballerini1232}, polar \SR{vibrated grains \cite{kumar2014flocking}}, and  Quincke rotors~\cite{bricard2013emergence}, can organize into states with polar (ferromagnetic) order and macroscopic mean motion - a flock.} 
\mcm{Apolar active particles in contrast form active fluids with nematic liquid crystalline order, where the rod-like entities have a common orientation, but no preferred direction. This more subtle type of orientational order, familiar from the physics of passive liquid crystals~\cite{chaikin2000principles}, has been observed in} active systems on many scales, from subcellular structures, such as the mitotic spindle that controls cell division, to suspensions of cytoskeletal filaments and associated motor proteins~\cite{sanchez2012spontaneous,zhang2020dynamics}, crawling bacteria~\cite{nishiguchi2017long,dell2018growing,yaman2018emergence}, epithelia~\cite{kawaguchi2017topological,saw2017topological,blanch2018turbulent}, monolayers of vibrated granular rods~\cite{narayan2007long} and even entire multicellular organisms~\cite{maroudas2021topological}. \mcm{Active nematics have no net mean flow, but display a complex self-sustained spatiotemporal dynamics that resembles the  streaming used by cells to continuously circulate their fluid content.}

\begin{figure}
    \centering
    \includegraphics[width=0.48\textwidth]{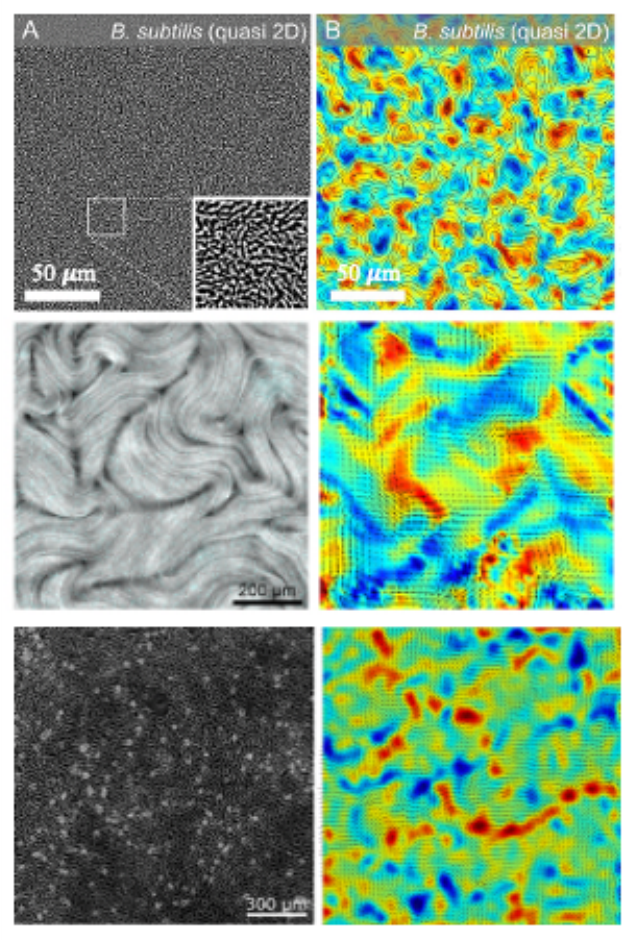}
    \caption{Experimental realizations of active turbulence, from top to bottom: a layer of \emph{B. subtilis} (adapted from Ref.~\cite{wensink2012meso}), a suspension of kinesin-powered microtubule bundles (courtesy of Linnea Lemma and Zvonimir Dogic), and a layer of MDCK cells (adapted from Ref.~\cite{blanch2018turbulent}). The left column shows images of the system while the right column displays snapshots of the local vorticity.}
    \label{fig:turbulence}
\end{figure}

The uniform ordered bulk states of both polar and nematic active fluids are generically unstable, at any activity~\cite{Simha2002}, to  self-sustained chaotic spatiotemporal flows or active  turbulence~\cite{alert2021active},
\mcm{ as exemplified in Fig.~\ref{fig:turbulence}. The vortical flows look qualitatively similar in both types of system, raising the question of how \SR{the symmetry of the orientational order affects} the behavior and scaling properties.}

The order parameter for the  flocking transition is  the mean polarization vector or self-propulsive velocity of the active entities, which of course has the same vectorial symmetry as the flow velocity of the fluid. One can then often describe the dynamics in terms of a single vector field through
the Toner-Tu equations~\cite{toner2005hydrodynamics} which marry the Navier-Stokes hydrodynamics of simple fluids to the relaxational dynamics of magnets. 


Due to the apolar nature of the ordered state, the nematic order parameter  is a tensor field
$Q_{ij}=\langle\sum_\alpha \nu^\alpha_i\nu^\alpha_j-\delta_{ij}/d\rangle=S(n_in_j-\delta_{ij}/d)$ that measures the \SR{uniaxial anisotropy of the distribution} 
of the orientation $\bm\nu^\alpha$ of individual molecules, with $d$ the system's dimensionality. The nematic is a fluid with a finite value of $S$ and spontaneously broken orientational symmetry along the director $\mathbf{n}$.

%
%
In the regime appropriate to most active fluids where inertia is negligible, \mcm{and assuming constant density, the dynamics of an active nematic is then described by the \SR{interplay of 
relaxation and flow-induced reorientation} of the $\mathbf{Q}$ tensor and force balance} encoded in the Stokes equation for the flow velocity $\mathbf{v}$, 
\begin{equation}
\eta\nabla^2\mathbf{v}-\bm\nabla p+\bm\nabla\cdot\bm\sigma^a=0\;,
\label{eq:Stokes}
\end{equation}
where we have for simplicity neglected  stresses due to liquid-crystalline elasticity. The pressure $p$ is determined by requiring incompressibility, $\bm\nabla\cdot\mathbf{v}=0$.
The  active stress $\bm\sigma^a=\alpha\mathbf{Q}$ describes the dipolar forces exerted by the active nematogens of the surrounding fluid. It is controlled by the activity $\alpha$ that incorporates the biomolecular 
\SR{parameters (such as the concentration of motor protein and of ATP) that drive active motions.} The sign of $\alpha$ depends on whether active forces are extensile ($\alpha<0$), as 
in  microtubule nematics, or contractile ($\alpha>0$), as may for instance be the case for certain actomyosin networks.
\mcm{In active nematics the instability of the ordered state  arises from the competition between energy input from active stress that drives disorienting flows and orientational relaxation that restores alignment. Due to the feedback between flow and orientational distortions, vortical flows are accompanied by proliferation of topological defects in the nematic texture.}

\paragraph{\mcm{Active turbulence.}} The chaotic spatio-temporal dynamics known as active turbulence is a \SR{phenomenon distinct} from the well-known inertial turbulence of high-Reynolds number fluids, perhaps more akin to the elastic turbulence observed twenty years ago in sheared polymer solutions~\cite{steinberg2021elastic}, where nonlinear elastic effects can destabilize laminar flow. The distinctive feature of active or bacterial turbulence is that the system is destabilized by stresses generated internally at short scales and then self-organized to larger scales through the interplay of interactions and dissipation. In other words, the spectrum of energy injection that drives the chaotic flows is not externally imposed but rather self-organized. This essentially arrests the energy cascade and leads to the selection of a length scale $\ell_a=\sqrt{K/|\alpha|}$ determined by the balance of active and elastic stresses (of strength determined by the nematic stiffness $K$) that controls both the spacing of topological defects in the nematic texture and the size of the flow vortices. As a result, these chaotic flows can be equivalently characterized by examining the statistics of flow vortices or that of the nematic disclinations. Numerical studies have established the emergence of power laws in the energy spectrum and other scalings analogous to, but different from, Kolmogorov's scaling of inertial turbulence~\cite{alert2021active}. A remarkable result has been the demonstration of the ``universality'' of the scaling properties of active nematic flows and associated defects in the nematic texture across systems and scales~\cite{giomi2015geometry}.  The study of turbulence using metrics from the theory of dynamical systems, such as Lyapunov exponents or topological entropy, is in its infancy. More work is needed to see if this is a useful approach. For a more extensive discussion of active turbulence, we point the reader to a  recent critical review of active turbulence 
~\cite{alert2021active}.

\paragraph{\mcm{Order parameters and inertia.}} An important open question is quantifying
the difference between chaotic active flows in polar and nematic fluids. For instance, in polar fluids nonlinear terms describing self-advection of the order parameter seem to be important in driving energy transfer across scales. 
How does this compare to the role of self-advection in inertial turbulence? In contrast, there is no self-advection in active nematics and advective nonlinearities seem to play a secondary role relative to flow alignment and flow-induced rotation of the order parameter in driving nonlinear flows. \mcm{Is this difference then important for explaining the vortex statistics in the two systems?}

 It has also recently \cite{chatterjee2019fluid} become clear that the inclusion of inertia transforms our understanding of extensile or ``pusher'' polar active suspensions. A new dimensionless control parameter $R \equiv \rho v_0^2/2 \sigma_0$, suggestive of an effective Reynolds number, emerges, where $\rho$ is the total mass density, $v_0$ the self-propulsion speed, and $\sigma_0$ the characteristc scale of active stress. Speed matters, and there are two in the problem. The instability advances with a finite velocity $(\sigma_0/\rho)^{1/2}$ in a treatment in which inertia ($\rho$) is taken into account, and if $v_0$ is large enough the suspension can outrun and thus eliminate the instability. The inescapable Stokesian instability of flocks in a fluid is simply the $R=0$ limit of a rich phase diagram, including a flocking transition driven by motility, separating a phase-turbulent but ordered flock from a defect-turbulent statistically isotropic state \cite{chatterjee2019fluid}.

\paragraph{\mcm{Role of dissipation and ``order from disorder''.}} \mcm{Another open challenge is understanding the role of different dissipative processes and associated length scales in mediating the onset of nonequilibrium dynamical steady states.} In many experimental realizations dissipation is controlled by the 
\SR{combination} of viscous stresses and frictional drag with an external inert medium, described by an additional force density $-\Gamma\mathbf{v}$ on the left hand side of Eq.~\eqref{eq:Stokes}. This introduces another length scale in the problem, the viscous screening length $\ell_v=\sqrt{\eta/\Gamma}$. Numerical work has indicated that the interplay between $\ell_v$, the active length $\ell_a$ and the nematic coherence length may result in vortex lattices, reminiscent of those familiar in type-II superconductors and Bose-Einstein condensates~\cite{doostmohammadi2016stabilization,doostmohammadi2018active}.
\mcm{More work is needed, however, to understand how the interplay between these two dissipation mechanisms may mediate the emergence of such ``order from disorder''.}

\paragraph{\mcm{Active nematics in three dimensions ($3D$).}} These systems are only beginning to be explored. Experimental realizations of $3D$ microtubule  nematics  have shown that in these systems turbulent flows are accompanied by the formation of neutral defect loops with zero topological charge. No charged 3D defects have been observed~\cite{duclos2020topological}. The structure of the topological defects generated by active flows in 3D and especially the connection between defects and flow structure remains to be explored. Recent numerical work has also shown that in $3D$ active fluids, spontaneous breaking of chiral symmetry leads to parity-violating Beltrami flows ~\cite{slomka2017spontaneous,slomka2018nature}. This points to the need to explore the role of chirality in active flows, in \emph{both} $3D$ and $2D$ ~\cite{soni2019odd}.

\paragraph{\mcm{Linking to interfacial instabilities.}} \mcm{Intriguing connections are also apparent between active mixtures and turbulent  multi-phase flows, where the interfaces between the coexisting phases play an important role in controlling the dynamics. Similarly, the largely unexplored interfacial properties of mixtures of  active liquid crystals and passive fluids should  control the highly nonequilibrium dynamics of these systems, with possible relevance to liquid-liquid phase separation in biological contexts~\cite{shin2017liquid}. Worth exploring are connections between active/passive interfaces and the interfaces found in staircases in density-stratified fluids~\cite{colm2020open} or in magnetized plasmas. }

\paragraph{\mcm{Need for experiments.}} Finally, even if theorists succeed at establishing classes of active turbulence distinguished by different scenarios of paths to chaotic flows and critical exponents, we will still need quantitative experimental tests of these predictions. This will require measurements of energy spectra in controlled settings, most likely to be carried out in synthetic active analogues or \emph{in vitro} systems. An important question of course is whether active turbulence plays a role in biology. The spontaneous flows, whether turbulent or orderly, observed in active nematics resemble cytoplasmic flows inside cells ~\cite{woodhouse2012spontaneous,kumar2014actomyosin}, suggesting that nature may use these mechanisms for stirring and transport to overcome slow passive diffusion process.

\section{Topological defects}
\label{sec:defects}

\mjb{Topological defects have been widely explored in both active and passive systems. What are the new features of \emph{active} defects and the challenges ahead?} \mcm{In active fluids spontaneous local currents turn the defects themselves into}  \mjb{self-propelled entities. 
The interacting defect gas then forms an active fluid 
with the possibilty of 
dynamically generated order of the defect themselves, such as polar (ferromagnetic) order on a larger scale than the microscopic nematogens, and nonreciprocal interactions between defects. Furthermore, the spontaneous pair creation of defects driven by activity means we need a {\em full} many-body theory to describe the interacting defect gas {--} this does not currently exist. }

\begin{figure}[h]
    \centering
    \includegraphics[width=0.45\textwidth]{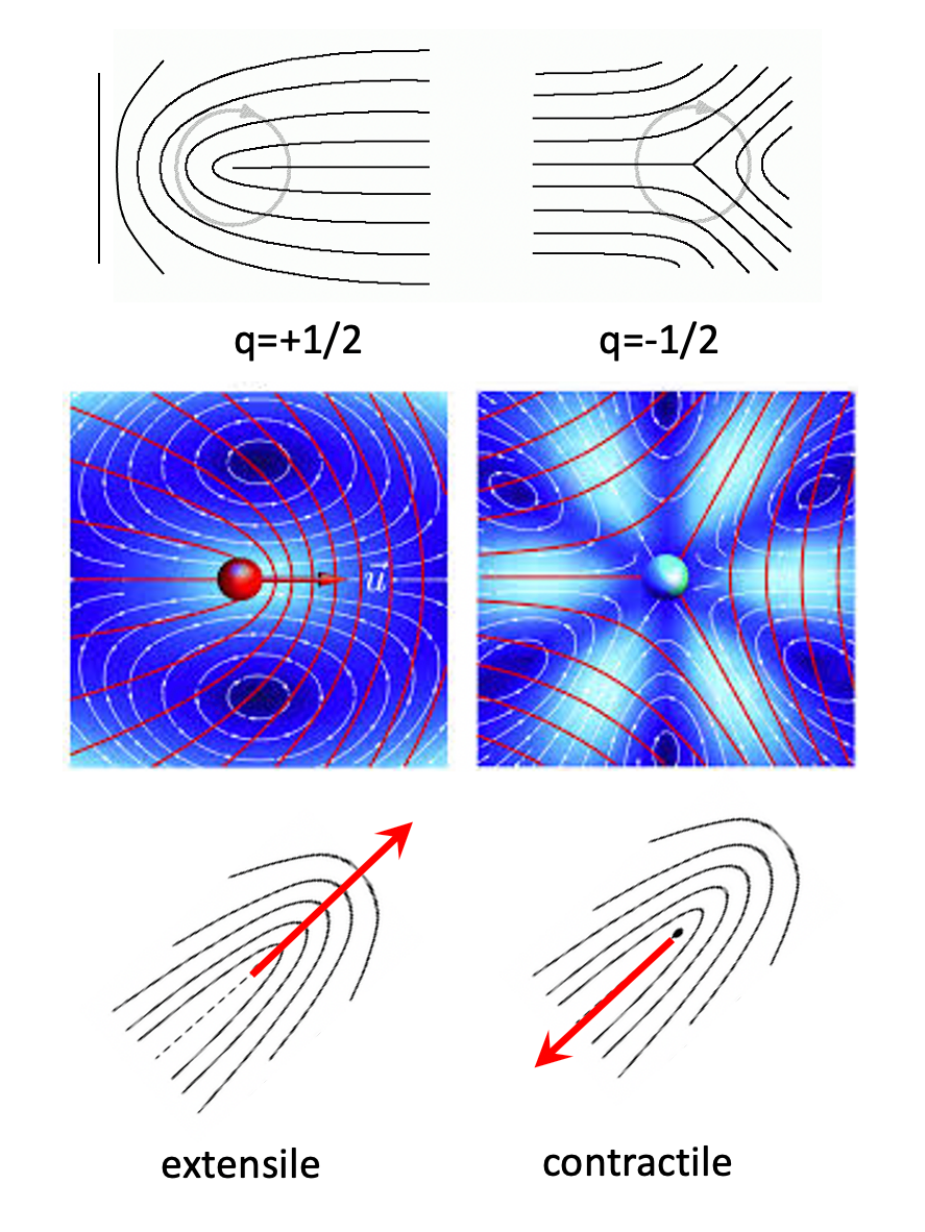}
    \caption{Topological defects in active nematic liquid crystals. Top row: configuration of the director texture for a $+1/2$ (left) and a $-1/2$ (right) defect. The topological charge $q$ is defined as the net rotation (in units of $2\pi$) of the order parameter as one encircles the defect. In the left frame the director rotates by $\pi$ in the same direction as the path is traversed, yielding $q=+1/2$. In the  right frame the director rotates by $\pi$ in the opposite direction as the path is traversed, hence $q=-1/2$. Middle row: the flow field around an isolated defect, adapted from Ref.~\cite{giomi2014defect}. The red lines are tangent to the director field. The blue background color represents the magnitude of the flow velocity and the white lines are the streamlines of the flow.  Bottom row: the red arrow show the self-propulsion speed of a $+1/2$ defect powered by flows induced by extensile (left) and contractile (right) active stresses.}
    \label{fig:defects}
\end{figure}

\paragraph{\mcm{Topological defects as a Coulomb gas in $2D$ passive matter.}} \mjb{To fill in with some background, }
topological defects are zeros of the order parameter field
classified by their topological charge (see Fig.~\ref{fig:defects} for definition.) In passive systems defects play a fundamental role in many two-dimensional phase transitions, the poster-child being the Berezinskii-Kosterlitz-Thouless (BKT) transition~\cite{kosterlitz1973ordering,chaikin2000principles}. First of all, the symmetry-breaking in the BKT transition is described by a subtle topological order in the defects themselves (in this case XY or U(1) vortices), rather than the spontaneous breaking of a global symmetry. At high temperature vortices  are randomly distributed, whereas below a critical temperature they form measurable vortex-antivortex bound pairs. The transition is then described as vortex unbinding.  Secondly, the defect degrees of freedom are the slowest, or rate-limiting,  degrees of freedom in the approach to the ordered phase. \mjb{The BKT transition is therefore fundamentally  formulated } as a many-body theory of a gas of defects \mjb{by mapping the defects onto interacting Coulomb charges.}
Similarly the half-integer strength disclinations of two-dimensional nematics control the development of nematic order in the isotropic to nematic (I-N) phase transition~\cite{stein1978kosterlitz}.

\paragraph{\mcm{Active defects as a gas of self-propelled Coulomb charges.}} For active nematics a striking new phenomenon emerges. Defect-antidefect pairs are spontaneously created and annihilated~\cite{sanchez2012spontaneous}. Furthermore the elementary +1/2 strength disclinations become \SR{self-propelled \cite{narayan2007long}}. 
When hydrodynamic flows are included, \mcm{self-propulsion originates} by virtue of a monopole moment of velocity at the defect cores arising from the induced dipolar fluid flow determined by the nematic texture of the defect~\cite{giomi2013defect,pismen2013dynamics} (Fig.~\ref{fig:defects}). The +1/2 defect has a two-fold symmetric comet-like pattern with a head and a tail that define a local polarization vector. The -1/2 strength anti-defect, in contrast, is not directly self-propelled by virtue of a three-fold symmetry in its nematic texture and vanishing fluid velocity at the defect core. The value of the self-propulsion velocity of a +1/2 defect is proportional to the activity, at least perturbatively.

There has been progress in describing active nematic flows in terms of the dynamics of the topological defects in the nematic texture by developing a description of defects as an interacting gas of active quasiparticles~\cite{shankar2018defect}. 
The self-propulsion of the $+1/2$ defect can oppose Coulomb attraction mediated by nematic elasticity. For sufficiently high activity a defect pair can even dynamically unbind, in an active version of the BKT transition~\cite{shankar2018defect}. The \mjb{existence of a finite activity threshold for the active BKT unbinding may seem surprising.} 
\SR{One might imagine that directed self-propulsion of the $+1/2$ defect can always overcome attraction} and there is no active nematic phase at all! A closer look though shows that defects do not separate ballistically. Noise-induced torques lead to rotational diffusion and the stabilization of an active nematic phase. 

\mjb{Existing work does not, however, tackle the important question of how defect pairs are unbound by active stresses. Answering this question will necessitate a microscopic calculation of the rate of defect pair creation and annihiliation as a function of activity and has so far been beyond reach. A complete theory of the active defect gas in fact requires a many body formulation that accounts for defect creation, 
just as electron-positron pair creation requires the formalism of quantum electrodynamics. 
It may also be that the polarization of defects is not an independent degree of freedom but instead emerges from the detailed dynamics of the set of locations of defects. Attempts to sort out this issue have raised the important question of when and to what extent one can fully describe the system's dynamics by tracking the defect motions, without also following the dynamics of the nematic texture, which could lag behind that of the defects ~\cite{zhang2020dynamics,vafa2020multi}.}

\paragraph{\SR{\it States of defect order.}} \mjb{Since the interacting defect gas is itself a mixture of active polar particles ($+1/2$) and passive particles ($-1/2$) with aligning torques, a natural question is whether active defects may organize in emergent ordered structures \SR{analogous to the blue phases in strongly chiral liquid crystals \cite{wright1989crystalline}}. Understanding the } possible local and global order of defect arrays in the active nematic phase remains an open challenge. Numerical work has suggested that defects and associated flow vortices may organize in ordered lattices when the  screening length controlling the interplay between viscous and frictional dissipation is comparable to the defect spacing~\cite{doostmohammadi2016stabilization}. A hydrodynamic description of the defect gas has shown that when flow is slaved to texture deformations, the spontaneous breaking of rotational symmetry leads to a collectively-moving defect-ordered polar liquid -- a flocking state of defects~\cite{shankar2019hydrodynamics}. This state emerges at activity levels beyond the turbulent state discussed earlier, a remarkable example of order from disorder. Polar order of $+1/2$ defects has been observed in  simulations of ``dry'' nematics~\footnote{\mcm{`Dry' active matter refers to particles moving in the absence of an ambient fluid, where dissipation is mainly controlled by frictional drag with an external medium. Conversely, `wet' active matter describes systems where dissipation is principally controlled by viscous stresses mediated by a suspending medium. The size of the viscous scale $\ell_v$ defined in Section~\ref{sec:turbulence}d relative to the system size controls the crossover between these two limits. When the medium is a fluid the fluid-mediated interactions are referred to as hydrodynamic interactions. }}~\cite{srivastava2016negative,putzig2016instabilities,patelli2019understanding}, \mjb{but seems to occur at intermediate activity}.
Theoretical models have also predicted \emph{local} nematic (i.e., antiparallel) order of the +1/2 defects~\cite{Pearce2020,Thijssen2020}, consistent with some recent experiments~\cite{Pearce2020}, but at odds with earlier work predicting~\cite{oza2016antipolar} and observing large-scale nematic ordering~\cite{decamp2015orientational}.  \mjb{Numerical solutions of continuum models have also indicated the possibility of ordered defect lattices with intriguing analogies to superconductors~\cite{doostmohammadi2016defect}. It is clear that}
more work is needed to sort out these conflicting results. The proper treatment of incompressibility, most likely the experimentally relevant case for ``wet'' systems \SR{under planar confinement}, remains a challenge in the modeling of defect dynamics.


\paragraph{\mcm{Defects in biology.}} \mjb{Topological defects are distinctive singularities in both the mathematical and physical sense, and as such  are natural places for functionalization or the generation of higher level structures. The unique local environment of a defect has been exploited in passive systems, using  place-exchange chemical reactions in which molecules bind directly and preferentially at defect sites \cite{stellacci2007divalent}. Similarly, biological structures have long been known to functionalize defects. The common \emph{Adenovirus40} (Ad40) has one fibril attached to each of the topologically minimal 12 pentons (pentagonal-shaped regions formed by proteins) of its viral capsid, creating 12 arms with which to latch on to target cells \cite{zandi2020virus}.}

\mjb{This field is flourishing.  There are now experimental demonstrations of the biological function of topological defects as loci of cell extrusion in cell layers~\cite{kawaguchi2017topological,saw2017topological,duclos2017topological}, seeds of multilayer formation in dense bacterial sheets~\cite{copenhagen2021topological,shimaya20213d} and mammalian cells~\cite{sarkar2021crisscross} and organizing centers of morphogenetic processes in multicellular organisms~\cite{maroudas2021topological}. }

\mjb{Dense monolayers of spindle-shaped Neural Progenitor Cells (NPCs) exhibit distinct nematic order with a finite concentration of $\pm 1/2$ defects. Nematic order is less pronounced in epithelial Madin-Darby Canine Kidney (MDCK) cell layers, but defects can also be tracked there. } In NPC layers  cells accumulate at $+1/2$ defects, leading to the formation of three-dimensional mounds -- a beautiful example of a nontrivial actively generated structure nucleated by a topological defect~\cite{kawaguchi2017topological}. At the $-1/2$ defects, in contrast, cells are depleted.  It was suggested that  an interplay between anisotropic friction and active stress is responsible for this source-sink behavior, but much remains to be clarified. In MDCK epithelia,  cells are vertically extruded at $+1/2$ defects and die -- a process known as apoptotic cell extrusion~\cite{saw2017topological}. Cell extrusion here seems to be driven \SR{not} by increased cell density (cell-crowding), but rather by the mechanical stresses associated with the texture of a defect.  That such a basic biological process as cell death is intertwined with the presence and formation of active nematic defects is both striking and powerful and could well have widespread applicability.


\mjb{Recent developments have also highlighted the role of nematic defects in the formation of multilayer structures in both bacteria~\cite{copenhagen2021topological,shimaya20213d} and NPCs~\cite{sarkar2021crisscross}. Especially intriguing is the suggestion put forward in Ref. \cite{sarkar2021crisscross} that the structure of nematic defects and the nature of active forces may together control the relative cell orientation in multilayered cell structures. Most of these experimental findings are at best qualitatively understood and await theoretical input. }



More recently it has been shown that in regenerating \emph{Hydra}, supra-cellular actin fibers also lead to active nematic order with significant morphological features developing at long-lived but dynamic $+1$ defects. Here the topology of the tissue is spherical and so, in the absence of topology change, the nematic texture is required to have a net topological charge of 2 fixed by the Euler characteristic of the sphere. This requirement has minimal solutions consisting of either four $+1/2$ defects, two $+1/2$s and one $+1$, or two $+1$s. Any additional number of defect-antidefect pairs also satisfy the constraint. In \emph{Hydra} two $+1$ defects form and nucleate the head and tail. This is interesting because the energy associated with defect formation goes as the square of the defect charge, hence the formation of a $+1$ defect  is energetically disfavored - the intrinsic elastic energy of a $+1$ defect is twice that of two $+1/2$s. Interactions between active $+1/2$ defects, however, lead to attractive forces that lead to tightly bound states of two $+1/2$ defects~\cite{shankar2018defect} and this state has the same far-field texture as a $+1$ defect.
Thus like-sign defect-defect attraction in active nematics is responsible for the head and tail of regenerating \emph{Hydra} -- amazing! Once again the distinctive environment in the neighborhood of an essential singularity drives fundamental morphological development. The quantitative exploration of the role of defects in morphogenesis is only just starting \cite{hoffmann2021defect}. \mjb{For instance, in the budding of tentacles in \emph{hydra}, do changes in the morphology of the actin supracellular structure with associated defect formation precede or follow cell accumulation at the budding site? And are there morphogenetic signaling proteins that drive cell accummulation?   Further work coupling theory and controlled experiments is needed to elucidate the interplay between structure, biochemical signaling, mechanics and curvature in controlling these processes.}

\mcm{Finally,  turbulent-like dynamics of scalar chemical fields associated with wave propagation patterns of signaling proteins was recently observed in the membrane of starfish egg cells~\cite{2020TzerHan_NatPhys}. The observed ``spiral defect chaos''  resembles spiral-wave patterns in reaction-diffusion systems and in the heart,  but the observation of such dynamics \emph{in vivo} and the identification of its role in biochemical signaling suggest the importantce  of biochemical patterns in controlling multicellular  organization and tissue mechanics in biological development. Understanding the statistics of such patterns in living systems and connecting them to the propagation of mechanical forces is an important area for exploration. }

\section{Harnessing active flows}
\label{sec:harness}

Passive liquid crystals have had enormous technological applications thanks to a detailed understanding of the role of boundary conditions and the response to external fields that allows remarkable control of these materials. The dream is to achieve equal control of active fluids so that self-organized active flows can be utilized, for instance, for microfluidic transport and tissue engineering, or for powering machines at the micro and nanoscale. Can we then achieve robust behaviors  by controlling  topology and geometry of active flows? \mcm{While we know that chaotic active flows can be tamed into stable laminar flows  through geometry and confinement~\cite{wioland2013confinement,duclos2018spontaneous}, we need to learn how to shape boundaries and vary boundary conditions to achieve specific control of flow in the bulk. Applications will require} a quantitative understanding of anchoring behavior, wetting and interfacial properties of active liquid crystals and their response to external fields. 

\begin{figure}[h]
    \centering
    \includegraphics[width=0.4\textwidth]{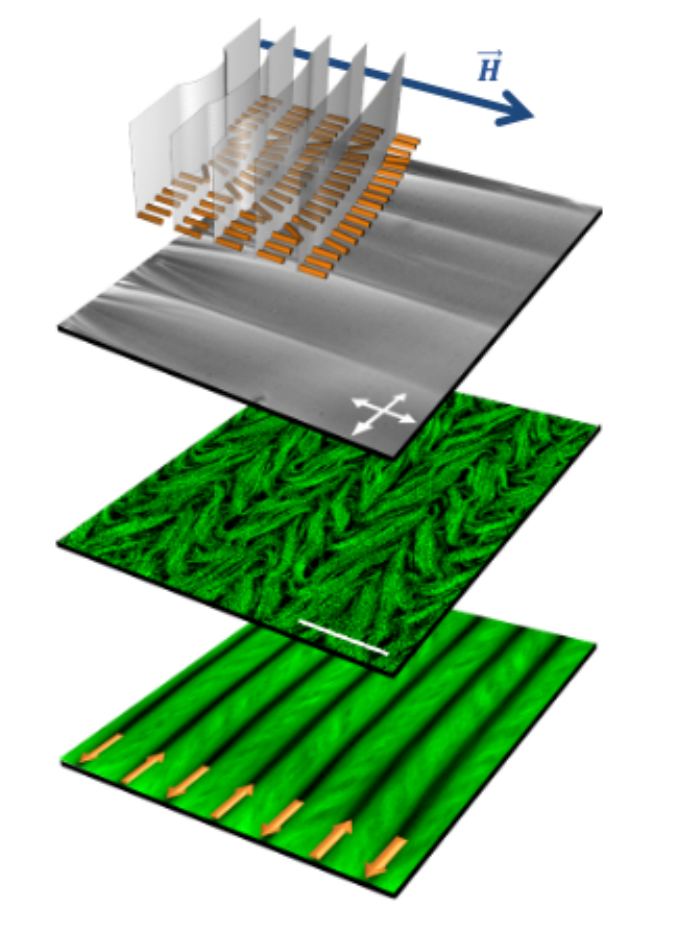}
    \caption{Aligning an active nematic with a magnetic field by interfacing it with a passive nematic layer. The figure, adapted from Ref.~\cite{guillamat2016control}, shows, from top to bottom, the configuration and an optical micrograph of the underlying passive liquid crystal aligned in the Smectic A (SmA) phase by application of a magnetic field (top);  a snapshot of the active nematic obtained by fluorescent confocal microscopy showing the alignment in ``kink walls''\mcm{(also  referred to ``arches'' in the literature)} induced by the coupling to the SmA structure of the passive layer,  and the time-average of the dynamical pattern in the active nematic layer, with arrows indicating antiparallel flow directions. (Scale bar: 100 $\mu$m)}
    \label{fig:Sagues}
\end{figure}

\paragraph{\mcm{Controlling flows with boundaries and geometry.}} While geometric confinement is relatively well explored, more recently it has been shown that spatial changes in Gaussian curvatures can   regulate specific defect structures and direct active flows~\cite{keber2014topology,sknepnek2015active,shankar2017topological,ellis2018curvature,henkes2018dynamical}.
The group of Sagu\'es has also demonstrated experimentally the possibility of achieving magnetic control of active liquid crystals by interfacing them with passive ones~\cite{guillamat2016control} (see Fig.~\ref{fig:Sagues}). The control of active fluids through patterned substrates is a promising direction where much remains to be explored both experimentally and theoretically.

\paragraph{\mcm{Controlling flows by patterning activity.}} An ambitious challenge is designing flows that cannot be achieved by  control of external boundaries. Progress in this direction has been achieved \mcm{by engineering active suspensions of cytoskeletal filaments crosslinked by optogenetically modified proteins, where the activity can be turned on and off with light illumination. This allows the  creation of controlled spatial patterns of active and passive regions within a given sample, and the temporal reconfiguration of such patterns with suitable light pulses~\cite{ross2019controlling,gong2020engineering,zhang2021spatiotemporal}.}  On the theoretical side, we are faced with the task of formulating a quantitative framework that will allow us to predict how we must sculpt activity in space and time to design specific active flows. This requires developing theoretical and experimental tools  to locally map out and quantify active stresses and connect stress to structure and flow.  Given active flows are directly coupled to topological defects in the orientational order, a complementary strategy focuses on using spatially inhomogeneous activity to confine and guide defects. Recent theoretical and experimental work has demonstrated that activity gradients engender pressure gradients that effectively act like electric fields on topological charge~\cite{shankar2019hydrodynamics,zhang2021spatiotemporal}. Further work is needed to understand how far the effect of activity extends both in length and time into the passive regions to achieve the design of emergent defect and flow states that exhibit both spatial and temporal organization. 
The ultimate challenge is to build materials that mimic biology, where the behavior is controlled by processes that are entirely internal to the system.  The next experimental step therefore will be to design a system that \mcm{can sense changes to flow patterns  and respond to such changes. }

\paragraph{\mcm{Other emerging strategies.}} Viscoelasticity of the suspending medium may also provide a mechanism for controlling active flows~\cite{saintillan2018rheology,emmanuel2020active}. Recent experimental and theoretical work has in fact shown that viscoelasticity not only can calm chaotic flows~\cite{bozorgi2014effects,hemingway2015active,hemingway2016viscoelastic,li2016collective}, but can also be used to simultaneously tune spatial \emph{and} temporal organization~\cite{liu2021viscoelastic}. This is distinct from the well-studied organization of interacting oscillators, as it provides a mechanism for the spontaneous organization of dissipative entities with no internal clock into macroscopic emergent states exhibiting sustained oscillations. 

Another intriguing direction is the use of active fluids for fluid-mediated computation strategies. The use of fluid networks for storing and transmitting information  is exploited in certain organisms, such as the slime mold \emph{Physarum polycephalum} which uses fluid networks to solve optimization problems, although the mechanisms through which it achieves this are yet to be understood~\cite{kramar2021encoding}. Pressure-driven microfluidic circuits have been employed to perform Boolean computation~\cite{prakash2007microfluidic}. The design of microfluidic devices powered by active fluids and capable of performing logical operations by exploiting the interplay of internal drive and constraints imposed by incompressibility has been suggested theoretically~\cite{woodhouse2017active}, but deserves further exploration.

\paragraph{\mcm{Role of theory.}} Finally, to answer many of the questions raised above, it will be necessary to carry out detailed coarse-graining to connect microscopic interactions to predictive descriptions at the macroscopic or continuum level. 
New approaches are being developed to translate large sets of experimental data on the microscopic dynamics of individual active entities into lower-dimensional models for the dynamics of a few coarse-grained fields. These range from mode decomposition approaches well-tested in condensed matter and turbulence theory~\cite{romeo2021learning} to inferring complex interactions  from the statistics of individual stochastic trajectories~\cite{bruckner2020inferring,bruckner2021learning}.
A new frontier is the use of machine learning for inferring the parameters of continuum models directly from experimental data. The power of this approach has been demonstrated for active microtubule nematics~\cite{colen2021machine} and bacteria~\cite{jeckel2019learning}, but promises to be a fruitful tool for modeling other active systems.

\section{Directing active assembly}
\label{sec:directing}

\begin{figure}[h]
    \centering
    \includegraphics[width=0.4\textwidth]{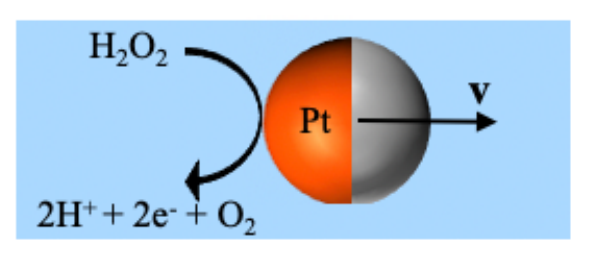}
    \caption{\mcm{The simplest example of an active colloid is a micron-size polystyrene bead half coated with platinum immersed in a hydrogen peroxide solution and powered by solvent concentration gradients that interact with the particle's surface. The platinum catalyzes the breakdown of H$_2$O$_2$ in oxygen and water, generating anisotropic concentration and charge currents that turn the micron-size bead into a swimmer. The speed of these synthetic swimmers is controlled by the concentration of H$_2$O$_2$ and is typically of order of tens of microns per second, comparable to that of flagellated bacteria like \emph{E. coli}.}
    }
    \label{fig:swimmer}
\end{figure}

Since the first catalytic microswimmer was engineered now almost 20 years ago~\cite{paxton2004catalytic} (see Fig.~\ref{fig:swimmer}), a broad class of ``machines'' powered by light or chemical sources has been developed, including more sophisticated  
catalytic colloids, enzymes, metabolic networks, Marangoni droplets, and Quincke rotors~\cite{Bechinger-RMP2018}.
In many of these, the swimming speed or degree of activity can be controlled with light or steered with external fields. 
\mcm{Thanks to the  unprecedented  control achieved in experiments, these active colloids, as they are broadly called,  provide an important testing ground for active matter theory and offer the potential for a variety of technological and biomedical applications, from microscale stirrers to targeted drug delivery~\cite{needleman2017active}. Fullfilling this potential will, however, require  a quantitative understanding of the propulsion mechanisms of these active particles and of their interactions, as well as the development of theories that connect  microscopic properties  to  emergent behavior.}


\paragraph{\mcm{Emergent organization of active colloids.}} Active colloids continuously draw energy from an ambient nonequilibrium medium, and their interactions are mediated by chemical phoresis and hydrodynamic flows~\cite{golestanian2005propulsion,Bechinger-RMP2018}. As a result, the \emph{effective} pair interactions are generally nonreciprocal~\cite{soto2014self,nasouri2020exact,you2020nonreciprocity,saha2020scalar}. Such nonreciprocity is especially important in active colloidal mixtures. Its consequences can be fascinating, and are discussed further below. Progress has been made in the classification of diffusiophoretic colloids by relating single-particle features and the symmetry and source of their pair interaction to the resulting variety of pair dynamics, such as bound dimers or orbiting pairs, versus scattering states ~\cite{Saha_2019}. An important open challenge is now establishing a quantitative connection between types of pair interactions and emergent behavior to formulate a classification of active colloids that relates their microscopic properties to their collective organization at large scale \mcm{and to map out the phase diagram of each class of particles.} Tackling this challenge will require systematic theoretical work closely coupled to experiments to relate the parameters of theoretical models to experimental ones.  

\begin{figure}[h]
    \centering
    \includegraphics[width=0.4\textwidth]{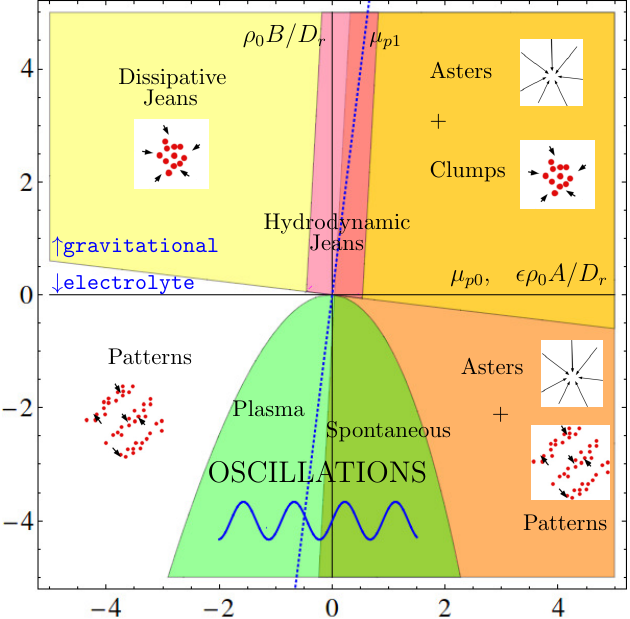}
    \caption{ Nonequilibrium phase diagram of polar active colloids in the plane spanned by adimensionalized ``chemotactic'' reorienting response coefficient $A$ \mcm{(horizontal axis)} and phoretic motility $B$ \mcm{(vertical axis)}. Reproduced from Ref. \cite{saha2014clusters}}
    \label{fig:colloid}
\end{figure}

Some progress has been made in this direction (Fig. \ref{fig:colloid}), but achieving a complete classification will require a better understanding of hydrodynamic interactions, \mcm{that is interactions mediated by fluid flow,} especially their role in driving coordination of swimmer orientation. For instance it is known that hydrodynamic interactions can hinder or arrest the motility-induced phase separation of scalar active active matter~\cite{matas2014hydrodynamic}. More generally, do flow-mediated couplings promote or hinder structure formation? What is the role of hydrodynamic interactions relative to  ones mediated by chemical fields? Can we even quantitatively distinguish the two? A quantitative understanding of the role of these competing mechanisms  is necessary in order to learn how to direct the assembly of such active particles to build smart and reconfigurable materials. A complementary approach, inspired by successes in metamaterial design,  is to tackle the inverse problem of configuring an active system and active interactions to obtain specific target structures  or target flows or to perform specific functions. This approach is still in its infancy \mcm{in the context of active matter}, although progress is being made by employing machine-learning methods~\cite{zhou2021machine}. This type of work may provide connections with methods developed in the robotics community.

\paragraph{\mcm{Retardation and memory effects.}} Most theoretical work so far has assumed that the dynamics of the colloidal particles is much slower than the time scale for diffusion of the chemical fuels that power it. In this limit the effective interactions generated by chemical fields (both self-interactions that drive the particle's motion and interparticle interactions) can be treated as instantaneous. Some work has considered the opposite limit of very slow chemical diffusion~\cite{gelimson2016multicellular}, but the intermediate situation likely to be relevant to many diffusiophoretic colloids where these time scales are comparable has not been addressed. 
Intermediate time scales of chemical diffusion can result in time-delayed interactions and memory effects, which may engender new emergent behavior, such as traveling and oscillating states.  A similar role is known to be played by viscoelasticity of the ambient medium. In both cases these effects act like an ``effective inertia'' on the dynamics, which could mediate emergent structures that exhibit \emph{both} spatial and temporal organization, as shown recently for bacteria swimming in a viscoelastic fluid~\cite{liu2021viscoelastic}. 

\paragraph{\mcm{Emergent organization from biochemical control.}} It is also tempting to draw an analogy between the chemically-driven dynamics of active colloids and biochemical patterns~\cite{2020TzerHan_NatPhys} as well as particle-like textures in a variety of active orientationally ordered systems \cite{husain2017emergent,sohn2019schools}.  Biochemical patterns are ubiquitous in biology and they control organization at both the subcellular and multi-cellular levels~\cite{2021Lecuit,2017Grill}. Spatiotemporal symmetry-breaking transitions in biochemical patterns are essential in triggering morphological changes during development, both at the unicellular and multicellular level. The realization of cell and tissue-scale deformations is achieved through intra-cellular force networks that translate localized biochemical signals into effective mechanical stresses that determine the global shape dynamics. Biochemical regulation has been studied extensively in the context of nonlinear chemical reaction networks~\cite{frey2020selforganisation}. It would be interesting to explore what biochemical and biophysical pattern formation, regulation and mechanochemical feeback mechanisms can teach us about active organization~\cite{Wigbers2021}. \mcm{The long term goal is to 
learn to regulate phoretic effects to control pattern formation in synthetic active systems.} \SR{The creation of polar flocks or apolar active liquid crystals made from anisotropic self-phoretic colloids is an interesting challenge.}


Finally, recent experiments have suggested the possibility of building active colloidal systems controlled by real-time feedback~\cite{khadka2018active,geiss2019brownian,lavergne2019group}. This is a first step towards the  design of responsive materials capable of sensing their environment and responding in organized and prescribed ways.  This work connects directly with advances made in the robotics community towards building smart flocks that can sense their environment \cite{dasgupta2020nanomotors} and adapt to it and has implications for the understanding of information flow ~\cite{loos2020irreversibility,2021Austin,Tang_2020}.

\section{Nonreciprocal interactions drive new emergent behavior}
\label{sec:NR}
Newton's third law establishes that pair interactions among parts of a mechanically isolated system are reciprocal: for every action there is an equal and opposite reaction. Such reciprocity applies to all systems where interactions can be derived from a Hamiltonian and governs all microscopic physical interactions.  
Reciprocity of interactions can also be seen as a consequence of detailed balance in a Markovian dynamics governed by a master equation. Just requiring a time-reversal-invariant steady state guarantees that the dynamics is downhill in the space described by a function of configurations that can be interpreted as energy. A natural definition of forces as gradients of this energy follows, ensuring reciprocity.  In physical systems nonreciprocity often emerges when effective interactions among mesoscopic parts of a system are mediated by a nonequilibrium medium, as in plasmas~\cite{ivlev2015statistical,kryuchkov2018dissipative} and mixtures of diffusiophoretic colloids~\cite{meredith2020predator,durve2018active}.
\mcm{Nonreciprocal interactions are ubiquitous in active and living systems that break detailed balance at the microscale,}
from social forces~\cite{helbing1995social,hong2011kuramoto} to promoter-inhibitor couplings
among cell types in developing organs and organisms~\cite{theveneau2013chase}, to antagonistic
interspecies interactions in bacteria~\cite{xiong2020flower} and prey-predator systems~\cite{tsyganov2003quasisoliton}.
\mcm{Understanding how nonreciprocity (NR) affects nonequilibrium phase transitions and  emergent states of active matter is a rapidly growing research focus in the field.}

\paragraph{\mcm{How do we define nonreciprocity?}} We believe it is important to first attempt to provide a  definition of nonreciprocity in a restricted context in which a notion of force operates and the evasion -- not violation of course -- of Newton's Third Law can be appreciated. Let us work with dynamical variables $\{x_a, a = 1, 2, \ldots, N\}$ with identical time-reversal signatures. These could be the positions or orientations or magnetic moments of $N$ particles labelled by $a$. Let us assume they undergo inertia-less dynamics governed purely by force balance in the presence of a dissipative medium with, for simplicity, a single damping constant $\Gamma$ and a noise $f_a$,  
\begin{equation} 
\label{eq:NR1}
\Gamma \frac{d x_a}{dt} = \sum_b F_{ab} +f_a\;,
\end{equation}
where $F_{ab}$ is the force on $a$ due to $b$.
If this were an equilibrium problem at temperature $T$, the noise $f_a$ would obey $\langle f_a(0) f_b(t) \rangle = 2 k_B T \Gamma \delta_{ab} \delta(t)$.
 The system is said to be nonreciprocal if $F_{ab} \neq - F_{ba}$. This is not a fundamental violation of Newton's Third Law, as the system is in contact with a damping medium. However, if the system were governed by an energy function that depended only on the relative values of the $\{x_a\}$, $F_{ab} + F_{ba}$ would necessarily be zero despite the presence of a medium that could take up the slack. Outside a context where the dynamics can be formulated in terms of forces or torques, a more general notion of nonreciprocity is still useful, in the form of an absence of $a \longleftrightarrow b$ symmetry in the sensitivities of $\dot{x}_a$ to changes in $x_b$: $dx_a/dt = C_{ab}x_b + \ldots$ with $C_{ab} \neq C_{ba}$ \cite{fruchart2021non}. We turn now to specific realizations of nonreciprocity that have been considered in the active matter context.

\paragraph{\mcm{Nonreciprocity and symmetry of the order parameter.}} The simplest description of phase separation of a binary mixture is through the classical Cahn-Hillard equations that describe the interdiffusive dynamics of two coupled conserved scalar concentration fields, $\phi_a$, for $a=A,B$, in one dimension~\cite{you2020nonreciprocity}. The evolution of each concentration  is governed by a $\phi^4$ field theory that allows for a spinodal instability according to Model B dynamics,
\begin{equation}
\partial_t\phi_a=\partial_x\left(\partial_x\frac{\delta F_a}{\delta\phi_a}+\kappa_{ab}\partial_x\phi_b\right)\;,
\label{eq:phia}
\end{equation}
with
\begin{equation}
F_a=\frac12\int_{\mathbf{r}}\left(\chi_a\phi_a^2+\frac{1}{6}\phi_a^4+\gamma_a(\partial_x\phi_a)^2\right)\;.
\label{eq:Fa}
\end{equation}
When the cross-mobilities of the two species are related by Onsager's relations \mcm{that guarantee the approach to a homogeneous state as required by diffusive equilibrium (here $\kappa_{ab}=\kappa_{ba}$),} such a system undergoes bulk phase separation from a homogeneous mixed phase into two coexisting phases, each rich in one of the species. Recent work has examined the effect of nonreciprocal cross-diffusion currents that cannot be derived from a free energy, leading to $\kappa_{AB}\not=\kappa_{BA}$~\cite{you2020nonreciprocity,saha2020scalar,frohoff2021suppression}. \mcm{Such a situation can arise, for instance, in mixtures of active colloids from the breaking of detailed balance in the microdynamics that is a defining feature of active systems. } It was shown that sufficiently strong nonreciprocity can set the phase-separated state into motion, resulting in traveling density waves that break orientational symmetry. This work has demonstrated that NR provides a generic mechanism for traveling, and possibly stable oscillatory  patterns, where the two components play a chase-and-run game with each other, eventually settling into a stable spatio-temporally modulated structure.  
NR can additionally arrest the phase separation, turning the spinodal decomposition into a Turing-type instability with length scale selection ~\cite{frohoff2021suppression}.
While the emergence of  \mcm{spatial and temporal patterns} is well known for instance in models of population dynamics with antagonistic or mutualistic reproduction rates~\cite{yanni2019drivers,curatolo2019engineering}, it is surprising in systems described by conserved scalar fields with purely diffusive dynamics \mcm{that intuitively is expected to give monotonic decay to a homogeneous state}. 

\begin{figure}[h]
    \centering
    \includegraphics[width=0.4\textwidth]{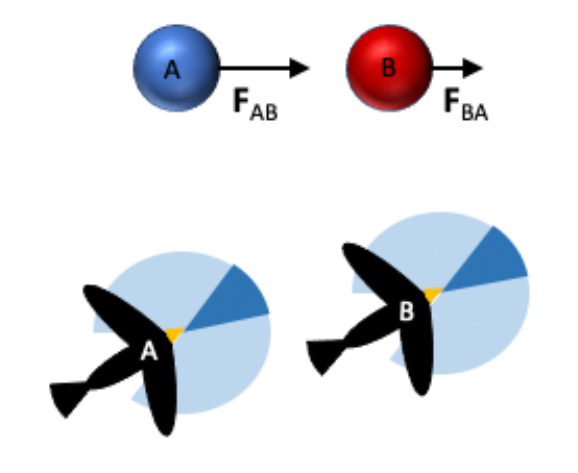}
    \caption{ Examples of nonreciprocal interactions. Top: active colloids can experience NR effective intercations due to  different surface reactivities. Here particle A is attracted to B, but B is repelled by A. Bottom: birds interact with other birds that are within their vision cone. Here bird A interacts with B which is within the vision cone of A, but B does not interact with A. }
    \label{fig:NR}
\end{figure}

Related work has examined the effect of nonreciprocity in a two-species Vicsek model characterized by coupled vector order parameters that encode the mean velocity of each species~\cite{fruchart2021non}. 
%
Nonreciprocity is introduced here by assuming that while species A aligns with both A and B, species B aligns with B, but antialigns with A. 
One then  finds a dynamical chiral state, with no equilibrium counterpart where the mean velocities of each species 
rotate either clockwise or counterclockwise, maintaining a constant phase difference - a vectorial analog
to the traveling states of phase separated scalar Cahn-Hillard fields~\cite{you2020nonreciprocity,frohoff2021suppression}. 

\mcm{In both examples discussed above NR introduces new  time-dependent collective states that  dynamically restore broken symmetries of the reciprocal system.  NR sets patterns of scalar fields into motion, effectively breaking polar symmetry, and endows antaginistically coupled polar flocks with handedness.  More generally, what are the consequences of NR in pattern-forming systems with scalar, polar or nematic symmetry? How does NR affect the emergent behavior and the nature of the phase transitions between states? Also to be explored is the role of boundary conditions that may play a special role in systems with nonreciprocal interactions.}
\mcm{Finally, the drift instability of scalar model described above, as well as the transition of the two-species Vicsek model appear to be examples of a general class of PT-breaking (P=parity, T=time reversal) transitions that appear  in open quantum systems with non-Hermitian Hamiltonian operators, resulting in time-periodic Floquet states~\cite{bender2007making,khemani2019brief,mcdonald2021non}. Exploring the connection between nonreciprocal active matter and non-Hermitian quantum mechanics  is an important emerging direction.}

\paragraph{\mcm{Nonreciprocity in active solids.}} The role of nonreciprocity in mesoscale effective interactions has also been highlighted in the context of the elasticity of materials where the interactions among the individual constituents are non-conservative~\cite{scheibner2020odd}. In this case nonreciprocity manifests itself in the form of antisymmetric (odd) elastic moduli that are needed to characterize the linear elasticity of isotropic solids, in addition to familiar shear and compressional moduli.  The presence of these odd response functions yields new nonequilibrium behavior, such as the ability of 
an inertia-less elastic solid to support elastic wave propagation, as well as auxetic response - when stretched, the material expands in the direction perpendicular to the applied force. Note that this last can arise in equilibrium systems as well, through a negative Poisson ratio. 

Even in the absence of chirality, and hence of odd elasticity, the effective pair interaction of orientable motile particles in an elastic medium is NR because it is mediated by the reorienting effect of the strain field that their motion generates. The result is a strongly non-mutual ``tactic'' \cite{o2020lamellar} interaction between two particles of the same type, distinguished only by which one lies ahead as defined by the direction in which it points. The particle at the rear acquires a purely mechanical ``stealth'' and can sense and move towards the one in front without signalling its presence~\cite{gupta2020active}. 

\paragraph{\mcm{Nonreciprocity, activity and information.}} 
%
It seems clear that nonreciprocity is generic in active matter, \mcm{but whether activity and nonreciprocity share the same fundamental origin}  is yet to be understood. Consider two coordinates $x_1$, $x_2$ in a system rendered active by maintaining a constant positive chemical potential difference $\Delta \mu$ between the reactants and products of a chemical reaction. Within the standard active-matter paradigm \cite{kruse2005generic,marchetti2013hydrodynamics,Ramaswamy_2017,dadhichi2018origins} active dynamics arises through chemomechanical cross-couplings $\zeta_i, \, i=1,2$: $\dot{x}_i = \zeta_i \Delta \mu + \ldots$ where the ellipsis denotes the passive part of the dynamics, and in general the $\zeta_i$ depend on $x_1,x_2$. Unless forbidden by additional symmetries, in general $\partial \zeta_i / \partial x_j \neq \partial \zeta_j / \partial x_i$ so the dynamics of $x_1,x_2$ should be nonreciprocal. 
Agent-based models like the iconic Vicsek model may be based on reciprocal interactions or alignment rules, but upon coarse-graining  are described by continuum equations  with macroscopic couplings that do not respect Onsager's relations, and hence break macroscopic nonreciprocity. 
Indeed, nonreciprocity rather than motility can be seen \cite{dadhichi2020nonmutual} to lie at the heart of the Vicsek/Toner-Tu models. If the orientation vector carried by a particle aligns more strongly with that of its neighbor ahead of it than with its neighbor behind it, where ahead and behind are defined with respect to the direction of the focal particle's orientation vector, the polar order parameter in the coarse-grained theory advects itself as if it were a velocity even if the particles are not motile \cite{dadhichi2020nonmutual}. The directed information transfer associated with this advective nonlinearity assures long-range order in two dimensions. \mcm{It would be useful to put the intuition described above on a firm footing by showing how nonreciprocity arises upon coarse-graining of reciprocal microscopic models.}

\mcm{Finally, NR may be the key to understanding directed information transmission in living systems, such as signalling in cell biology or communication in social environments. In synthetic active matter and robotics, NR interactions can be used for control of time-delayed feedback, memory and  information flow. The understanding of these processes is an emerging direction with far reaching imnplications from biology to engineering.}  

\section{Chiral active matter} 
\label{sec:chiral}
An object is conventionally termed chiral \cite{kelvin1894molecular,pasteur1848relations,lubensky1998chirality,harris1999molecular} if it cannot be superimposed on its image in a plane mirror by means of rigid motions. Chirality can enter through structure, as in a helix in three dimensions or a scalene triangle in two, or dynamically, as in a spinning object such as a rotary molecular motor. In $2D$ chirality is uniquely defined with respect to an axis normal to the plane in which the system lives, such as the clockwise (CW) or counterclockwise (CCW) sense of rotation of a spinner. In $3D$, however, the handedness of an object depends, quite literally, on one's point of view \cite{efrati2014orientation}. Indeed an object that is three-dimensionally \textit{achiral} in an absolute sense can nonetheless display chiral behavior in its dynamical response about a given axis \cite{dietler2020chirality}. Chirality is inescapable in biology, and living matter is active, so it is natural to explore the interplay of chirality and activity. 

\paragraph{\mcm{Chirality is an asymmetry}} \SR{Chirality describes} the \textit{absence} of a symmetry. Asymmetry, in active systems, begets spontaneous motion and governs its direction. 
\SR{A pragmatic approach to chiral active systems thus emerges, 
through time-reversal-breaking stresses and currents constructed from local fields and their gradients, with an odd number of appearances of the Levi-Civita tensor. 
Chirality pertains to spatial, and activity to temporal, asymmetry. One reason to study the two together is that in} passive systems chirality tends not to reveal its presence in long-wavelength mechanical properties -- for example, the elastic and hydrodynamic properties of cholesteric liquid crystals at equilibrium map exactly to those of smectics \cite{lubensky1972hydrodynamics,radzihovsky2011nonlinear}. In \textit{achiral} systems with \textit{translational} order, active forces introduce terms whose form superficially resembles those already present in the corresponding passive systems \cite{maitra2019oriented}, or created by static external fields \cite{adhyapak2013live,kole2020active}. Taken together, however, the effects of chirality and activity can reinforce each other, with surprising consequences. Some of these can be seen in the sampling of results below, but we expect much richness from this interplay in future studies of chiral active matter.

\paragraph{\SR{Turners and spinners: dry chiral active matter}} 
We begin with agent-based models without an ambient fluid. 
\SR{Motile particles whose heading turns at a constant rate 
\cite{lowen2016chirality,liebchen2017collective} provide a simple realization of chiral, polar active matter, 
displaying enhanced order and, for rapid enough turning rate, micro-phase separation} into coherently rotating domains with scale set by the turning radius. Persistently spinning particles allow the study of activity without translational motility or even an axis of alignment. The particles could be actuated by a motor \cite{vanZuiden12919} or a rotating field \cite{aragones2016elasticity}, in which case chirality is a consequence of rotation, or they could convert incoherent energy input into rotary motion by virtue of their chiral shape \cite{Tsai2005chiral}. These active but non-motile spinners give rise to a rich range of phenomena. Spinners in a dense passive monolayer display a long-ranged interaction whose character changes as the layer changes from fluid to solid \cite{aragones2016elasticity}, while two-dimensional crystalline phases of spinners display one-way propagating edge currents \cite{vanZuiden12919}. 

A potentially important question of principle in these dry chiral active systems concerns how much, and under what circumstances, the circle-walker system, which has a local polar order parameter but no long-range vector order, differs from the spinner system. How do properties vary upon tuning the turning direction from clockwise to counterclockwise as one crosses the point of infinite turning radius where a uniform flock intervenes? In a more speculative vein: can elementary active-matter models with chirality offer novel approaches to exploring chiral discrimination and proofreading -- life-or-death issues during protein synthesis in the cell \cite{ahmad2013mechanism}? In this connection, recent experiments \cite{arora2021emergent} on a mixture of left- and right-turning motile ellipsoids are noteworthy for placing in evidence an active mechanism for stereoselection, through the formation of a preponderance of motile achiral dimers and a smaller fraction of spinning chiral dimers. 

\paragraph{\SR{Living chiral fluids}} 
The description of `wet' chiral active systems \mcm{(see note of page 5)} originates in \cite{furthauer2012active} for bulk and \cite{furthauer2013active} for thin-film settings. Ref. \cite{furthauer2012active} considers suspensions of torque dipoles consisting of a pair of oppositely directed point torques perpendicular or parallel to their separation vector, as well as chiral force dipoles made of a pair of oppositely directed point forces perpendicular to their separation vector, compensated by a torque monopole to ensure zero total torque. The resulting extended hydrodynamic equations, including an angular momentum density, predict that the intrinsic rotation and the vorticity can differ even in steady state, and that a confined active chiral fluid with polar order can produce macroscopic \SR{shear \cite{markovich2019chiral}. 
The spontaneous rotation of chiral active fluids yields an escape route \cite{maitra2019spontaneous} from the generic instability of \cite{simha2002hydrodynamic}.}

Arguably the most far-reaching implications of \mcm{wet} chiral active hydrodynamics \SR{are for developmental biology, specifically 
on the emergence of macroscopic left-right asymmetry in the development of a multicellular organism.  
Experiments on the development of the nematode \emph{Caenorhabditis elegans}, informed by the coarse-grained theory of a thin film of chiral active fluid, have shown that counter-rotating flows arising from cytoskeletal stresses and torques at the scale of a cell lead to asymmetry at the scale of the embryo, in particular cell lineages \cite{naganathan2014active,pimpale2020cell}. Major open directions include identifying the (macro)molecular players responsible for the operative torque generation, and the mechanism that organizes these coherently to yield the macroscopic torques and flows.
}

\begin{figure}
    \centering
    \includegraphics[width=0.48\textwidth]{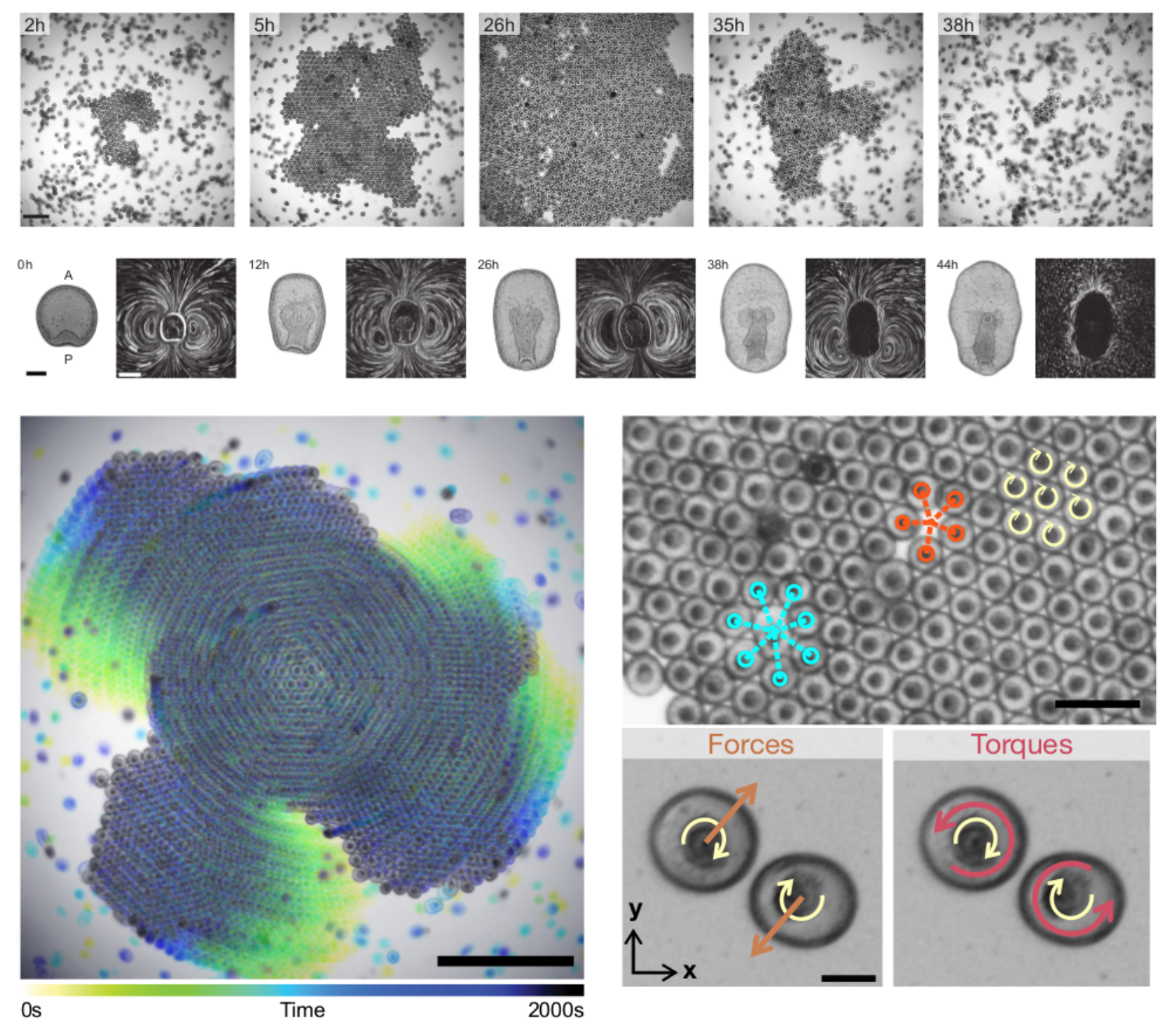}
    \caption{Developing starfish embryos self-organize into living chiral crystals. Time sequence of still images showing crystal assembly and dissolution (t = 0 hours corresponds to the onset of clustering, scale bar 1\,mm). Embryo morphology and flow fields change with developmental time (Shape scale bar, 100\,$\mu$m. Flow field scale bar, 200\,$\mu$m). Embryos assembled in a crystal perform a global collective rotation (Scale bar, 2\,mm). Spinning embryos (yellow arrows) in the crystal form a hexagonal lattice, containing 5-fold (red) and 7-fold (cyan) defects (Scale bar, 0.5\,mm). Spinning embryos exchange forces (brown arrows) and torques (red arrows) due to hydrodynamic interactions. Adapted from Ref.~\cite{tan2021development}}
    \label{fig:LLC}
\end{figure}

A dramatic example of chiral organization in biology 
is provided by recent experiments on dense collections of starfish embryos at an interface \cite{tan2021development}. 
Over the course of their natural development, thousands of swimming embryos come together to form living chiral crystal (LCC) structures that persist for many hours (see Fig.~\ref{fig:LLC}). The self-assembly, dynamics, and dissolution of these LCCs are controlled entirely by the embryos’ internal developmental program. Starfish embryos are inherently chiral, as they spin about their animal pole axis in a handed manner. When self-organized into a cluster, a fraction of each embryo’s torque is transferred to the whole crystal, resulting in a global cluster rotation. Perhaps more importantly, the chiral spinning motions also lead to transverse interactions and torque exchanges between embryo pairs. Because of the nonreciprocal nature of these interactions LCCs can support self-sustained chiral waves and shear cycles, similar to those recently predicted in odd elastic materials, \nf{providing evidence for the importance of nonreciprocality in multi-organismal living matter. Since many multicellular systems naturally break time-reversal and chiral symmetries in a manner similar to the starfish embryo system, this work can open up an exciting avenue in search of odd properties in biophysical systems.}

\paragraph{\mcm{Translational order, bulk versus boundary and dimenionality.}} 
Odd elasticity \mcm{of active \emph{solids}}~\cite{scheibner2020odd} is an especially dramatic manifestation of the combined effects of chirality and activity. \SR{A linear elastic tensor $C_{abcd}$ relates stress to strain: $\sigma_{ab} = C_{abcd} U_{cd}$, 
but the absence of an energy function liberates $C_{abcd}$ from the constraint of symmetry under $ab \leftrightarrow cd$}. 
\SR{Among the consequences are work extraction from the active solid in quasistatic cycles, and propagating} modes in the nominally inertia-free regime.

Active chiral systems with \textit{one-dimensional} translational order -- active cholesterics \cite{kole2020active} -- display a unique nonreciprocal effect: gradients of layer curvature evoke a response in the perpendicular in-plane direction, like an “odd” Laplace pressure gradient. Whereas the odd elastic force density of two-dimensional chiral active \textit{solids} reflects an antisymmetric contribution to the linear relation between stress and strain, this force density arises even when the strain is zero. If such a system goes through a Helfrich-Hurault \cite{helfrich1970deformation,hurault1973static} undulational instability or its active counterpart \cite{adhyapak2013live,whitfield2017hydrodynamic}, this effect, odder than odd elasticity, produces a columnar array of fluid-flow vortices, with an ``antiferromagnetic'' spatial pattern of vorticity \cite{kole2020active}.


In these translationally ordered systems the combined effects of chirality and activity manifest themselves in the bulk. By contrast, the most striking features of chiral active \textit{fluids} seem to lie in their edge modes \cite{van2016spatiotemporal}. Understanding the fundamental reason for this contrast in behaviors, and an exploration of possible connections to the expulsion of chirality to the edge in a layer of three-dimensionally chiral particles \cite{gibaud2017achiral} are interesting open directions. 

Lastly, studies of chiral active systems have largely focused on two dimensions, where chiral effects can be viewed in the simple CW/CCW dichotomy which links naturally to persistent currents. Understanding how activity and chirality combine in three-dimensional systems, especially in the absence of a preferred direction with respect to which to project to two dimensions, is a challenge. It is clear that chirality provides unusual opportunities for the manifestation of active effects on large spatial scales, and that explorations of the interplay of activity and chirality will be a major theme in the study of active systems for years to come. 

\section{Thermodynamics of active matter}
\label{sec:thermo}

Life’s ability to exploit energy across scales is remarkable. Living systems, operating far from equilibrium, can harness energy at the molecular scale through ATP hydrolysis and dissipate it on much larger spatiotemporal scales. 
Energy dissipation results in emergence of self-organized structures that span the entire length of a cell, such as the actomyosin cortex~\cite{chugh2018actin,tan2018self} or mitotic spindle~\cite{sawin1992mitotic,brugues2014physical}. On the scale of populations, suspensions of bacteria, cells in tissues and flocks of birds can form remarkable swirling patterns, due to their nonequilibrium dynamics. \nf{These examples and all active matter systems in general are ``open'' from the thermodynamic point of view. How do information, energy and entropy flow and transform due to interactions with the system's environment or within the system itself? 
 \SR{
Efforts towards a fundamental understanding of the physical and information-theoretic dynamics of these systems, and its exploitation to discover novel design principles, are another exciting frontier in active matter.}
}

\nf{The current efforts are \mcm{focused on answering two questions:} how far active matter systems are from equilibrium and what can we do far from equilibrium.}

\paragraph{\nf{How far from equilibrium is active matter?}}
\nf{Work on nonequilibrium thermodynamics has had major successes in building the thermodynamics of far-from equilibrium systems, with universal fluctuation theorems and other model-free results that deeply constrain the probability distributions for quantities like applied or extracted work and entropy production~\cite{seifert2012stochastic}. In recent years progress has been made in applying these concepts to active matter.}

\nf{An important challenge is developing measures of dissipation and irreversibility that allow us to distinguish between active and passive systems, as well as to quantify the difference between active systems that are driven out of equilibrium by internal processes and systems driven out of equilibrium by externally applied forces and perturbations. Quantifying dissipation will open new avenues for probing self-organization principles in these far-from-equilibrium systems. Below we  introduce three frameworks from nonequilibrium statistical physics that are promising candiadtes for providing insight for this very exciting endeavor.}

\paragraph{\nf{Departure from fluctuation dissipation theorem.}}
A consequence of the time-reversal symmetry of equilibrium is the fluctuation-dissipation theorem (FDT). Put simply, for a small perturbation at frequency $\omega$ the system’s response will be linear and completely characterized by the generalized susceptibility $\chi''(\omega)$. Near equilibrium, time-reversal symmetry relates this response to the intrinsic thermal fluctuations characterized by the power spectrum $S(\omega)$ (Fourier transform of a correlation function) through the FDT:
\begin{equation}\label{eq:FDT}
S(\omega)=\frac{k_{\rm B}T}{\omega}\chi''(\omega).
\end{equation}

Here, $k_{\rm B}$ is Boltzmann's constant and $T$ is the temperature of the surroundings. A major consequence has been in refining our understanding of the material coefficients that determine how spatial inhomogeneities in near-equilibrium macroscopic systems relax via hydrodynamic transport. The resulting predictions, known as Green-Kubo relations, equate these macroscopic transport coefficients to the microscopic equilibrium correlation functions of local current observables. This is central to our theoretical description of weakly nonequilibrium systems and underlies a number of experimental techniques for probing materials properties, such as microrheology and light scattering.
\nf{Active systems, however, are nonequilibrium and far from equilibrium the FDT becomes an inequality. One of the interesting directions is identifying classes of perturbations whose response verifies an equilibrium-like fluctuation-response equality. This in principle will allow us to extract linearized hydrodynamic transport equations around homogenous nonequilibrium steady-states~\cite{chun2021nonequilibrium,han2020statistical}.}

\nf{A signature of every nonequilibrium system is current, for instance heat flux down a temperature gradient. Recently, a new kind of nonequilibrium principle - a thermodynamic uncertainty relation - has been proposed that demonstrates how energy dissipation continues to constraint current fluctuations far from equilibrium~\cite{2020Horowitz,2019Li}. This novel principle relates the heat dissipated and the variance of the current fluctuations, \mcm{offering a remarkable bound on  response coefficients out of equilibrium akin to the equilibrium} fluctuation-dissipation theorem. \mcm{The thermodynamic uncertainty relation} has been implemented to quantify work and dissipation cycles within emergent strain waves in chiral active matter~\cite{tan2021development}. \mcm{Exploring its implications in other active context and formulating other general relations of this type are important open challenges.}}


Since all equilibrium systems satisfy the FDT, an observed departure in the absence of visible external forcing signals an underlying active process. This basic principle has been utilized to reveal nonequilibrium activity~\cite{mizuno2007nonequilibrium}. 
Often, deviations from the FDT are characterized by introducing a frequency-dependent effective temperature through the relation (cf.~Eq.~\eqref{eq:FDT}):
\begin{equation}
T(\omega)=\frac{1}{k_{\rm B}}\frac{\omega S(\omega)}{\chi''(\omega)}.
\end{equation}

One might suspect that this relation reveals frequencies (timescales) relevant to the nonequilibrium behavior~\cite{Wang2016entropy,dadhichi2018origins}. There is, however, no general principle that allows us to use this quantity to investigate the underlying microscopic mechanism. Insight is usually gained in this context in conjunction with modeling~\cite{mizuno2007nonequilibrium,Levine2009,prost2009generalized,fakhri2014high}. \SR{Note that $T(\omega)$ can even be negative \cite{dinis2012fluctuation}.}

\nf{Thus, the development of new model-independent frameworks that can be used to identify the scales of energy dissipation is crucial for a mechanistic understanding of nonequilibrium processes in active matter systems.}

\paragraph{\nf{Thermodynamic arrow of time.}} 
Nonequilibrium thermodynamics makes precise quantitative predictions about how time-reversal asymmetry and energy utilization (or dissipation) are manifested in nonequilibrium fluctuations ~\cite{seifert2012stochastic}. 
Some of this theory’s most prominent results are the fluctuation theorems, a collection of symmetries of the fluctuations of thermodynamic quantities such as the heat flow between a system and its environment.
\nf{The fluctuation theorems have proven to be a very powerful tool to gain information from small systems where traditional thermodynamics \mcm{does not apply \cite{Liphardt1832}.}  Whether these frameworks can be applied to multiscale, complex systems such as active matter to provide novel insight is yet unkown and an exciting direction~\cite{2019Pietzonka}.} 

A cornerstone of stochastic thermodynamics is a quantitative connection between physical energy dissipation and the statistical irreversibility (arrow of time) of the dynamics.

\begin{figure}
    \centering
    \includegraphics[width=0.48\textwidth]{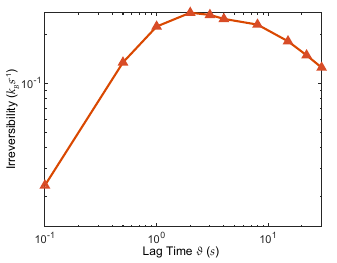}
    \caption{Experimentally measured irreversibility metric (see Eq.~\eqref{eq:relent}) as a function of measurement frequency shows a peak around the timescale of nonequilibrium activity. Adapted from Ref.~\cite{Tan2021KLD}.}
    \label{fig:KLD}
\end{figure}

To be specific, imagine we make a sequence of measurements of a physical observable every $\tau$ seconds, and collect them into a list, or trajectory, $\gamma_\tau=(x_\tau,x_{2\tau},\dots x_{n\tau})$.  These could be anything from the position of a particle to densities. Thermal fluctuations will make the measurements noisy, and in each realization of this experiment we obtain a different sequence of outcomes, which we characterize with a probability distribution ${\mathcal P}(\gamma_\tau)$. In fact it is possible, if not uncommon, to observe a previous sequence in the exact reverse order, $\tilde\gamma_\tau=(x_{n\tau},\dots, x_{2\tau}, x_{\tau})$, which occurs with probability ${\mathcal P}(\tilde\gamma_\tau)$. Stochastic thermodynamics teaches  us that these fluctuations actually are not just incoherent noise, but in fact constrain the physical heat dissipation (or energy consumption) ${\dot Q}_{\rm diss}$ of the nonequilibrium system. Formally, we compare the probabilities of observing any sequence of measurements and its reverse through a information-theoretic metric of distinguishability~\cite{Kawai2007, Roldan2012}, which is called the relative entropy rate, ${\dot D}_\tau=\lim_{n\to\infty}(1/n)\int {\mathcal P}(\gamma_\tau)\ln[{\mathcal P}(\gamma_\tau)/{\mathcal P}({\tilde\gamma}_\tau)]{\mathcal D}[\gamma_\tau]$:
\begin{equation}\label{eq:relent}
{\dot Q}_{\rm diss} \ge  k_{\rm B}T{\dot D}_{\tau}\;.
\end{equation}

where $k_{\rm B}$ is Boltzmann's constant and $T$ is the temperature of the surroundings.
As the relative entropy ${\dot D}_\tau$ measures how distinguishable the processes is from its reverse, we call it the \emph{irreversibility}. In other words, it quantifies the direction of the arrow of time. Equation~\eqref{eq:relent} reveals a fundamental  relationship between how irreversible a processes is---obtained from passive measurements of the dynamics---to the rate of energy consumption. This applies to any observable and nearly any nonequilibrium steady state. The irreversibility metric has proven to be a robust experimental and computational tool to detect nonequilibrium activity even in the absence of observable flows~\cite{Kawai2007,Parrondo2009,Gomez-Marin2008a,Horowitz2009b,Roldan2012,roldan2018arrow,Martinez2018}.

\nf{But can we apply this powerful framework to complex many body active matter systems?
This is a very exciting direction and very recently it has been demonstrated experimentally that one can extract quite a bit more \mcm{information} from this fundamental principle by using a new method, namely multiscale statistical irreversibility, which \mcm{can yield} the scales of energy consumption in active systems. It is observed that the measured irreversibility changes with measurement frequency~\cite{Tan2021KLD},in a manner correlated with a characteristic time scale of the underlying energy consuming process (See Fig. \ref{fig:KLD}). Thus, by analyzing statistical irreversibility over different time and length-scales we can discern the characteristic features of how energy is used in active matter without building a model.} 
And this is just the beginning. Correlating such data-analysis techniques with observations of function and structure can offer a principled method to characterize energy dynamics in complex matter. This also raises the question whether there is a scale at which dissipation is maximum and whether it is possible to use effective equilibrium descriptions beyond these scales~\cite{2016Fodor,Nardini2017}. 

\paragraph{\nf{What we can do far from equilibrium?}}
\nf{Can we push beyond these quantification of distance from equilibrium toward exploiting time-varying interactions, fluctuations, phase space structures in novel ways to enable the generation of useful work and engineered energetic and entropic transformations~\cite{2019Tociu}\mcm{?}} 

\nf{
Some recent efforts focus on exploring what types of steady states are efficiently assembled through engineered dissipation~\cite{Fodor_2020}.
How does dissipation results in self-organization and maintenance of spatiotemporal patterns in active matter~\cite{2018Falasco}?
Can we control activity to enable optimal nonequilibrium environmental energetic and information transfer? 
Does dissipation  engineering  enable programmable active matter? This will be a first step towards developing thermodynamically efficient methods for actively modulating the phase-space structure of active matter systems to enable adaptive control and learning.}

Looking forward, it will be exciting to see how these frameworks can provide a quantitative understanding of how thermodynamics dictates structure in active matter and function in living systems.
\\
\section{From Living Cells to biological  tissue}
\label{sec:cells}

Living cells are active entities capable of a number of self-driven mechanical functions, such as shape changes, motility and division. Through interactions and coupling to the environment they assemble into biological tissue, forming organs and organisms with entirely new emergent behavior. An important open question in active matter physics is  developing a predictive continuum theory of living matter that relates subcellular and cell-scale processes to adaptive mechanics at the tissue scale.   A key challenge is relating the coefficients of the sought after continuum models to both the parameters of mesoscopic  models, such as Vertex, Voronoi or multi-phase field models, and to quantities controlled in experiments. A measure of success of the theory will be its ability to identify classes of  molecular signaling mechanisms that define specific effective material parameters capable of characterizing  behaviors at the organ and organism scale. 

\paragraph{\nf{Coupling between multiple fields underlies dynamics of tissues.}}
\mcm{An approach to developing such a theory finds its inspiration in a successful paradigm in developmental biology. The essence of this  }  paradigm  is  the notion that spatial and temporal concentration patterns of diffusable chemicals known as morphogens specify the organization of cells into emergent structures~\cite{rogers2011morphogen,2021Lecuit,2017Grill}. In recent years the traditional notion of scalar concentration fields has acquired broader scope as it has become evident that organization at the tissue scale can often be described in terms of the dynamics of continuum \emph{supracellular} fields that often include orientational degrees of freedom and define the tissue's mechanical \SR{behavior~\cite{prost2015active,streichan2018global,morris2019active}. An} important example is Planar Cell Polarity that describes the tendency of epithelial cells to polarize due to anisotropic protein distribution within a given cell~\cite{struhl2012dissecting}. Cell-cell interactions can then  coordinate such polarity at the tissue level, resulting in large scale tissue anisotropy usefully described in terms of spatial and temporal variation of a continuum vector field. Such an approach provides the opportunity to take advantage of the machinery of the physics of flocking, and it acquires quantitative power when the continuum field can be related to specific cellular processes. 

\begin{figure}
    \centering
    \includegraphics[width=0.48\textwidth]{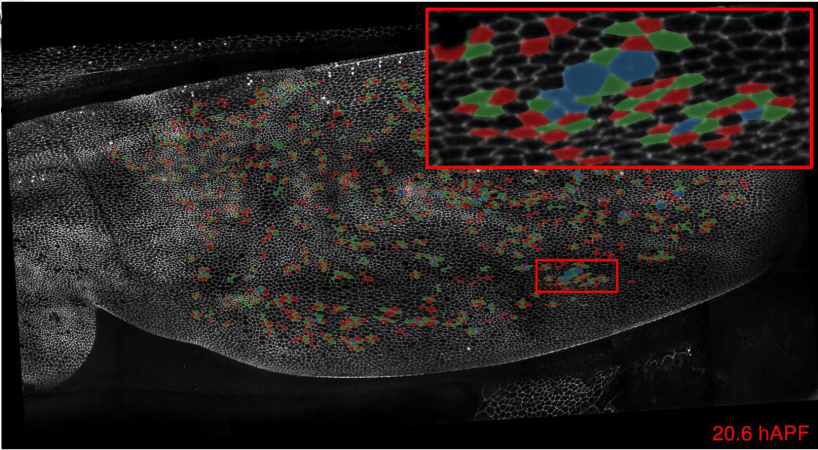}
    \caption{Morphogenesis of the \emph{Drosophila} pupal wing. The cells in the tissue undergo shape changes, cell divisions, cell rearrangements and cell extrusions during wing morphogenesis. Colors denote cell contact dynamics during tissue morphogenesis. Adapted from Ref.~\cite{2016Eaton}}
    \label{fig:Tissue}
\end{figure}

\mcm{Another application of these ideas to living matter is motivated by the recent } observation of nematic order in a variety of biological settings, \mcm{as discussed in Section~\ref{sec:defects}.} In some cases, as in layers of spindle-shapes progenitor neural cells~\cite{kawaguchi2017topological}, the individual cells are clearly elongated and nematic textures can readily be associated with the arrangements of cellular shapes, \mcm{suggesting that orientational order may be driven by crowding through the interplay of steric and entropic effects, much like in equilibrium.} In other systems, such as the \emph{Drosophila} embryo~\cite{streichan2018global} or the freshwater organism \emph{Hydra}~\cite{maroudas2021topological}, nematic order is evident in the organization of \emph{supracellular} myosin or actin fibers. \mcm{Less clear are the origin and signature of nematic order in  epithelial layers, such as MDCK cells~\cite{saw2017topological}. In general, more work is needed to understand what may be the mechanical or biochemical processes that drive and control the formation of nematic textures and the biological role of such orientational order.}


\paragraph{\nf{Forces shaping tissues: from intracellular to cellular.}}
A related question concerns the nature of active forces in biological tissue. Individual cells crawl on substrates by generating contractile active stresses through the actomyosin machinery of their cytoskeleton. Recent experiments have indicated, however, that when such cells organize in  confluent epithelia, interactions mediated by e-cadherins result in pulling forces exerted by cells on their neighbors, with extensile stresses at the tissue scale~\cite{balasubramaniam2021investigating}. What are the relative roles of polar traction forces exerted by cells on a substrate  or the surrounding medium and  cell-cell ``tractions’’ in controlling  the extensile/contractile and polar/apolar nature of  active stresses in tissue and the resulting modes of collective cell dynamics?  

\paragraph{\nf{Spatiotemporally varying material properties.}}
\mcm{A} key limitation of the continuum modeling approach lies in the assumption of fixed material properties of tissues,
which is encoded in the choice of a particular constitutive law. Tissues are able to adapt their mechanical response to
perturbations (both external and internal) and are characterized by multiple relaxation times. This demands a rheological
model capable of capturing both active solid-like and fluid-like behavior in different regimes of stress response and to
dynamically transition between the two. In other words any rheological model of tissue mechanics must incorporate the active feedbacks between cellular mechanics,
polarized motility, and the regulatory biochemistry of actomyosin contractility. These couplings play an essential role in the
transmission of spatial information in large cell monolayers, which are often mediated by travelling waves, pulses, and a tug of war between cell-cell
and cell-substrate forces~\cite{serra2012mechanical,alert2020physical}.
While some recent progress has been made on incorporating these couplings in continuum models~\cite{duclut2021nonlinear}, informed by  studies of mesoscopic models and by experiments, formulating an adaptive rheological model of tissue remains an open challenge. This is further complicated by the fact that living tissue is also capable of adapting its mechanical state in response to changes in environment, through feedback loops in a way that has so far largely eluded predictive theoretical descriptions. 

\paragraph{\nf{Form meets fuction.}}
There are also situations where epithelial cells organize in remarkable orderly patterns that seem to be essential to the functioning of many tissues. Examples are photoreceptor cells in the eyes~\cite{pujic2004retinal}, the hexagonal cell packings in the wing of developing \emph{Drosophila}~\cite{classen2005hexagonal} (see Fig. \ref{fig:Tissue}), and the remarkable rectangular cell lattice observed in the development of the freshwater shrimp \emph{Parhyale}~\cite{sun2019amphipod}. The mechanisms that control the  development of such regular, ordered epithelial cell packings remain only partially understood. Do protein anisotropies existing at the single-cell or subcellular level control the emergence of ordered structures, as for instance suggested in models of the eye retina~\cite{salbreux2012coupling},  or do  regular cellular arrangements emerge spontaneously from cell-cell interactions? What is the role of growth and growth anisotropy in organizing ordered cell packings?
\mcm{Answering these questions may also help inform and guide new pathways of active assembly for the design of fucntional materials.}

\paragraph{\nf{Beyond  broken symmetries.}}
Finally, in condensed matter physics  the notions of broken symmetry and conservation laws provide powerful principles for the identification of  coarse-grained fields that allow the formulation of  predictive continuum descriptions of complex phenomena. So far the active matter community has largely been using the same ideas to formulate continuum models of active and living systems~\cite{prost2015active}. But living matter develops, divides, repairs itself, adapts to its environment, and evolves to perform specific functions. Its hierarchically organized constituents often compete for fixed pools of resources: for instance, in wound healing, the same actin pool may drive protrusive cell motility and cell contraction~\cite{ajeti2019wound}). Can we identify general principles that may guide us in constructing field theories for this more complex type of matter, where spatial and temporal responses are often coupled and feed back onto each other? 
{How do we build in to our coarse-grained theories the fact that we are dealing with systems that have emerged from an evolutionary process, that they correspond to evolutionarily stable strategies? For example: when constructing a generic model of a physical system, one would never insist that parameter values should be poised at a threshold separating two qualitatively distinct behaviors. Such phase-transition points correspond to unstable fixed points of the renormalization group, so that this parameter choice would amount to non-generic fine-tuning. But evidence from diverse living systems \cite{camalet2000auditory,bialek2012statistical,krotov2014morphogenesis,milewski2017homeostatic} suggests that tuning of this type may be an emergent result of evolution. What is the influence of noise on the regulatory feedbacks that control cellular organization?}\\

\section{The Future is Active}
\label{sec:outlook}

The field of active matter continues to evolve rapidly and to establish connections and relevance to many areas of science. In this article we have presented a biased selection of topics where we expect significant progress will be made in the coming years. 
\mcm{There are many other important emerging directions that have been omitted. An example is  the study of the interplay of motility and information transmission with   models of active agents that change their state upon interaction with each other~\cite{paoluzzi2020information,norambuena2020understanding,zhao2021contagion}, as relevant for instance to epidemics spreading.
Another is systems in which the active forcing can be viewed as maintaining temperature difference between two species of particles ~\cite{ganai2014chromosome,joanny-2015}.}

The field of active matter started out by marrying the fluid dynamics of swimmers with the field theory of pattern formation and phase transitions. It has now acquired its own identity as a powerful framework for the description of spontaneous organization in both the living and the engineered worlds on a vast range of scales. The role of topology has become evident in active systems and has pointed to the possibility of deep connections with the statistical physics of open quantum systems~\cite{shankar2020topological,fruchart2021non}. 

\begin{figure}
    \centering
    \includegraphics[width=0.5\textwidth]{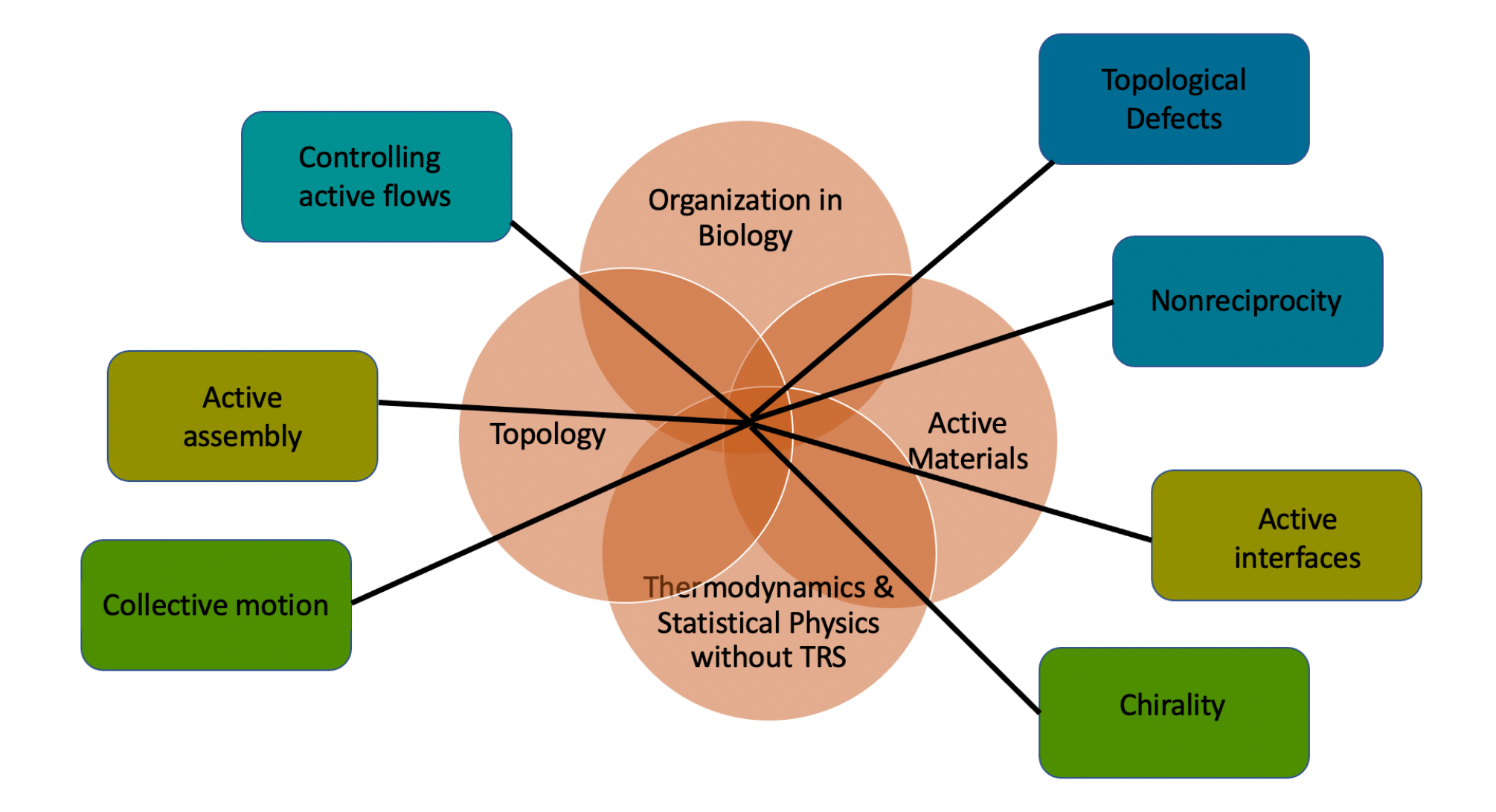}
    \caption{\mcm{A pictorial depiction of the connections among the various aspects of active matter physics discussed in this article.}}
    \label{fig:chart}
\end{figure}

\mcm{It has also become evident that the  topics discussed here are all interconnected, as displayed in Fig.~\ref{fig:chart}, with many of these connectinons still awaiting  quantitative exploration. 
The field of active matter was born from the physicist's ambition to use statistical physics and hydrodynamics to describe collective motion in the living world. The active matter framework has now had important successes in capturing examples of organization in living matter on scales from subnuclear to oceanic. }
The overall dream is to develop a predictive theory that will allow us to design the hierarchical organization of active agents or ``machines'' into larger scale machines tuned to perform specific functions and to adapt to the task at hand. While nature does this every day, we are still far from achieving this goal, but continue to make consistent progress.

\begin{acknowledgments}
The authors thank all participants \mjb{in} \emph{Active20}, the KITP Program on \emph{Symmetry, Thermodynamics and Topology in Active Matter}, for their contribution to this article through presentations and discussions.  The KITP program was supported by the NSF through grant PHY-1748958. The authors received additional support from the National Science Foundation under Grant No.~DMR-2041459 (MCM), the National Science Foundation through the Materials Science and Engineering Center at UC Santa Barbara, DMR-1720256 (iSuperSeed) (MCM and MJB). NF acknowledges National Science Foundation CAREER award, PHYS-1848247. SR acknowledges support from a J C Bose Fellowship of the SERB, India, and from the Tata Education and Development Trust.
\end{acknowledgments}

%


\begin{thebibliography}{236}%
\makeatletter
\providecommand \@ifxundefined [1]{%
 \@ifx{#1\undefined}
}%
\providecommand \@ifnum [1]{%
 \ifnum #1\expandafter \@firstoftwo
 \else \expandafter \@secondoftwo
 \fi
}%
\providecommand \@ifx [1]{%
 \ifx #1\expandafter \@firstoftwo
 \else \expandafter \@secondoftwo
 \fi
}%
\providecommand \natexlab [1]{#1}%
\providecommand \enquote  [1]{``#1''}%
\providecommand \bibnamefont  [1]{#1}%
\providecommand \bibfnamefont [1]{#1}%
\providecommand \citenamefont [1]{#1}%
\providecommand \href@noop [0]{\@secondoftwo}%
\providecommand \href [0]{\begingroup \@sanitize@url \@href}%
\providecommand \@href[1]{\@@startlink{#1}\@@href}%
\providecommand \@@href[1]{\endgroup#1\@@endlink}%
\providecommand \@sanitize@url [0]{\catcode `\\12\catcode `\$12\catcode
  `\&12\catcode `\#12\catcode `\^12\catcode `\_12\catcode `\%12\relax}%
\providecommand \@@startlink[1]{}%
\providecommand \@@endlink[0]{}%
\providecommand \url  [0]{\begingroup\@sanitize@url \@url }%
\providecommand \@url [1]{\endgroup\@href {#1}{\urlprefix }}%
\providecommand \urlprefix  [0]{URL }%
\providecommand \Eprint [0]{\href }%
\providecommand \doibase [0]{http://dx.doi.org/}%
\providecommand \selectlanguage [0]{\@gobble}%
\providecommand \bibinfo  [0]{\@secondoftwo}%
\providecommand \bibfield  [0]{\@secondoftwo}%
\providecommand \translation [1]{[#1]}%
\providecommand \BibitemOpen [0]{}%
\providecommand \bibitemStop [0]{}%
\providecommand \bibitemNoStop [0]{.\EOS\space}%
\providecommand \EOS [0]{\spacefactor3000\relax}%
\providecommand \BibitemShut  [1]{\csname bibitem#1\endcsname}%
\let\auto@bib@innerbib\@empty
\bibitem [{\citenamefont {Vicsek}\ \emph {et~al.}(1995)\citenamefont {Vicsek},
  \citenamefont {Czir{\'o}k}, \citenamefont {Ben-Jacob}, \citenamefont
  {Cohen},\ and\ \citenamefont {Shochet}}]{vicsek1995novel}%
  \BibitemOpen
  \bibfield  {author} {\bibinfo {author} {\bibfnamefont {Tam{\'a}s}\
  \bibnamefont {Vicsek}}, \bibinfo {author} {\bibfnamefont {Andr{\'a}s}\
  \bibnamefont {Czir{\'o}k}}, \bibinfo {author} {\bibfnamefont {Eshel}\
  \bibnamefont {Ben-Jacob}}, \bibinfo {author} {\bibfnamefont {Inon}\
  \bibnamefont {Cohen}}, \ and\ \bibinfo {author} {\bibfnamefont {Ofer}\
  \bibnamefont {Shochet}},\ }\bibfield  {title} {\enquote {\bibinfo {title}
  {Novel type of phase transition in a system of self-driven particles},}\
  }\href {\doibase 10.1103/PhysRevLett.75.1226} {\bibfield  {journal} {\bibinfo
   {journal} {Physical Review Letters}\ }\textbf {\bibinfo {volume} {75}},\
  \bibinfo {pages} {1226} (\bibinfo {year} {1995})}\BibitemShut {NoStop}%
\bibitem [{\citenamefont {Toner}\ and\ \citenamefont
  {Tu}(1995)}]{toner1995long}%
  \BibitemOpen
  \bibfield  {author} {\bibinfo {author} {\bibfnamefont {John}\ \bibnamefont
  {Toner}}\ and\ \bibinfo {author} {\bibfnamefont {Yuhai}\ \bibnamefont {Tu}},\
  }\bibfield  {title} {\enquote {\bibinfo {title} {Long-range order in a
  two-dimensional dynamical xy model: how birds fly together},}\ }\href
  {\doibase 10.1103/PhysRevLett.75.4326} {\bibfield  {journal} {\bibinfo
  {journal} {Physical Review Letters}\ }\textbf {\bibinfo {volume} {75}},\
  \bibinfo {pages} {4326} (\bibinfo {year} {1995})}\BibitemShut {NoStop}%
\bibitem [{\citenamefont {Reynolds}(1987)}]{reynolds1987flocks}%
  \BibitemOpen
  \bibfield  {author} {\bibinfo {author} {\bibfnamefont {Craig~W}\ \bibnamefont
  {Reynolds}},\ }\bibfield  {title} {\enquote {\bibinfo {title} {Flocks, herds
  and schools: A distributed behavioral model},}\ }in\ \href@noop {} {\emph
  {\bibinfo {booktitle} {Proceedings of the 14th annual conference on Computer
  graphics and interactive techniques}}}\ (\bibinfo {year} {1987})\ pp.\
  \bibinfo {pages} {25--34}\BibitemShut {NoStop}%
\bibitem [{\citenamefont {Aoki}(1982)}]{Aoki1982simulation}%
  \BibitemOpen
  \bibfield  {author} {\bibinfo {author} {\bibfnamefont {Ichiro}\ \bibnamefont
  {Aoki}},\ }\bibfield  {title} {\enquote {\bibinfo {title} {A simulation study
  on the schooling mechanism in fish.}}\ }\href
  {https://www.jstage.jst.go.jp/article/suisan1932/48/8/48_8_1081/_article}
  {\bibfield  {journal} {\bibinfo  {journal} {NIPPON SUISAN GAKKAISHI (Bulletin
  of the Japanese Society of Scientific Fisheries)}\ }\textbf {\bibinfo
  {volume} {48}},\ \bibinfo {pages} {1081--1088} (\bibinfo {year}
  {1982})}\BibitemShut {NoStop}%
\bibitem [{\citenamefont {Partridge}(1982)}]{partridge1982structure}%
  \BibitemOpen
  \bibfield  {author} {\bibinfo {author} {\bibfnamefont {Brian~L}\ \bibnamefont
  {Partridge}},\ }\bibfield  {title} {\enquote {\bibinfo {title} {The structure
  and function of fish schools},}\ }\href
  {https://www.scientificamerican.com/article/the-structure-and-function-of-fish/}
  {\bibfield  {journal} {\bibinfo  {journal} {Scientific American}\ }\textbf
  {\bibinfo {volume} {246}},\ \bibinfo {pages} {114--123} (\bibinfo {year}
  {1982})}\BibitemShut {NoStop}%
\bibitem [{\citenamefont {Ramaswamy}\ and\ \citenamefont
  {Simha}(2006)}]{ramaswamy2006mechanics}%
  \BibitemOpen
  \bibfield  {author} {\bibinfo {author} {\bibfnamefont {Sriram}\ \bibnamefont
  {Ramaswamy}}\ and\ \bibinfo {author} {\bibfnamefont {R~Aditi}\ \bibnamefont
  {Simha}},\ }\bibfield  {title} {\enquote {\bibinfo {title} {The mechanics of
  active matter: Broken-symmetry hydrodynamics of motile particles and granular
  layers},}\ }\href
  {https://www.sciencedirect.com/science/article/abs/pii/S0038109806004765}
  {\bibfield  {journal} {\bibinfo  {journal} {Solid state communications}\
  }\textbf {\bibinfo {volume} {139}},\ \bibinfo {pages} {617--622} (\bibinfo
  {year} {2006})}\BibitemShut {NoStop}%
\bibitem [{\citenamefont {Prost}\ and\ \citenamefont
  {Bruinsma}(1996)}]{prost1996shape}%
  \BibitemOpen
  \bibfield  {author} {\bibinfo {author} {\bibfnamefont {J}~\bibnamefont
  {Prost}}\ and\ \bibinfo {author} {\bibfnamefont {R}~\bibnamefont
  {Bruinsma}},\ }\bibfield  {title} {\enquote {\bibinfo {title} {Shape
  fluctuations of active membranes},}\ }\href {\doibase
  10.1209/epl/i1996-00340-1} {\bibfield  {journal} {\bibinfo  {journal} {EPL
  (Europhysics Letters)}\ }\textbf {\bibinfo {volume} {33}},\ \bibinfo {pages}
  {321} (\bibinfo {year} {1996})}\BibitemShut {NoStop}%
\bibitem [{\citenamefont {Ramaswamy}\ \emph {et~al.}(2000)\citenamefont
  {Ramaswamy}, \citenamefont {Toner},\ and\ \citenamefont
  {Prost}}]{ramaswamy2000nonequilibrium}%
  \BibitemOpen
  \bibfield  {author} {\bibinfo {author} {\bibfnamefont {Sriram}\ \bibnamefont
  {Ramaswamy}}, \bibinfo {author} {\bibfnamefont {John}\ \bibnamefont {Toner}},
  \ and\ \bibinfo {author} {\bibfnamefont {Jacques}\ \bibnamefont {Prost}},\
  }\bibfield  {title} {\enquote {\bibinfo {title} {Nonequilibrium fluctuations,
  traveling waves, and instabilities in active membranes},}\ }\href
  {https://journals.aps.org/prl/abstract/10.1103/PhysRevLett.84.3494}
  {\bibfield  {journal} {\bibinfo  {journal} {Physical review letters}\
  }\textbf {\bibinfo {volume} {84}},\ \bibinfo {pages} {3494} (\bibinfo {year}
  {2000})}\BibitemShut {NoStop}%
\bibitem [{\citenamefont {Darnell}\ \emph {et~al.}(1990)\citenamefont
  {Darnell}, \citenamefont {Lodish},\ and\ \citenamefont
  {Baltimore}}]{darnell1990molecular}%
  \BibitemOpen
  \bibfield  {author} {\bibinfo {author} {\bibfnamefont {James}\ \bibnamefont
  {Darnell}}, \bibinfo {author} {\bibfnamefont {Harvey}\ \bibnamefont
  {Lodish}}, \ and\ \bibinfo {author} {\bibfnamefont {David}\ \bibnamefont
  {Baltimore}},\ }\href@noop {} {\emph {\bibinfo {title} {Molecular cell
  biology.}}},\ \bibinfo {number} {QH581. 2 D22 1990}\ (\bibinfo {year}
  {1990})\BibitemShut {NoStop}%
\bibitem [{\citenamefont {Finlayson}\ and\ \citenamefont
  {Scriven}(1969)}]{finlayson1969convective}%
  \BibitemOpen
  \bibfield  {author} {\bibinfo {author} {\bibfnamefont {BA}~\bibnamefont
  {Finlayson}}\ and\ \bibinfo {author} {\bibfnamefont {LE}~\bibnamefont
  {Scriven}},\ }\bibfield  {title} {\enquote {\bibinfo {title} {Convective
  instability by active stress},}\ }\href
  {https://royalsocietypublishing.org/doi/abs/10.1098/rspa.1969.0071}
  {\bibfield  {journal} {\bibinfo  {journal} {Proceedings of the Royal Society
  of London. A. Mathematical and Physical Sciences}\ }\textbf {\bibinfo
  {volume} {310}},\ \bibinfo {pages} {183--219} (\bibinfo {year}
  {1969})}\BibitemShut {NoStop}%
\bibitem [{\citenamefont {Ramaswamy}(2010)}]{ramaswamy2010mechanics}%
  \BibitemOpen
  \bibfield  {author} {\bibinfo {author} {\bibfnamefont {Sriram}\ \bibnamefont
  {Ramaswamy}},\ }\bibfield  {title} {\enquote {\bibinfo {title} {The mechanics
  and statistics of active matter},}\ }\href@noop {} {\bibfield  {journal}
  {\bibinfo  {journal} {Annu. Rev. Condens. Matter Phys.}\ }\textbf {\bibinfo
  {volume} {1}},\ \bibinfo {pages} {323--345} (\bibinfo {year}
  {2010})}\BibitemShut {NoStop}%
\bibitem [{\citenamefont {Marchetti}\ \emph {et~al.}(2013)\citenamefont
  {Marchetti}, \citenamefont {Joanny}, \citenamefont {Ramaswamy}, \citenamefont
  {Liverpool}, \citenamefont {Prost}, \citenamefont {Rao},\ and\ \citenamefont
  {Simha}}]{marchetti2013hydrodynamics}%
  \BibitemOpen
  \bibfield  {author} {\bibinfo {author} {\bibfnamefont {M~Cristina}\
  \bibnamefont {Marchetti}}, \bibinfo {author} {\bibfnamefont
  {Jean-Fran{\c{c}}ois}\ \bibnamefont {Joanny}}, \bibinfo {author}
  {\bibfnamefont {Sriram}\ \bibnamefont {Ramaswamy}}, \bibinfo {author}
  {\bibfnamefont {Tanniemola~B}\ \bibnamefont {Liverpool}}, \bibinfo {author}
  {\bibfnamefont {Jacques}\ \bibnamefont {Prost}}, \bibinfo {author}
  {\bibfnamefont {Madan}\ \bibnamefont {Rao}}, \ and\ \bibinfo {author}
  {\bibfnamefont {R~Aditi}\ \bibnamefont {Simha}},\ }\bibfield  {title}
  {\enquote {\bibinfo {title} {Hydrodynamics of soft active matter},}\ }\href
  {\doibase 10.1103/RevModPhys.85.1143} {\bibfield  {journal} {\bibinfo
  {journal} {Reviews of Modern Physics}\ }\textbf {\bibinfo {volume} {85}},\
  \bibinfo {pages} {1143} (\bibinfo {year} {2013})}\BibitemShut {NoStop}%
\bibitem [{\citenamefont {Vicsek}\ and\ \citenamefont
  {Zafeiris}(2012)}]{vicsek2012collective}%
  \BibitemOpen
  \bibfield  {author} {\bibinfo {author} {\bibfnamefont {Tam{\'a}s}\
  \bibnamefont {Vicsek}}\ and\ \bibinfo {author} {\bibfnamefont {Anna}\
  \bibnamefont {Zafeiris}},\ }\bibfield  {title} {\enquote {\bibinfo {title}
  {Collective motion},}\ }\href@noop {} {\bibfield  {journal} {\bibinfo
  {journal} {Physics reports}\ }\textbf {\bibinfo {volume} {517}},\ \bibinfo
  {pages} {71--140} (\bibinfo {year} {2012})}\BibitemShut {NoStop}%
\bibitem [{\citenamefont {Poujade}\ \emph {et~al.}(2007)\citenamefont
  {Poujade}, \citenamefont {Grasland-Mongrain}, \citenamefont {Hertzog},
  \citenamefont {Jouanneau}, \citenamefont {Chavrier}, \citenamefont {Ladoux},
  \citenamefont {Buguin},\ and\ \citenamefont {Silberzan}}]{Poujade15988}%
  \BibitemOpen
  \bibfield  {author} {\bibinfo {author} {\bibfnamefont {M.}~\bibnamefont
  {Poujade}}, \bibinfo {author} {\bibfnamefont {E.}~\bibnamefont
  {Grasland-Mongrain}}, \bibinfo {author} {\bibfnamefont {A.}~\bibnamefont
  {Hertzog}}, \bibinfo {author} {\bibfnamefont {J.}~\bibnamefont {Jouanneau}},
  \bibinfo {author} {\bibfnamefont {P.}~\bibnamefont {Chavrier}}, \bibinfo
  {author} {\bibfnamefont {B.}~\bibnamefont {Ladoux}}, \bibinfo {author}
  {\bibfnamefont {A.}~\bibnamefont {Buguin}}, \ and\ \bibinfo {author}
  {\bibfnamefont {P.}~\bibnamefont {Silberzan}},\ }\bibfield  {title} {\enquote
  {\bibinfo {title} {Collective migration of an epithelial monolayer in
  response to a model wound},}\ }\href {\doibase 10.1073/pnas.0705062104}
  {\bibfield  {journal} {\bibinfo  {journal} {Proceedings of the National
  Academy of Sciences}\ }\textbf {\bibinfo {volume} {104}},\ \bibinfo {pages}
  {15988--15993} (\bibinfo {year} {2007})},\ \Eprint
  {http://arxiv.org/abs/https://www.pnas.org/content/104/41/15988.full.pdf}
  {https://www.pnas.org/content/104/41/15988.full.pdf} \BibitemShut {NoStop}%
\bibitem [{\citenamefont {Ballerini}\ \emph {et~al.}(2008)\citenamefont
  {Ballerini}, \citenamefont {Cabibbo}, \citenamefont {Candelier},
  \citenamefont {Cavagna}, \citenamefont {Cisbani}, \citenamefont {Giardina},
  \citenamefont {Lecomte}, \citenamefont {Orlandi}, \citenamefont {Parisi},
  \citenamefont {Procaccini}, \citenamefont {Viale},\ and\ \citenamefont
  {Zdravkovic}}]{Ballerini1232}%
  \BibitemOpen
  \bibfield  {author} {\bibinfo {author} {\bibfnamefont {M.}~\bibnamefont
  {Ballerini}}, \bibinfo {author} {\bibfnamefont {N.}~\bibnamefont {Cabibbo}},
  \bibinfo {author} {\bibfnamefont {R.}~\bibnamefont {Candelier}}, \bibinfo
  {author} {\bibfnamefont {A.}~\bibnamefont {Cavagna}}, \bibinfo {author}
  {\bibfnamefont {E.}~\bibnamefont {Cisbani}}, \bibinfo {author} {\bibfnamefont
  {I.}~\bibnamefont {Giardina}}, \bibinfo {author} {\bibfnamefont
  {V.}~\bibnamefont {Lecomte}}, \bibinfo {author} {\bibfnamefont
  {A.}~\bibnamefont {Orlandi}}, \bibinfo {author} {\bibfnamefont
  {G.}~\bibnamefont {Parisi}}, \bibinfo {author} {\bibfnamefont
  {A.}~\bibnamefont {Procaccini}}, \bibinfo {author} {\bibfnamefont
  {M.}~\bibnamefont {Viale}}, \ and\ \bibinfo {author} {\bibfnamefont
  {V.}~\bibnamefont {Zdravkovic}},\ }\bibfield  {title} {\enquote {\bibinfo
  {title} {Interaction ruling animal collective behavior depends on topological
  rather than metric distance: Evidence from a field study},}\ }\href {\doibase
  10.1073/pnas.0711437105} {\bibfield  {journal} {\bibinfo  {journal}
  {Proceedings of the National Academy of Sciences}\ }\textbf {\bibinfo
  {volume} {105}},\ \bibinfo {pages} {1232--1237} (\bibinfo {year} {2008})},\
  \Eprint
  {http://arxiv.org/abs/https://www.pnas.org/content/105/4/1232.full.pdf}
  {https://www.pnas.org/content/105/4/1232.full.pdf} \BibitemShut {NoStop}%
\bibitem [{\citenamefont {Helbing}\ \emph {et~al.}(2000)\citenamefont
  {Helbing}, \citenamefont {Farkas},\ and\ \citenamefont
  {Vicsek}}]{Helbing2000}%
  \BibitemOpen
  \bibfield  {author} {\bibinfo {author} {\bibfnamefont {Dirk}\ \bibnamefont
  {Helbing}}, \bibinfo {author} {\bibfnamefont {Ill\'es}\ \bibnamefont
  {Farkas}}, \ and\ \bibinfo {author} {\bibfnamefont {Tam\'as}\ \bibnamefont
  {Vicsek}},\ }\bibfield  {title} {\enquote {\bibinfo {title} {Simulating
  dynamical features of escape panic},}\ }\href {\doibase 10.1038/35035023}
  {\bibfield  {journal} {\bibinfo  {journal} {Nature}\ }\textbf {\bibinfo
  {volume} {407}},\ \bibinfo {pages} {487--490} (\bibinfo {year}
  {2000})}\BibitemShut {NoStop}%
\bibitem [{\citenamefont {Bain}\ and\ \citenamefont
  {Bartolo}(2019)}]{Bain2019}%
  \BibitemOpen
  \bibfield  {author} {\bibinfo {author} {\bibfnamefont {Nicolas}\ \bibnamefont
  {Bain}}\ and\ \bibinfo {author} {\bibfnamefont {Denis}\ \bibnamefont
  {Bartolo}},\ }\bibfield  {title} {\enquote {\bibinfo {title} {Dynamic
  response and hydrodynamics of polarized crowds},}\ }\href {\doibase
  10.1126/science.aat9891} {\bibfield  {journal} {\bibinfo  {journal}
  {Science}\ }\textbf {\bibinfo {volume} {363}},\ \bibinfo {pages} {46--49}
  (\bibinfo {year} {2019})}\BibitemShut {NoStop}%
\bibitem [{\citenamefont {Narayan}\ \emph {et~al.}(2007)\citenamefont
  {Narayan}, \citenamefont {Ramaswamy},\ and\ \citenamefont
  {Menon}}]{narayan2007long}%
  \BibitemOpen
  \bibfield  {author} {\bibinfo {author} {\bibfnamefont {Vijay}\ \bibnamefont
  {Narayan}}, \bibinfo {author} {\bibfnamefont {Sriram}\ \bibnamefont
  {Ramaswamy}}, \ and\ \bibinfo {author} {\bibfnamefont {Narayanan}\
  \bibnamefont {Menon}},\ }\bibfield  {title} {\enquote {\bibinfo {title}
  {Long-lived giant number fluctuations in a swarming granular nematic},}\
  }\href {\doibase 10.1126/science.1140414} {\bibfield  {journal} {\bibinfo
  {journal} {Science}\ }\textbf {\bibinfo {volume} {317}},\ \bibinfo {pages}
  {105--108} (\bibinfo {year} {2007})}\BibitemShut {NoStop}%
\bibitem [{\citenamefont {Deseigne}\ \emph {et~al.}(2010)\citenamefont
  {Deseigne}, \citenamefont {Dauchot},\ and\ \citenamefont
  {Chat{\'e}}}]{deseigne2010collective}%
  \BibitemOpen
  \bibfield  {author} {\bibinfo {author} {\bibfnamefont {Julien}\ \bibnamefont
  {Deseigne}}, \bibinfo {author} {\bibfnamefont {Olivier}\ \bibnamefont
  {Dauchot}}, \ and\ \bibinfo {author} {\bibfnamefont {Hugues}\ \bibnamefont
  {Chat{\'e}}},\ }\bibfield  {title} {\enquote {\bibinfo {title} {Collective
  motion of vibrated polar disks},}\ }\href
  {https://journals.aps.org/prl/abstract/10.1103/PhysRevLett.105.098001}
  {\bibfield  {journal} {\bibinfo  {journal} {Physical review letters}\
  }\textbf {\bibinfo {volume} {105}},\ \bibinfo {pages} {098001} (\bibinfo
  {year} {2010})}\BibitemShut {NoStop}%
\bibitem [{\citenamefont {Kumar}\ \emph
  {et~al.}(2014{\natexlab{a}})\citenamefont {Kumar}, \citenamefont {Soni},
  \citenamefont {Ramaswamy},\ and\ \citenamefont {Sood}}]{kumar2014flocking}%
  \BibitemOpen
  \bibfield  {author} {\bibinfo {author} {\bibfnamefont {Nitin}\ \bibnamefont
  {Kumar}}, \bibinfo {author} {\bibfnamefont {Harsh}\ \bibnamefont {Soni}},
  \bibinfo {author} {\bibfnamefont {Sriram}\ \bibnamefont {Ramaswamy}}, \ and\
  \bibinfo {author} {\bibfnamefont {AK}~\bibnamefont {Sood}},\ }\bibfield
  {title} {\enquote {\bibinfo {title} {Flocking at a distance in active
  granular matter},}\ }\href {https://www.nature.com/articles/ncomms5688}
  {\bibfield  {journal} {\bibinfo  {journal} {Nature communications}\ }\textbf
  {\bibinfo {volume} {5}},\ \bibinfo {pages} {1--9} (\bibinfo {year}
  {2014}{\natexlab{a}})}\BibitemShut {NoStop}%
\bibitem [{\citenamefont {Golestanian}(2019)}]{golestanian2019phoretic}%
  \BibitemOpen
  \bibfield  {author} {\bibinfo {author} {\bibfnamefont {Ramin}\ \bibnamefont
  {Golestanian}},\ }\bibfield  {title} {\enquote {\bibinfo {title} {Phoretic
  active matter},}\ }\href {https://arxiv.org/abs/1909.03747} {\bibfield
  {journal} {\bibinfo  {journal} {arXiv preprint arXiv:1909.03747}\ } (\bibinfo
  {year} {2019})}\BibitemShut {NoStop}%
\bibitem [{\citenamefont {Bechinger}\ \emph {et~al.}(2016)\citenamefont
  {Bechinger}, \citenamefont {Di~Leonardo}, \citenamefont {L\"owen},
  \citenamefont {Reichhardt}, \citenamefont {Volpe},\ and\ \citenamefont
  {Volpe}}]{Bechinger-RMP2018}%
  \BibitemOpen
  \bibfield  {author} {\bibinfo {author} {\bibfnamefont {Clemens}\ \bibnamefont
  {Bechinger}}, \bibinfo {author} {\bibfnamefont {Roberto}\ \bibnamefont
  {Di~Leonardo}}, \bibinfo {author} {\bibfnamefont {Hartmut}\ \bibnamefont
  {L\"owen}}, \bibinfo {author} {\bibfnamefont {Charles}\ \bibnamefont
  {Reichhardt}}, \bibinfo {author} {\bibfnamefont {Giorgio}\ \bibnamefont
  {Volpe}}, \ and\ \bibinfo {author} {\bibfnamefont {Giovanni}\ \bibnamefont
  {Volpe}},\ }\bibfield  {title} {\enquote {\bibinfo {title} {Active particles
  in complex and crowded environments},}\ }\href {\doibase
  10.1103/RevModPhys.88.045006} {\bibfield  {journal} {\bibinfo  {journal}
  {Rev. Mod. Phys.}\ }\textbf {\bibinfo {volume} {88}},\ \bibinfo {pages}
  {045006} (\bibinfo {year} {2016})}\BibitemShut {NoStop}%
\bibitem [{\citenamefont {Simha}\ and\ \citenamefont
  {Ramaswamy}(2002{\natexlab{a}})}]{Simha2002}%
  \BibitemOpen
  \bibfield  {author} {\bibinfo {author} {\bibfnamefont {R.~Aditi}\
  \bibnamefont {Simha}}\ and\ \bibinfo {author} {\bibfnamefont {Sriram}\
  \bibnamefont {Ramaswamy}},\ }\bibfield  {title} {\enquote {\bibinfo {title}
  {Hydrodynamic fluctuations and instabilities in ordered suspensions of
  self-propelled particles},}\ }\href {\doibase 10.1103/PhysRevLett.89.058101}
  {\bibfield  {journal} {\bibinfo  {journal} {Phys. Rev. Lett.}\ }\textbf
  {\bibinfo {volume} {89}},\ \bibinfo {pages} {058101} (\bibinfo {year}
  {2002}{\natexlab{a}})}\BibitemShut {NoStop}%
\bibitem [{\citenamefont {Voituriez}\ \emph {et~al.}(2005)\citenamefont
  {Voituriez}, \citenamefont {Joanny},\ and\ \citenamefont
  {Prost}}]{voituriez2005spontaneous}%
  \BibitemOpen
  \bibfield  {author} {\bibinfo {author} {\bibfnamefont {R}~\bibnamefont
  {Voituriez}}, \bibinfo {author} {\bibfnamefont {Jean-Fran{\c{c}}ois}\
  \bibnamefont {Joanny}}, \ and\ \bibinfo {author} {\bibfnamefont {Jacques}\
  \bibnamefont {Prost}},\ }\bibfield  {title} {\enquote {\bibinfo {title}
  {Spontaneous flow transition in active polar gels},}\ }\href@noop {}
  {\bibfield  {journal} {\bibinfo  {journal} {EPL (Europhysics Letters)}\
  }\textbf {\bibinfo {volume} {70}},\ \bibinfo {pages} {404} (\bibinfo {year}
  {2005})}\BibitemShut {NoStop}%
\bibitem [{\citenamefont {Wensink}\ \emph {et~al.}(2012)\citenamefont
  {Wensink}, \citenamefont {Dunkel}, \citenamefont {Heidenreich}, \citenamefont
  {Drescher}, \citenamefont {Goldstein}, \citenamefont {L{\"o}wen},\ and\
  \citenamefont {Yeomans}}]{wensink2012meso}%
  \BibitemOpen
  \bibfield  {author} {\bibinfo {author} {\bibfnamefont {Henricus~H}\
  \bibnamefont {Wensink}}, \bibinfo {author} {\bibfnamefont {J{\"o}rn}\
  \bibnamefont {Dunkel}}, \bibinfo {author} {\bibfnamefont {Sebastian}\
  \bibnamefont {Heidenreich}}, \bibinfo {author} {\bibfnamefont {Knut}\
  \bibnamefont {Drescher}}, \bibinfo {author} {\bibfnamefont {Raymond~E}\
  \bibnamefont {Goldstein}}, \bibinfo {author} {\bibfnamefont {Hartmut}\
  \bibnamefont {L{\"o}wen}}, \ and\ \bibinfo {author} {\bibfnamefont {Julia~M}\
  \bibnamefont {Yeomans}},\ }\bibfield  {title} {\enquote {\bibinfo {title}
  {Meso-scale turbulence in living fluids},}\ }\href@noop {} {\bibfield
  {journal} {\bibinfo  {journal} {Proceedings of the National Academy of
  Sciences}\ }\textbf {\bibinfo {volume} {109}},\ \bibinfo {pages}
  {14308--14313} (\bibinfo {year} {2012})}\BibitemShut {NoStop}%
\bibitem [{\citenamefont {Bricard}\ \emph {et~al.}(2013)\citenamefont
  {Bricard}, \citenamefont {Caussin}, \citenamefont {Desreumaux}, \citenamefont
  {Dauchot},\ and\ \citenamefont {Bartolo}}]{bricard2013emergence}%
  \BibitemOpen
  \bibfield  {author} {\bibinfo {author} {\bibfnamefont {Antoine}\ \bibnamefont
  {Bricard}}, \bibinfo {author} {\bibfnamefont {Jean-Baptiste}\ \bibnamefont
  {Caussin}}, \bibinfo {author} {\bibfnamefont {Nicolas}\ \bibnamefont
  {Desreumaux}}, \bibinfo {author} {\bibfnamefont {Olivier}\ \bibnamefont
  {Dauchot}}, \ and\ \bibinfo {author} {\bibfnamefont {Denis}\ \bibnamefont
  {Bartolo}},\ }\bibfield  {title} {\enquote {\bibinfo {title} {Emergence of
  macroscopic directed motion in populations of motile colloids},}\ }\href@noop
  {} {\bibfield  {journal} {\bibinfo  {journal} {Nature}\ }\textbf {\bibinfo
  {volume} {503}},\ \bibinfo {pages} {95--98} (\bibinfo {year}
  {2013})}\BibitemShut {NoStop}%
\bibitem [{\citenamefont {Chaikin}\ and\ \citenamefont
  {Lubensky}(2000)}]{chaikin2000principles}%
  \BibitemOpen
  \bibfield  {author} {\bibinfo {author} {\bibfnamefont {Paul~M}\ \bibnamefont
  {Chaikin}}\ and\ \bibinfo {author} {\bibfnamefont {Tom~C}\ \bibnamefont
  {Lubensky}},\ }\href@noop {} {\emph {\bibinfo {title} {Principles of
  condensed matter physics}}}\ (\bibinfo  {publisher} {Cambridge university
  press},\ \bibinfo {year} {2000})\BibitemShut {NoStop}%
\bibitem [{\citenamefont {Sanchez}\ \emph {et~al.}(2012)\citenamefont
  {Sanchez}, \citenamefont {Chen}, \citenamefont {DeCamp}, \citenamefont
  {Heymann},\ and\ \citenamefont {Dogic}}]{sanchez2012spontaneous}%
  \BibitemOpen
  \bibfield  {author} {\bibinfo {author} {\bibfnamefont {Tim}\ \bibnamefont
  {Sanchez}}, \bibinfo {author} {\bibfnamefont {Daniel~TN}\ \bibnamefont
  {Chen}}, \bibinfo {author} {\bibfnamefont {Stephen~J}\ \bibnamefont
  {DeCamp}}, \bibinfo {author} {\bibfnamefont {Michael}\ \bibnamefont
  {Heymann}}, \ and\ \bibinfo {author} {\bibfnamefont {Zvonimir}\ \bibnamefont
  {Dogic}},\ }\bibfield  {title} {\enquote {\bibinfo {title} {Spontaneous
  motion in hierarchically assembled active matter},}\ }\href {\doibase
  10.1038/nature11591} {\bibfield  {journal} {\bibinfo  {journal} {Nature}\
  }\textbf {\bibinfo {volume} {491}},\ \bibinfo {pages} {431} (\bibinfo {year}
  {2012})}\BibitemShut {NoStop}%
\bibitem [{\citenamefont {Zhang}\ \emph {et~al.}(2020)\citenamefont {Zhang},
  \citenamefont {Deserno}, \citenamefont {Tu} \emph
  {et~al.}}]{zhang2020dynamics}%
  \BibitemOpen
  \bibfield  {author} {\bibinfo {author} {\bibfnamefont {Yi-Heng}\ \bibnamefont
  {Zhang}}, \bibinfo {author} {\bibfnamefont {Markus}\ \bibnamefont {Deserno}},
  \bibinfo {author} {\bibfnamefont {Zhan-Chun}\ \bibnamefont {Tu}},  \emph
  {et~al.},\ }\bibfield  {title} {\enquote {\bibinfo {title} {Dynamics of
  active nematic defects on the surface of a sphere},}\ }\href@noop {}
  {\bibfield  {journal} {\bibinfo  {journal} {Physical Review E}\ }\textbf
  {\bibinfo {volume} {102}},\ \bibinfo {pages} {012607} (\bibinfo {year}
  {2020})}\BibitemShut {NoStop}%
\bibitem [{\citenamefont {Nishiguchi}\ \emph {et~al.}(2017)\citenamefont
  {Nishiguchi}, \citenamefont {Nagai}, \citenamefont {Chat{\'e}},\ and\
  \citenamefont {Sano}}]{nishiguchi2017long}%
  \BibitemOpen
  \bibfield  {author} {\bibinfo {author} {\bibfnamefont {Daiki}\ \bibnamefont
  {Nishiguchi}}, \bibinfo {author} {\bibfnamefont {Ken~H}\ \bibnamefont
  {Nagai}}, \bibinfo {author} {\bibfnamefont {Hugues}\ \bibnamefont
  {Chat{\'e}}}, \ and\ \bibinfo {author} {\bibfnamefont {Masaki}\ \bibnamefont
  {Sano}},\ }\bibfield  {title} {\enquote {\bibinfo {title} {Long-range nematic
  order and anomalous fluctuations in suspensions of swimming filamentous
  bacteria},}\ }\href {\doibase 10.1103/PhysRevE.95.020601} {\bibfield
  {journal} {\bibinfo  {journal} {Physical Review E}\ }\textbf {\bibinfo
  {volume} {95}},\ \bibinfo {pages} {020601} (\bibinfo {year}
  {2017})}\BibitemShut {NoStop}%
\bibitem [{\citenamefont {Dell'Arciprete}\ \emph {et~al.}(2018)\citenamefont
  {Dell'Arciprete}, \citenamefont {Blow}, \citenamefont {Brown}, \citenamefont
  {Farrell}, \citenamefont {Lintuvuori}, \citenamefont {McVey}, \citenamefont
  {Marenduzzo},\ and\ \citenamefont {Poon}}]{dell2018growing}%
  \BibitemOpen
  \bibfield  {author} {\bibinfo {author} {\bibfnamefont {D}~\bibnamefont
  {Dell'Arciprete}}, \bibinfo {author} {\bibfnamefont {ML}~\bibnamefont
  {Blow}}, \bibinfo {author} {\bibfnamefont {AT}~\bibnamefont {Brown}},
  \bibinfo {author} {\bibfnamefont {FDC}\ \bibnamefont {Farrell}}, \bibinfo
  {author} {\bibfnamefont {Juho~S}\ \bibnamefont {Lintuvuori}}, \bibinfo
  {author} {\bibfnamefont {AF}~\bibnamefont {McVey}}, \bibinfo {author}
  {\bibfnamefont {D}~\bibnamefont {Marenduzzo}}, \ and\ \bibinfo {author}
  {\bibfnamefont {Wilson~CK}\ \bibnamefont {Poon}},\ }\bibfield  {title}
  {\enquote {\bibinfo {title} {A growing bacterial colony in two dimensions as
  an active nematic},}\ }\href@noop {} {\bibfield  {journal} {\bibinfo
  {journal} {Nature communications}\ }\textbf {\bibinfo {volume} {9}},\
  \bibinfo {pages} {4190} (\bibinfo {year} {2018})}\BibitemShut {NoStop}%
\bibitem [{\citenamefont {Yaman}\ \emph {et~al.}(2019)\citenamefont {Yaman},
  \citenamefont {Demir}, \citenamefont {Vetter},\ and\ \citenamefont
  {Kocabas}}]{yaman2018emergence}%
  \BibitemOpen
  \bibfield  {author} {\bibinfo {author} {\bibfnamefont {Yusuf~Ilker}\
  \bibnamefont {Yaman}}, \bibinfo {author} {\bibfnamefont {Esin}\ \bibnamefont
  {Demir}}, \bibinfo {author} {\bibfnamefont {Roman}\ \bibnamefont {Vetter}}, \
  and\ \bibinfo {author} {\bibfnamefont {Askin}\ \bibnamefont {Kocabas}},\
  }\bibfield  {title} {\enquote {\bibinfo {title} {Emergence of active nematics
  in chaining bacterial biofilms},}\ }\href@noop {} {\bibfield  {journal}
  {\bibinfo  {journal} {Nature communications}\ }\textbf {\bibinfo {volume}
  {10}},\ \bibinfo {pages} {2285} (\bibinfo {year} {2019})}\BibitemShut
  {NoStop}%
\bibitem [{\citenamefont {Kawaguchi}\ \emph {et~al.}(2017)\citenamefont
  {Kawaguchi}, \citenamefont {Kageyama},\ and\ \citenamefont
  {Sano}}]{kawaguchi2017topological}%
  \BibitemOpen
  \bibfield  {author} {\bibinfo {author} {\bibfnamefont {Kyogo}\ \bibnamefont
  {Kawaguchi}}, \bibinfo {author} {\bibfnamefont {Ryoichiro}\ \bibnamefont
  {Kageyama}}, \ and\ \bibinfo {author} {\bibfnamefont {Masaki}\ \bibnamefont
  {Sano}},\ }\bibfield  {title} {\enquote {\bibinfo {title} {Topological
  defects control collective dynamics in neural progenitor cell cultures},}\
  }\href {\doibase 10.1038/nature22321} {\bibfield  {journal} {\bibinfo
  {journal} {Nature}\ }\textbf {\bibinfo {volume} {545}},\ \bibinfo {pages}
  {327} (\bibinfo {year} {2017})}\BibitemShut {NoStop}%
\bibitem [{\citenamefont {Saw}\ \emph {et~al.}(2017)\citenamefont {Saw},
  \citenamefont {Doostmohammadi}, \citenamefont {Nier}, \citenamefont
  {Kocgozlu}, \citenamefont {Thampi}, \citenamefont {Toyama}, \citenamefont
  {Marcq}, \citenamefont {Lim}, \citenamefont {Yeomans},\ and\ \citenamefont
  {Ladoux}}]{saw2017topological}%
  \BibitemOpen
  \bibfield  {author} {\bibinfo {author} {\bibfnamefont {Thuan~Beng}\
  \bibnamefont {Saw}}, \bibinfo {author} {\bibfnamefont {Amin}\ \bibnamefont
  {Doostmohammadi}}, \bibinfo {author} {\bibfnamefont {Vincent}\ \bibnamefont
  {Nier}}, \bibinfo {author} {\bibfnamefont {Leyla}\ \bibnamefont {Kocgozlu}},
  \bibinfo {author} {\bibfnamefont {Sumesh}\ \bibnamefont {Thampi}}, \bibinfo
  {author} {\bibfnamefont {Yusuke}\ \bibnamefont {Toyama}}, \bibinfo {author}
  {\bibfnamefont {Philippe}\ \bibnamefont {Marcq}}, \bibinfo {author}
  {\bibfnamefont {Chwee~Teck}\ \bibnamefont {Lim}}, \bibinfo {author}
  {\bibfnamefont {Julia~M}\ \bibnamefont {Yeomans}}, \ and\ \bibinfo {author}
  {\bibfnamefont {Benoit}\ \bibnamefont {Ladoux}},\ }\bibfield  {title}
  {\enquote {\bibinfo {title} {Topological defects in epithelia govern cell
  death and extrusion},}\ }\href {\doibase 10.1038/nature21718} {\bibfield
  {journal} {\bibinfo  {journal} {Nature}\ }\textbf {\bibinfo {volume} {544}},\
  \bibinfo {pages} {212} (\bibinfo {year} {2017})}\BibitemShut {NoStop}%
\bibitem [{\citenamefont {Blanch-Mercader}\ \emph {et~al.}(2018)\citenamefont
  {Blanch-Mercader}, \citenamefont {Yashunsky}, \citenamefont {Garcia},
  \citenamefont {Duclos}, \citenamefont {Giomi},\ and\ \citenamefont
  {Silberzan}}]{blanch2018turbulent}%
  \BibitemOpen
  \bibfield  {author} {\bibinfo {author} {\bibfnamefont {C}~\bibnamefont
  {Blanch-Mercader}}, \bibinfo {author} {\bibfnamefont {V}~\bibnamefont
  {Yashunsky}}, \bibinfo {author} {\bibfnamefont {S}~\bibnamefont {Garcia}},
  \bibinfo {author} {\bibfnamefont {G}~\bibnamefont {Duclos}}, \bibinfo
  {author} {\bibfnamefont {L}~\bibnamefont {Giomi}}, \ and\ \bibinfo {author}
  {\bibfnamefont {P}~\bibnamefont {Silberzan}},\ }\bibfield  {title} {\enquote
  {\bibinfo {title} {Turbulent dynamics of epithelial cell cultures},}\ }\href
  {\doibase 10.1103/PhysRevLett.120.208101} {\bibfield  {journal} {\bibinfo
  {journal} {Physical Review Letters}\ }\textbf {\bibinfo {volume} {120}},\
  \bibinfo {pages} {208101} (\bibinfo {year} {2018})}\BibitemShut {NoStop}%
\bibitem [{\citenamefont {Maroudas-Sacks}\ \emph {et~al.}(2021)\citenamefont
  {Maroudas-Sacks}, \citenamefont {Garion}, \citenamefont {Shani-Zerbib},
  \citenamefont {Livshits}, \citenamefont {Braun},\ and\ \citenamefont
  {Keren}}]{maroudas2021topological}%
  \BibitemOpen
  \bibfield  {author} {\bibinfo {author} {\bibfnamefont {Yonit}\ \bibnamefont
  {Maroudas-Sacks}}, \bibinfo {author} {\bibfnamefont {Liora}\ \bibnamefont
  {Garion}}, \bibinfo {author} {\bibfnamefont {Lital}\ \bibnamefont
  {Shani-Zerbib}}, \bibinfo {author} {\bibfnamefont {Anton}\ \bibnamefont
  {Livshits}}, \bibinfo {author} {\bibfnamefont {Erez}\ \bibnamefont {Braun}},
  \ and\ \bibinfo {author} {\bibfnamefont {Kinneret}\ \bibnamefont {Keren}},\
  }\bibfield  {title} {\enquote {\bibinfo {title} {Topological defects in the
  nematic order of actin fibres as organization centres of hydra
  morphogenesis},}\ }\href@noop {} {\bibfield  {journal} {\bibinfo  {journal}
  {Nature Physics}\ }\textbf {\bibinfo {volume} {17}},\ \bibinfo {pages}
  {251--259} (\bibinfo {year} {2021})}\BibitemShut {NoStop}%
\bibitem [{\citenamefont {Alert}\ \emph {et~al.}(2021)\citenamefont {Alert},
  \citenamefont {Casademunt},\ and\ \citenamefont {Joanny}}]{alert2021active}%
  \BibitemOpen
  \bibfield  {author} {\bibinfo {author} {\bibfnamefont {Ricard}\ \bibnamefont
  {Alert}}, \bibinfo {author} {\bibfnamefont {Jaume}\ \bibnamefont
  {Casademunt}}, \ and\ \bibinfo {author} {\bibfnamefont {Jean-Fran{\c{c}}ois}\
  \bibnamefont {Joanny}},\ }\bibfield  {title} {\enquote {\bibinfo {title}
  {Active turbulence},}\ }\href@noop {} {\bibfield  {journal} {\bibinfo
  {journal} {arXiv preprint arXiv:2104.02122}\ } (\bibinfo {year}
  {2021})}\BibitemShut {NoStop}%
\bibitem [{\citenamefont {Toner}\ \emph {et~al.}(2005)\citenamefont {Toner},
  \citenamefont {Tu},\ and\ \citenamefont
  {Ramaswamy}}]{toner2005hydrodynamics}%
  \BibitemOpen
  \bibfield  {author} {\bibinfo {author} {\bibfnamefont {John}\ \bibnamefont
  {Toner}}, \bibinfo {author} {\bibfnamefont {Yuhai}\ \bibnamefont {Tu}}, \
  and\ \bibinfo {author} {\bibfnamefont {Sriram}\ \bibnamefont {Ramaswamy}},\
  }\bibfield  {title} {\enquote {\bibinfo {title} {Hydrodynamics and phases of
  flocks},}\ }\href@noop {} {\bibfield  {journal} {\bibinfo  {journal} {Annals
  of Physics}\ }\textbf {\bibinfo {volume} {318}},\ \bibinfo {pages} {170--244}
  (\bibinfo {year} {2005})}\BibitemShut {NoStop}%
\bibitem [{\citenamefont {Steinberg}(2021)}]{steinberg2021elastic}%
  \BibitemOpen
  \bibfield  {author} {\bibinfo {author} {\bibfnamefont {Victor}\ \bibnamefont
  {Steinberg}},\ }\bibfield  {title} {\enquote {\bibinfo {title} {Elastic
  turbulence: an experimental view on inertialess random flow},}\ }\href@noop
  {} {\bibfield  {journal} {\bibinfo  {journal} {Annual Review of Fluid
  Mechanics}\ }\textbf {\bibinfo {volume} {53}},\ \bibinfo {pages} {27--58}
  (\bibinfo {year} {2021})}\BibitemShut {NoStop}%
\bibitem [{\citenamefont {Giomi}(2015)}]{giomi2015geometry}%
  \BibitemOpen
  \bibfield  {author} {\bibinfo {author} {\bibfnamefont {Luca}\ \bibnamefont
  {Giomi}},\ }\bibfield  {title} {\enquote {\bibinfo {title} {Geometry and
  topology of turbulence in active nematics},}\ }\href {\doibase
  10.1103/PhysRevX.5.031003} {\bibfield  {journal} {\bibinfo  {journal}
  {Physical Review X}\ }\textbf {\bibinfo {volume} {5}},\ \bibinfo {pages}
  {031003} (\bibinfo {year} {2015})}\BibitemShut {NoStop}%
\bibitem [{\citenamefont {Chatterjee}\ \emph {et~al.}(2019)\citenamefont
  {Chatterjee}, \citenamefont {Rana}, \citenamefont {Simha}, \citenamefont
  {Perlekar},\ and\ \citenamefont {Ramaswamy}}]{chatterjee2019fluid}%
  \BibitemOpen
  \bibfield  {author} {\bibinfo {author} {\bibfnamefont {Rayan}\ \bibnamefont
  {Chatterjee}}, \bibinfo {author} {\bibfnamefont {Navdeep}\ \bibnamefont
  {Rana}}, \bibinfo {author} {\bibfnamefont {R~Aditi}\ \bibnamefont {Simha}},
  \bibinfo {author} {\bibfnamefont {Prasad}\ \bibnamefont {Perlekar}}, \ and\
  \bibinfo {author} {\bibfnamefont {Sriram}\ \bibnamefont {Ramaswamy}},\
  }\bibfield  {title} {\enquote {\bibinfo {title} {Fluid flocks with
  inertia},}\ }\href {https://arxiv.org/abs/1907.03492v3} {\bibfield  {journal}
  {\bibinfo  {journal} {arXiv preprint arXiv:1907.03492v3}\ } (\bibinfo {year}
  {2019})},\ \bibinfo {note} {{P}hys Rev X, accepted}\BibitemShut {NoStop}%
\bibitem [{\citenamefont {Doostmohammadi}\ \emph
  {et~al.}(2016{\natexlab{a}})\citenamefont {Doostmohammadi}, \citenamefont
  {Adamer}, \citenamefont {Thampi},\ and\ \citenamefont
  {Yeomans}}]{doostmohammadi2016stabilization}%
  \BibitemOpen
  \bibfield  {author} {\bibinfo {author} {\bibfnamefont {Amin}\ \bibnamefont
  {Doostmohammadi}}, \bibinfo {author} {\bibfnamefont {Michael~F}\ \bibnamefont
  {Adamer}}, \bibinfo {author} {\bibfnamefont {Sumesh~P}\ \bibnamefont
  {Thampi}}, \ and\ \bibinfo {author} {\bibfnamefont {Julia~M}\ \bibnamefont
  {Yeomans}},\ }\bibfield  {title} {\enquote {\bibinfo {title} {Stabilization
  of active matter by flow-vortex lattices and defect ordering},}\ }\href
  {\doibase 10.1038/ncomms10557} {\bibfield  {journal} {\bibinfo  {journal}
  {Nature communications}\ }\textbf {\bibinfo {volume} {7}},\ \bibinfo {pages}
  {10557} (\bibinfo {year} {2016}{\natexlab{a}})}\BibitemShut {NoStop}%
\bibitem [{\citenamefont {Doostmohammadi}\ \emph {et~al.}(2018)\citenamefont
  {Doostmohammadi}, \citenamefont {Ign{\'e}s-Mullol}, \citenamefont {Yeomans},\
  and\ \citenamefont {Sagu{\'e}s}}]{doostmohammadi2018active}%
  \BibitemOpen
  \bibfield  {author} {\bibinfo {author} {\bibfnamefont {Amin}\ \bibnamefont
  {Doostmohammadi}}, \bibinfo {author} {\bibfnamefont {Jordi}\ \bibnamefont
  {Ign{\'e}s-Mullol}}, \bibinfo {author} {\bibfnamefont {Julia~M}\ \bibnamefont
  {Yeomans}}, \ and\ \bibinfo {author} {\bibfnamefont {Francesc}\ \bibnamefont
  {Sagu{\'e}s}},\ }\bibfield  {title} {\enquote {\bibinfo {title} {Active
  nematics},}\ }\href {\doibase 0.1038/s41467-018-05666-8} {\bibfield
  {journal} {\bibinfo  {journal} {Nature communications}\ }\textbf {\bibinfo
  {volume} {9}},\ \bibinfo {pages} {3246} (\bibinfo {year} {2018})}\BibitemShut
  {NoStop}%
\bibitem [{\citenamefont {Duclos}\ \emph {et~al.}(2020)\citenamefont {Duclos},
  \citenamefont {Adkins}, \citenamefont {Banerjee}, \citenamefont {Peterson},
  \citenamefont {Varghese}, \citenamefont {Kolvin}, \citenamefont {Baskaran},
  \citenamefont {Pelcovits}, \citenamefont {Powers}, \citenamefont {Baskaran}
  \emph {et~al.}}]{duclos2020topological}%
  \BibitemOpen
  \bibfield  {author} {\bibinfo {author} {\bibfnamefont {Guillaume}\
  \bibnamefont {Duclos}}, \bibinfo {author} {\bibfnamefont {Raymond}\
  \bibnamefont {Adkins}}, \bibinfo {author} {\bibfnamefont {Debarghya}\
  \bibnamefont {Banerjee}}, \bibinfo {author} {\bibfnamefont {Matthew~SE}\
  \bibnamefont {Peterson}}, \bibinfo {author} {\bibfnamefont {Minu}\
  \bibnamefont {Varghese}}, \bibinfo {author} {\bibfnamefont {Itamar}\
  \bibnamefont {Kolvin}}, \bibinfo {author} {\bibfnamefont {Arvind}\
  \bibnamefont {Baskaran}}, \bibinfo {author} {\bibfnamefont {Robert~A}\
  \bibnamefont {Pelcovits}}, \bibinfo {author} {\bibfnamefont {Thomas~R}\
  \bibnamefont {Powers}}, \bibinfo {author} {\bibfnamefont {Aparna}\
  \bibnamefont {Baskaran}},  \emph {et~al.},\ }\bibfield  {title} {\enquote
  {\bibinfo {title} {Topological structure and dynamics of three-dimensional
  active nematics},}\ }\href@noop {} {\bibfield  {journal} {\bibinfo  {journal}
  {Science}\ }\textbf {\bibinfo {volume} {367}},\ \bibinfo {pages} {1120--1124}
  (\bibinfo {year} {2020})}\BibitemShut {NoStop}%
\bibitem [{\citenamefont {S{\l}omka}\ and\ \citenamefont
  {Dunkel}(2017)}]{slomka2017spontaneous}%
  \BibitemOpen
  \bibfield  {author} {\bibinfo {author} {\bibfnamefont {Jonasz}\ \bibnamefont
  {S{\l}omka}}\ and\ \bibinfo {author} {\bibfnamefont {J{\"o}rn}\ \bibnamefont
  {Dunkel}},\ }\bibfield  {title} {\enquote {\bibinfo {title} {Spontaneous
  mirror-symmetry breaking induces inverse energy cascade in 3d active
  fluids},}\ }\href@noop {} {\bibfield  {journal} {\bibinfo  {journal}
  {Proceedings of the National Academy of Sciences}\ }\textbf {\bibinfo
  {volume} {114}},\ \bibinfo {pages} {2119--2124} (\bibinfo {year}
  {2017})}\BibitemShut {NoStop}%
\bibitem [{\citenamefont {S{\l}omka}\ \emph {et~al.}(2018)\citenamefont
  {S{\l}omka}, \citenamefont {Suwara},\ and\ \citenamefont
  {Dunkel}}]{slomka2018nature}%
  \BibitemOpen
  \bibfield  {author} {\bibinfo {author} {\bibfnamefont {Jonasz}\ \bibnamefont
  {S{\l}omka}}, \bibinfo {author} {\bibfnamefont {Piotr}\ \bibnamefont
  {Suwara}}, \ and\ \bibinfo {author} {\bibfnamefont {J{\"o}rn}\ \bibnamefont
  {Dunkel}},\ }\bibfield  {title} {\enquote {\bibinfo {title} {The nature of
  triad interactions in active turbulence},}\ }\href@noop {} {\bibfield
  {journal} {\bibinfo  {journal} {Journal of Fluid Mechanics}\ }\textbf
  {\bibinfo {volume} {841}},\ \bibinfo {pages} {702--731} (\bibinfo {year}
  {2018})}\BibitemShut {NoStop}%
\bibitem [{\citenamefont {Soni}\ \emph {et~al.}(2019)\citenamefont {Soni},
  \citenamefont {Bililign}, \citenamefont {Magkiriadou}, \citenamefont
  {Sacanna}, \citenamefont {Bartolo}, \citenamefont {Shelley},\ and\
  \citenamefont {Irvine}}]{soni2019odd}%
  \BibitemOpen
  \bibfield  {author} {\bibinfo {author} {\bibfnamefont {Vishal}\ \bibnamefont
  {Soni}}, \bibinfo {author} {\bibfnamefont {Ephraim~S}\ \bibnamefont
  {Bililign}}, \bibinfo {author} {\bibfnamefont {Sofia}\ \bibnamefont
  {Magkiriadou}}, \bibinfo {author} {\bibfnamefont {Stefano}\ \bibnamefont
  {Sacanna}}, \bibinfo {author} {\bibfnamefont {Denis}\ \bibnamefont
  {Bartolo}}, \bibinfo {author} {\bibfnamefont {Michael~J}\ \bibnamefont
  {Shelley}}, \ and\ \bibinfo {author} {\bibfnamefont {William~TM}\
  \bibnamefont {Irvine}},\ }\bibfield  {title} {\enquote {\bibinfo {title} {The
  odd free surface flows of a colloidal chiral fluid},}\ }\href@noop {}
  {\bibfield  {journal} {\bibinfo  {journal} {Nature Physics}\ }\textbf
  {\bibinfo {volume} {15}},\ \bibinfo {pages} {1188--1194} (\bibinfo {year}
  {2019})}\BibitemShut {NoStop}%
\bibitem [{\citenamefont {Shin}\ and\ \citenamefont
  {Brangwynne}(2017)}]{shin2017liquid}%
  \BibitemOpen
  \bibfield  {author} {\bibinfo {author} {\bibfnamefont {Yongdae}\ \bibnamefont
  {Shin}}\ and\ \bibinfo {author} {\bibfnamefont {Clifford~P}\ \bibnamefont
  {Brangwynne}},\ }\bibfield  {title} {\enquote {\bibinfo {title} {Liquid phase
  condensation in cell physiology and disease},}\ }\href@noop {} {\bibfield
  {journal} {\bibinfo  {journal} {Science}\ }\textbf {\bibinfo {volume} {357}}
  (\bibinfo {year} {2017})}\BibitemShut {NoStop}%
\bibitem [{\citenamefont {Caulfield}(2020)}]{colm2020open}%
  \BibitemOpen
  \bibfield  {author} {\bibinfo {author} {\bibfnamefont {Colm-cille~P}\
  \bibnamefont {Caulfield}},\ }\bibfield  {title} {\enquote {\bibinfo {title}
  {Open questions in turbulent stratified mixing: Do we even know what we do
  not know?}}\ }\href@noop {} {\bibfield  {journal} {\bibinfo  {journal}
  {Physical Review Fluids}\ }\textbf {\bibinfo {volume} {5}},\ \bibinfo {pages}
  {110518} (\bibinfo {year} {2020})}\BibitemShut {NoStop}%
\bibitem [{\citenamefont {Woodhouse}\ and\ \citenamefont
  {Goldstein}(2012)}]{woodhouse2012spontaneous}%
  \BibitemOpen
  \bibfield  {author} {\bibinfo {author} {\bibfnamefont {Francis~G}\
  \bibnamefont {Woodhouse}}\ and\ \bibinfo {author} {\bibfnamefont {Raymond~E}\
  \bibnamefont {Goldstein}},\ }\bibfield  {title} {\enquote {\bibinfo {title}
  {Spontaneous circulation of confined active suspensions},}\ }\href {\doibase
  10.1103/PhysRevLett.109.168105} {\bibfield  {journal} {\bibinfo  {journal}
  {Physical review letters}\ }\textbf {\bibinfo {volume} {109}},\ \bibinfo
  {pages} {168105} (\bibinfo {year} {2012})}\BibitemShut {NoStop}%
\bibitem [{\citenamefont {Kumar}\ \emph
  {et~al.}(2014{\natexlab{b}})\citenamefont {Kumar}, \citenamefont {Maitra},
  \citenamefont {Sumit}, \citenamefont {Ramaswamy},\ and\ \citenamefont
  {Shivashankar}}]{kumar2014actomyosin}%
  \BibitemOpen
  \bibfield  {author} {\bibinfo {author} {\bibfnamefont {Abhishek}\
  \bibnamefont {Kumar}}, \bibinfo {author} {\bibfnamefont {Ananyo}\
  \bibnamefont {Maitra}}, \bibinfo {author} {\bibfnamefont {Madhuresh}\
  \bibnamefont {Sumit}}, \bibinfo {author} {\bibfnamefont {Sriram}\
  \bibnamefont {Ramaswamy}}, \ and\ \bibinfo {author} {\bibfnamefont
  {GV}~\bibnamefont {Shivashankar}},\ }\bibfield  {title} {\enquote {\bibinfo
  {title} {Actomyosin contractility rotates the cell nucleus},}\ }\href
  {https://doi.org/10.1038/srep03781} {\bibfield  {journal} {\bibinfo
  {journal} {Scientific reports}\ }\textbf {\bibinfo {volume} {4}},\ \bibinfo
  {pages} {1--7} (\bibinfo {year} {2014}{\natexlab{b}})}\BibitemShut {NoStop}%
\bibitem [{\citenamefont {Kosterlitz}\ and\ \citenamefont
  {Thouless}(1973)}]{kosterlitz1973ordering}%
  \BibitemOpen
  \bibfield  {author} {\bibinfo {author} {\bibfnamefont {John~Michael}\
  \bibnamefont {Kosterlitz}}\ and\ \bibinfo {author} {\bibfnamefont
  {David~James}\ \bibnamefont {Thouless}},\ }\bibfield  {title} {\enquote
  {\bibinfo {title} {Ordering, metastability and phase transitions in
  two-dimensional systems},}\ }\href@noop {} {\bibfield  {journal} {\bibinfo
  {journal} {Journal of Physics C: Solid State Physics}\ }\textbf {\bibinfo
  {volume} {6}},\ \bibinfo {pages} {1181} (\bibinfo {year} {1973})}\BibitemShut
  {NoStop}%
\bibitem [{\citenamefont {Stein}(1978)}]{stein1978kosterlitz}%
  \BibitemOpen
  \bibfield  {author} {\bibinfo {author} {\bibfnamefont {DL}~\bibnamefont
  {Stein}},\ }\bibfield  {title} {\enquote {\bibinfo {title}
  {Kosterlitz-thouless phase transitions in two-dimensional liquid crystals},}\
  }\href@noop {} {\bibfield  {journal} {\bibinfo  {journal} {Physical Review
  B}\ }\textbf {\bibinfo {volume} {18}},\ \bibinfo {pages} {2397} (\bibinfo
  {year} {1978})}\BibitemShut {NoStop}%
\bibitem [{\citenamefont {Giomi}\ \emph {et~al.}(2013)\citenamefont {Giomi},
  \citenamefont {Bowick}, \citenamefont {Ma},\ and\ \citenamefont
  {Marchetti}}]{giomi2013defect}%
  \BibitemOpen
  \bibfield  {author} {\bibinfo {author} {\bibfnamefont {Luca}\ \bibnamefont
  {Giomi}}, \bibinfo {author} {\bibfnamefont {Mark~J}\ \bibnamefont {Bowick}},
  \bibinfo {author} {\bibfnamefont {Xu}~\bibnamefont {Ma}}, \ and\ \bibinfo
  {author} {\bibfnamefont {M~Cristina}\ \bibnamefont {Marchetti}},\ }\bibfield
  {title} {\enquote {\bibinfo {title} {Defect annihilation and proliferation in
  active nematics},}\ }\href {\doibase 10.1103/PhysRevLett.110.228101}
  {\bibfield  {journal} {\bibinfo  {journal} {Physical Review Letters}\
  }\textbf {\bibinfo {volume} {110}},\ \bibinfo {pages} {228101} (\bibinfo
  {year} {2013})}\BibitemShut {NoStop}%
\bibitem [{\citenamefont {Pismen}(2013)}]{pismen2013dynamics}%
  \BibitemOpen
  \bibfield  {author} {\bibinfo {author} {\bibfnamefont {LM}~\bibnamefont
  {Pismen}},\ }\bibfield  {title} {\enquote {\bibinfo {title} {Dynamics of
  defects in an active nematic layer},}\ }\href@noop {} {\bibfield  {journal}
  {\bibinfo  {journal} {Physical Review E}\ }\textbf {\bibinfo {volume} {88}},\
  \bibinfo {pages} {050502} (\bibinfo {year} {2013})}\BibitemShut {NoStop}%
\bibitem [{\citenamefont {Shankar}\ \emph {et~al.}(2018)\citenamefont
  {Shankar}, \citenamefont {Ramaswamy}, \citenamefont {Marchetti},\ and\
  \citenamefont {Bowick}}]{shankar2018defect}%
  \BibitemOpen
  \bibfield  {author} {\bibinfo {author} {\bibfnamefont {Suraj}\ \bibnamefont
  {Shankar}}, \bibinfo {author} {\bibfnamefont {Sriram}\ \bibnamefont
  {Ramaswamy}}, \bibinfo {author} {\bibfnamefont {M~Cristina}\ \bibnamefont
  {Marchetti}}, \ and\ \bibinfo {author} {\bibfnamefont {Mark~J}\ \bibnamefont
  {Bowick}},\ }\bibfield  {title} {\enquote {\bibinfo {title} {Defect unbinding
  in active nematics},}\ }\href {\doibase 10.1103/PhysRevLett.121.108002}
  {\bibfield  {journal} {\bibinfo  {journal} {Physical Review Letters}\
  }\textbf {\bibinfo {volume} {121}},\ \bibinfo {pages} {108002} (\bibinfo
  {year} {2018})}\BibitemShut {NoStop}%
\bibitem [{\citenamefont {Vafa}\ \emph {et~al.}(2020)\citenamefont {Vafa},
  \citenamefont {Bowick}, \citenamefont {Marchetti},\ and\ \citenamefont
  {Shraiman}}]{vafa2020multi}%
  \BibitemOpen
  \bibfield  {author} {\bibinfo {author} {\bibfnamefont {Farzan}\ \bibnamefont
  {Vafa}}, \bibinfo {author} {\bibfnamefont {Mark~J}\ \bibnamefont {Bowick}},
  \bibinfo {author} {\bibfnamefont {M~Cristina}\ \bibnamefont {Marchetti}}, \
  and\ \bibinfo {author} {\bibfnamefont {Boris~I}\ \bibnamefont {Shraiman}},\
  }\bibfield  {title} {\enquote {\bibinfo {title} {Multi-defect dynamics in
  active nematics},}\ }\href@noop {} {\bibfield  {journal} {\bibinfo  {journal}
  {arXiv preprint arXiv:2007.02947}\ } (\bibinfo {year} {2020})}\BibitemShut
  {NoStop}%
\bibitem [{\citenamefont {Giomi}\ \emph {et~al.}(2014)\citenamefont {Giomi},
  \citenamefont {Bowick}, \citenamefont {Mishra}, \citenamefont {Sknepnek},\
  and\ \citenamefont {Cristina~Marchetti}}]{giomi2014defect}%
  \BibitemOpen
  \bibfield  {author} {\bibinfo {author} {\bibfnamefont {Luca}\ \bibnamefont
  {Giomi}}, \bibinfo {author} {\bibfnamefont {Mark~J}\ \bibnamefont {Bowick}},
  \bibinfo {author} {\bibfnamefont {Prashant}\ \bibnamefont {Mishra}}, \bibinfo
  {author} {\bibfnamefont {Rastko}\ \bibnamefont {Sknepnek}}, \ and\ \bibinfo
  {author} {\bibfnamefont {M}~\bibnamefont {Cristina~Marchetti}},\ }\bibfield
  {title} {\enquote {\bibinfo {title} {Defect dynamics in active nematics},}\
  }\href {\doibase 10.1098/rsta.2013.0365} {\bibfield  {journal} {\bibinfo
  {journal} {Philosophical Transactions of the Royal Society A: Mathematical,
  Physical and Engineering Sciences}\ }\textbf {\bibinfo {volume} {372}},\
  \bibinfo {pages} {20130365} (\bibinfo {year} {2014})}\BibitemShut {NoStop}%
\bibitem [{\citenamefont {Wright}\ and\ \citenamefont
  {Mermin}(1989)}]{wright1989crystalline}%
  \BibitemOpen
  \bibfield  {author} {\bibinfo {author} {\bibfnamefont {David~C}\ \bibnamefont
  {Wright}}\ and\ \bibinfo {author} {\bibfnamefont {N~David}\ \bibnamefont
  {Mermin}},\ }\bibfield  {title} {\enquote {\bibinfo {title} {Crystalline
  liquids: the blue phases},}\ }\href
  {https://journals.aps.org/rmp/abstract/10.1103/RevModPhys.61.385} {\bibfield
  {journal} {\bibinfo  {journal} {Reviews of Modern physics}\ }\textbf
  {\bibinfo {volume} {61}},\ \bibinfo {pages} {385} (\bibinfo {year}
  {1989})}\BibitemShut {NoStop}%
\bibitem [{\citenamefont {Shankar}\ and\ \citenamefont
  {Marchetti}(2019)}]{shankar2019hydrodynamics}%
  \BibitemOpen
  \bibfield  {author} {\bibinfo {author} {\bibfnamefont {Suraj}\ \bibnamefont
  {Shankar}}\ and\ \bibinfo {author} {\bibfnamefont {M~Cristina}\ \bibnamefont
  {Marchetti}},\ }\bibfield  {title} {\enquote {\bibinfo {title} {Hydrodynamics
  of active defects: from order to chaos to defect ordering},}\ }\href
  {\doibase 10.1103/PhysRevLett.124.028002} {\bibfield  {journal} {\bibinfo
  {journal} {Physical Review X}\ }\textbf {\bibinfo {volume} {9}},\ \bibinfo
  {pages} {041047} (\bibinfo {year} {2019})}\BibitemShut {NoStop}%
\bibitem [{\citenamefont {Srivastava}\ \emph {et~al.}(2016)\citenamefont
  {Srivastava}, \citenamefont {Mishra},\ and\ \citenamefont
  {Marchetti}}]{srivastava2016negative}%
  \BibitemOpen
  \bibfield  {author} {\bibinfo {author} {\bibfnamefont {Pragya}\ \bibnamefont
  {Srivastava}}, \bibinfo {author} {\bibfnamefont {Prashant}\ \bibnamefont
  {Mishra}}, \ and\ \bibinfo {author} {\bibfnamefont {M~Cristina}\ \bibnamefont
  {Marchetti}},\ }\bibfield  {title} {\enquote {\bibinfo {title} {Negative
  stiffness and modulated states in active nematics},}\ }\href {\doibase
  10.1039/C6SM01493C} {\bibfield  {journal} {\bibinfo  {journal} {Soft matter}\
  }\textbf {\bibinfo {volume} {12}},\ \bibinfo {pages} {8214--8225} (\bibinfo
  {year} {2016})}\BibitemShut {NoStop}%
\bibitem [{\citenamefont {Putzig}\ \emph {et~al.}(2016)\citenamefont {Putzig},
  \citenamefont {Redner}, \citenamefont {Baskaran},\ and\ \citenamefont
  {Baskaran}}]{putzig2016instabilities}%
  \BibitemOpen
  \bibfield  {author} {\bibinfo {author} {\bibfnamefont {Elias}\ \bibnamefont
  {Putzig}}, \bibinfo {author} {\bibfnamefont {Gabriel~S}\ \bibnamefont
  {Redner}}, \bibinfo {author} {\bibfnamefont {Arvind}\ \bibnamefont
  {Baskaran}}, \ and\ \bibinfo {author} {\bibfnamefont {Aparna}\ \bibnamefont
  {Baskaran}},\ }\bibfield  {title} {\enquote {\bibinfo {title} {Instabilities,
  defects, and defect ordering in an overdamped active nematic},}\ }\href
  {\doibase 10.1039/C6SM00268D} {\bibfield  {journal} {\bibinfo  {journal}
  {Soft Matter}\ }\textbf {\bibinfo {volume} {12}},\ \bibinfo {pages}
  {3854--3859} (\bibinfo {year} {2016})}\BibitemShut {NoStop}%
\bibitem [{\citenamefont {Patelli}\ \emph {et~al.}(2019)\citenamefont
  {Patelli}, \citenamefont {Djafer-Cherif}, \citenamefont {Aranson},
  \citenamefont {Bertin},\ and\ \citenamefont
  {Chat{\'e}}}]{patelli2019understanding}%
  \BibitemOpen
  \bibfield  {author} {\bibinfo {author} {\bibfnamefont {Aurelio}\ \bibnamefont
  {Patelli}}, \bibinfo {author} {\bibfnamefont {Ilyas}\ \bibnamefont
  {Djafer-Cherif}}, \bibinfo {author} {\bibfnamefont {Igor~S}\ \bibnamefont
  {Aranson}}, \bibinfo {author} {\bibfnamefont {Eric}\ \bibnamefont {Bertin}},
  \ and\ \bibinfo {author} {\bibfnamefont {Hugues}\ \bibnamefont {Chat{\'e}}},\
  }\bibfield  {title} {\enquote {\bibinfo {title} {Understanding dense active
  nematics from microscopic models},}\ }\href@noop {} {\bibfield  {journal}
  {\bibinfo  {journal} {Physical review letters}\ }\textbf {\bibinfo {volume}
  {123}},\ \bibinfo {pages} {258001} (\bibinfo {year} {2019})}\BibitemShut
  {NoStop}%
\bibitem [{\citenamefont {Pearce}\ \emph {et~al.}(2020)\citenamefont {Pearce},
  \citenamefont {Nambisan}, \citenamefont {Ellis}, \citenamefont
  {Fernandez-Nieves},\ and\ \citenamefont {Giomi}}]{Pearce2020}%
  \BibitemOpen
  \bibfield  {author} {\bibinfo {author} {\bibfnamefont {D.~J.~G.}\
  \bibnamefont {Pearce}}, \bibinfo {author} {\bibfnamefont {J.}~\bibnamefont
  {Nambisan}}, \bibinfo {author} {\bibfnamefont {P.~W.}\ \bibnamefont {Ellis}},
  \bibinfo {author} {\bibfnamefont {A.}~\bibnamefont {Fernandez-Nieves}}, \
  and\ \bibinfo {author} {\bibfnamefont {L.}~\bibnamefont {Giomi}},\ }\bibfield
   {title} {\enquote {\bibinfo {title} {Scale-free defect ordering in passive
  and active nematics},}\ }\href@noop {} {\bibfield  {journal} {\bibinfo
  {journal} {arXiv preprint arXiv:2004.13704}\ } (\bibinfo {year}
  {2020})}\BibitemShut {NoStop}%
\bibitem [{\citenamefont {Thijssen}\ \emph {et~al.}(2020)\citenamefont
  {Thijssen}, \citenamefont {Nejad},\ and\ \citenamefont
  {Yeomans}}]{Thijssen2020}%
  \BibitemOpen
  \bibfield  {author} {\bibinfo {author} {\bibfnamefont {Kristian}\
  \bibnamefont {Thijssen}}, \bibinfo {author} {\bibfnamefont {Mehrana~R.}\
  \bibnamefont {Nejad}}, \ and\ \bibinfo {author} {\bibfnamefont {Julia~M.}\
  \bibnamefont {Yeomans}},\ }\bibfield  {title} {\enquote {\bibinfo {title}
  {Large scale ordering of active defects},}\ }\href@noop {} {\bibfield
  {journal} {\bibinfo  {journal} {arXiv preprint arXiv:2005.01164}\ } (\bibinfo
  {year} {2020})}\BibitemShut {NoStop}%
\bibitem [{\citenamefont {Oza}\ and\ \citenamefont
  {Dunkel}(2016)}]{oza2016antipolar}%
  \BibitemOpen
  \bibfield  {author} {\bibinfo {author} {\bibfnamefont {Anand~U}\ \bibnamefont
  {Oza}}\ and\ \bibinfo {author} {\bibfnamefont {J{\"o}rn}\ \bibnamefont
  {Dunkel}},\ }\bibfield  {title} {\enquote {\bibinfo {title} {Antipolar
  ordering of topological defects in active liquid crystals},}\ }\href
  {\doibase 10.1088/1367-2630/18/9/093006} {\bibfield  {journal} {\bibinfo
  {journal} {New Journal of Physics}\ }\textbf {\bibinfo {volume} {18}},\
  \bibinfo {pages} {093006} (\bibinfo {year} {2016})}\BibitemShut {NoStop}%
\bibitem [{\citenamefont {DeCamp}\ \emph {et~al.}(2015)\citenamefont {DeCamp},
  \citenamefont {Redner}, \citenamefont {Baskaran}, \citenamefont {Hagan},\
  and\ \citenamefont {Dogic}}]{decamp2015orientational}%
  \BibitemOpen
  \bibfield  {author} {\bibinfo {author} {\bibfnamefont {Stephen~J}\
  \bibnamefont {DeCamp}}, \bibinfo {author} {\bibfnamefont {Gabriel~S}\
  \bibnamefont {Redner}}, \bibinfo {author} {\bibfnamefont {Aparna}\
  \bibnamefont {Baskaran}}, \bibinfo {author} {\bibfnamefont {Michael~F}\
  \bibnamefont {Hagan}}, \ and\ \bibinfo {author} {\bibfnamefont {Zvonimir}\
  \bibnamefont {Dogic}},\ }\bibfield  {title} {\enquote {\bibinfo {title}
  {Orientational order of motile defects in active nematics},}\ }\href
  {\doibase 10.1038/nmat4387} {\bibfield  {journal} {\bibinfo  {journal}
  {Nature materials}\ }\textbf {\bibinfo {volume} {14}},\ \bibinfo {pages}
  {1110} (\bibinfo {year} {2015})}\BibitemShut {NoStop}%
\bibitem [{\citenamefont {Doostmohammadi}\ \emph
  {et~al.}(2016{\natexlab{b}})\citenamefont {Doostmohammadi}, \citenamefont
  {Thampi},\ and\ \citenamefont {Yeomans}}]{doostmohammadi2016defect}%
  \BibitemOpen
  \bibfield  {author} {\bibinfo {author} {\bibfnamefont {Amin}\ \bibnamefont
  {Doostmohammadi}}, \bibinfo {author} {\bibfnamefont {Sumesh~P}\ \bibnamefont
  {Thampi}}, \ and\ \bibinfo {author} {\bibfnamefont {Julia~M}\ \bibnamefont
  {Yeomans}},\ }\bibfield  {title} {\enquote {\bibinfo {title} {Defect-mediated
  morphologies in growing cell colonies},}\ }\href@noop {} {\bibfield
  {journal} {\bibinfo  {journal} {Physical review letters}\ }\textbf {\bibinfo
  {volume} {117}},\ \bibinfo {pages} {048102} (\bibinfo {year}
  {2016}{\natexlab{b}})}\BibitemShut {NoStop}%
\bibitem [{\citenamefont {DeVries}\ \emph {et~al.}(2007)\citenamefont
  {DeVries}, \citenamefont {Brunnbauer}, \citenamefont {Hu}, \citenamefont
  {Jackson}, \citenamefont {Long}, \citenamefont {Neltner}, \citenamefont
  {Uzun}, \citenamefont {Wunsch},\ and\ \citenamefont
  {Stellacci}}]{stellacci2007divalent}%
  \BibitemOpen
  \bibfield  {author} {\bibinfo {author} {\bibfnamefont {A.~Gretchen}\
  \bibnamefont {DeVries}}, \bibinfo {author} {\bibfnamefont {Markus}\
  \bibnamefont {Brunnbauer}}, \bibinfo {author} {\bibfnamefont {Ying}\
  \bibnamefont {Hu}}, \bibinfo {author} {\bibfnamefont {M.~Alicia}\
  \bibnamefont {Jackson}}, \bibinfo {author} {\bibfnamefont {Brenda}\
  \bibnamefont {Long}}, \bibinfo {author} {\bibfnamefont {T.~Brian}\
  \bibnamefont {Neltner}}, \bibinfo {author} {\bibfnamefont {Oktay}\
  \bibnamefont {Uzun}}, \bibinfo {author} {\bibfnamefont {H.~Benjamin}\
  \bibnamefont {Wunsch}}, \ and\ \bibinfo {author} {\bibfnamefont {Francesco}\
  \bibnamefont {Stellacci}},\ }\bibfield  {title} {\enquote {\bibinfo {title}
  {Divalent metal nanoparticles},}\ }\href {\doibase 10.1126/science.1133162}
  {\bibfield  {journal} {\bibinfo  {journal} {Science}\ }\textbf {\bibinfo
  {volume} {315}},\ \bibinfo {pages} {358} (\bibinfo {year}
  {2007})}\BibitemShut {NoStop}%
\bibitem [{\citenamefont {Zandi}\ \emph {et~al.}(2020)\citenamefont {Zandi},
  \citenamefont {Dragnea}, \citenamefont {Travesset},\ and\ \citenamefont
  {Podgornik}}]{zandi2020virus}%
  \BibitemOpen
  \bibfield  {author} {\bibinfo {author} {\bibfnamefont {Roya}\ \bibnamefont
  {Zandi}}, \bibinfo {author} {\bibfnamefont {Bogdan}\ \bibnamefont {Dragnea}},
  \bibinfo {author} {\bibfnamefont {Alex}\ \bibnamefont {Travesset}}, \ and\
  \bibinfo {author} {\bibfnamefont {Rudolf}\ \bibnamefont {Podgornik}},\
  }\bibfield  {title} {\enquote {\bibinfo {title} {On virus growth and form},}\
  }\href@noop {} {\bibfield  {journal} {\bibinfo  {journal} {Physics Reports}\
  }\textbf {\bibinfo {volume} {847}},\ \bibinfo {pages} {1--102} (\bibinfo
  {year} {2020})}\BibitemShut {NoStop}%
\bibitem [{\citenamefont {Duclos}\ \emph {et~al.}(2017)\citenamefont {Duclos},
  \citenamefont {Erlenk{\"a}mper}, \citenamefont {Joanny},\ and\ \citenamefont
  {Silberzan}}]{duclos2017topological}%
  \BibitemOpen
  \bibfield  {author} {\bibinfo {author} {\bibfnamefont {Guillaume}\
  \bibnamefont {Duclos}}, \bibinfo {author} {\bibfnamefont {Christoph}\
  \bibnamefont {Erlenk{\"a}mper}}, \bibinfo {author} {\bibfnamefont
  {Jean-Fran{\c{c}}ois}\ \bibnamefont {Joanny}}, \ and\ \bibinfo {author}
  {\bibfnamefont {Pascal}\ \bibnamefont {Silberzan}},\ }\bibfield  {title}
  {\enquote {\bibinfo {title} {Topological defects in confined populations of
  spindle-shaped cells},}\ }\href {\doibase 10.1038/nphys3876} {\bibfield
  {journal} {\bibinfo  {journal} {Nature Physics}\ }\textbf {\bibinfo {volume}
  {13}},\ \bibinfo {pages} {58} (\bibinfo {year} {2017})}\BibitemShut {NoStop}%
\bibitem [{\citenamefont {Copenhagen}\ \emph {et~al.}(2021)\citenamefont
  {Copenhagen}, \citenamefont {Alert}, \citenamefont {Wingreen},\ and\
  \citenamefont {Shaevitz}}]{copenhagen2021topological}%
  \BibitemOpen
  \bibfield  {author} {\bibinfo {author} {\bibfnamefont {Katherine}\
  \bibnamefont {Copenhagen}}, \bibinfo {author} {\bibfnamefont {Ricard}\
  \bibnamefont {Alert}}, \bibinfo {author} {\bibfnamefont {Ned~S}\ \bibnamefont
  {Wingreen}}, \ and\ \bibinfo {author} {\bibfnamefont {Joshua~W}\ \bibnamefont
  {Shaevitz}},\ }\bibfield  {title} {\enquote {\bibinfo {title} {Topological
  defects promote layer formation in myxococcus xanthus colonies},}\
  }\href@noop {} {\bibfield  {journal} {\bibinfo  {journal} {Nature Physics}\
  }\textbf {\bibinfo {volume} {17}},\ \bibinfo {pages} {211--215} (\bibinfo
  {year} {2021})}\BibitemShut {NoStop}%
\bibitem [{\citenamefont {Shimaya}\ and\ \citenamefont
  {Takeuchi}(2021)}]{shimaya20213d}%
  \BibitemOpen
  \bibfield  {author} {\bibinfo {author} {\bibfnamefont {Takuro}\ \bibnamefont
  {Shimaya}}\ and\ \bibinfo {author} {\bibfnamefont {Kazumasa~A}\ \bibnamefont
  {Takeuchi}},\ }\bibfield  {title} {\enquote {\bibinfo {title} {3d-induced
  polar order and topological defects in growing bacterial populations},}\
  }\href {https://arxiv.org/abs/2106.10954} {\bibfield  {journal} {\bibinfo
  {journal} {arXiv preprint arXiv:2106.10954}\ } (\bibinfo {year}
  {2021})}\BibitemShut {NoStop}%
\bibitem [{\citenamefont {Sarkar}\ \emph {et~al.}(2021)\citenamefont {Sarkar},
  \citenamefont {Yashunsky}, \citenamefont {Br{\'e}zin}, \citenamefont
  {Mercader}, \citenamefont {Aryaksama}, \citenamefont {Lacroix}, \citenamefont
  {Risler}, \citenamefont {Joanny},\ and\ \citenamefont
  {Silberzan}}]{sarkar2021crisscross}%
  \BibitemOpen
  \bibfield  {author} {\bibinfo {author} {\bibfnamefont {Trinish}\ \bibnamefont
  {Sarkar}}, \bibinfo {author} {\bibfnamefont {Victor}\ \bibnamefont
  {Yashunsky}}, \bibinfo {author} {\bibfnamefont {Louis}\ \bibnamefont
  {Br{\'e}zin}}, \bibinfo {author} {\bibfnamefont {Carles~Blanch}\ \bibnamefont
  {Mercader}}, \bibinfo {author} {\bibfnamefont {Thibault}\ \bibnamefont
  {Aryaksama}}, \bibinfo {author} {\bibfnamefont {Mathilde}\ \bibnamefont
  {Lacroix}}, \bibinfo {author} {\bibfnamefont {Thomas}\ \bibnamefont
  {Risler}}, \bibinfo {author} {\bibfnamefont {Jean-Fran{\c{c}}ois}\
  \bibnamefont {Joanny}}, \ and\ \bibinfo {author} {\bibfnamefont {Pascal}\
  \bibnamefont {Silberzan}},\ }\bibfield  {title} {\enquote {\bibinfo {title}
  {Crisscross multilayering of cell sheets},}\ }\href@noop {} {\bibfield
  {journal} {\bibinfo  {journal} {bioRxiv}\ } (\bibinfo {year}
  {2021})}\BibitemShut {NoStop}%
\bibitem [{\citenamefont {Hoffmann}\ \emph {et~al.}(2021)\citenamefont
  {Hoffmann}, \citenamefont {Carenza}, \citenamefont {Eckert},\ and\
  \citenamefont {Giomi}}]{hoffmann2021defect}%
  \BibitemOpen
  \bibfield  {author} {\bibinfo {author} {\bibfnamefont {Ludwig~A}\
  \bibnamefont {Hoffmann}}, \bibinfo {author} {\bibfnamefont {Livio~N}\
  \bibnamefont {Carenza}}, \bibinfo {author} {\bibfnamefont {Julia}\
  \bibnamefont {Eckert}}, \ and\ \bibinfo {author} {\bibfnamefont {Luca}\
  \bibnamefont {Giomi}},\ }\bibfield  {title} {\enquote {\bibinfo {title}
  {Defect-mediated morphogenesis},}\ }\href {https://arxiv.org/abs/2105.15200}
  {\bibfield  {journal} {\bibinfo  {journal} {arXiv preprint arXiv:2105.15200}\
  } (\bibinfo {year} {2021})}\BibitemShut {NoStop}%
\bibitem [{\citenamefont {Tan}\ \emph {et~al.}(2020)\citenamefont {Tan},
  \citenamefont {Liu}, \citenamefont {Miller}, \citenamefont {Tekant},
  \citenamefont {Dunkel},\ and\ \citenamefont {Fakhri}}]{2020TzerHan_NatPhys}%
  \BibitemOpen
  \bibfield  {author} {\bibinfo {author} {\bibfnamefont {Tzer~Han}\
  \bibnamefont {Tan}}, \bibinfo {author} {\bibfnamefont {Jinghui}\ \bibnamefont
  {Liu}}, \bibinfo {author} {\bibfnamefont {Pearson~W.}\ \bibnamefont
  {Miller}}, \bibinfo {author} {\bibfnamefont {Melis}\ \bibnamefont {Tekant}},
  \bibinfo {author} {\bibfnamefont {J{\"o}rn}\ \bibnamefont {Dunkel}}, \ and\
  \bibinfo {author} {\bibfnamefont {Nikta}\ \bibnamefont {Fakhri}},\ }\bibfield
   {title} {\enquote {\bibinfo {title} {Topological turbulence in the membrane
  of a living cell},}\ }\href {\doibase 10.1038/s41567-020-0841-9} {\bibfield
  {journal} {\bibinfo  {journal} {Nat. Phys.}\ }\textbf {\bibinfo {volume}
  {16}},\ \bibinfo {pages} {657--662} (\bibinfo {year} {2020})}\BibitemShut
  {NoStop}%
\bibitem [{\citenamefont {Wioland}\ \emph {et~al.}(2013)\citenamefont
  {Wioland}, \citenamefont {Woodhouse}, \citenamefont {Dunkel}, \citenamefont
  {Kessler},\ and\ \citenamefont {Goldstein}}]{wioland2013confinement}%
  \BibitemOpen
  \bibfield  {author} {\bibinfo {author} {\bibfnamefont {Hugo}\ \bibnamefont
  {Wioland}}, \bibinfo {author} {\bibfnamefont {Francis~G}\ \bibnamefont
  {Woodhouse}}, \bibinfo {author} {\bibfnamefont {J{\"o}rn}\ \bibnamefont
  {Dunkel}}, \bibinfo {author} {\bibfnamefont {John~O}\ \bibnamefont
  {Kessler}}, \ and\ \bibinfo {author} {\bibfnamefont {Raymond~E}\ \bibnamefont
  {Goldstein}},\ }\bibfield  {title} {\enquote {\bibinfo {title} {Confinement
  stabilizes a bacterial suspension into a spiral vortex},}\ }\href {\doibase
  10.1103/PhysRevLett.110.268102} {\bibfield  {journal} {\bibinfo  {journal}
  {Physical Review Letters}\ }\textbf {\bibinfo {volume} {110}},\ \bibinfo
  {pages} {268102} (\bibinfo {year} {2013})}\BibitemShut {NoStop}%
\bibitem [{\citenamefont {Duclos}\ \emph {et~al.}(2018)\citenamefont {Duclos},
  \citenamefont {Blanch-Mercader}, \citenamefont {Yashunsky}, \citenamefont
  {Salbreux}, \citenamefont {Joanny}, \citenamefont {Prost},\ and\
  \citenamefont {Silberzan}}]{duclos2018spontaneous}%
  \BibitemOpen
  \bibfield  {author} {\bibinfo {author} {\bibfnamefont {G}~\bibnamefont
  {Duclos}}, \bibinfo {author} {\bibfnamefont {C}~\bibnamefont
  {Blanch-Mercader}}, \bibinfo {author} {\bibfnamefont {V}~\bibnamefont
  {Yashunsky}}, \bibinfo {author} {\bibfnamefont {G}~\bibnamefont {Salbreux}},
  \bibinfo {author} {\bibfnamefont {J-F}\ \bibnamefont {Joanny}}, \bibinfo
  {author} {\bibfnamefont {J}~\bibnamefont {Prost}}, \ and\ \bibinfo {author}
  {\bibfnamefont {P}~\bibnamefont {Silberzan}},\ }\bibfield  {title} {\enquote
  {\bibinfo {title} {Spontaneous shear flow in confined cellular nematics},}\
  }\href@noop {} {\bibfield  {journal} {\bibinfo  {journal} {Nature physics}\
  }\textbf {\bibinfo {volume} {14}},\ \bibinfo {pages} {728} (\bibinfo {year}
  {2018})}\BibitemShut {NoStop}%
\bibitem [{\citenamefont {Keber}\ \emph {et~al.}(2014)\citenamefont {Keber},
  \citenamefont {Loiseau}, \citenamefont {Sanchez}, \citenamefont {DeCamp},
  \citenamefont {Giomi}, \citenamefont {Bowick}, \citenamefont {Marchetti},
  \citenamefont {Dogic},\ and\ \citenamefont {Bausch}}]{keber2014topology}%
  \BibitemOpen
  \bibfield  {author} {\bibinfo {author} {\bibfnamefont {Felix~C}\ \bibnamefont
  {Keber}}, \bibinfo {author} {\bibfnamefont {Etienne}\ \bibnamefont
  {Loiseau}}, \bibinfo {author} {\bibfnamefont {Tim}\ \bibnamefont {Sanchez}},
  \bibinfo {author} {\bibfnamefont {Stephen~J}\ \bibnamefont {DeCamp}},
  \bibinfo {author} {\bibfnamefont {Luca}\ \bibnamefont {Giomi}}, \bibinfo
  {author} {\bibfnamefont {Mark~J}\ \bibnamefont {Bowick}}, \bibinfo {author}
  {\bibfnamefont {M~Cristina}\ \bibnamefont {Marchetti}}, \bibinfo {author}
  {\bibfnamefont {Zvonimir}\ \bibnamefont {Dogic}}, \ and\ \bibinfo {author}
  {\bibfnamefont {Andreas~R}\ \bibnamefont {Bausch}},\ }\bibfield  {title}
  {\enquote {\bibinfo {title} {Topology and dynamics of active nematic
  vesicles},}\ }\href {\doibase 10.1126/science.1254784} {\bibfield  {journal}
  {\bibinfo  {journal} {Science}\ }\textbf {\bibinfo {volume} {345}},\ \bibinfo
  {pages} {1135--1139} (\bibinfo {year} {2014})}\BibitemShut {NoStop}%
\bibitem [{\citenamefont {Sknepnek}\ and\ \citenamefont
  {Henkes}(2015)}]{sknepnek2015active}%
  \BibitemOpen
  \bibfield  {author} {\bibinfo {author} {\bibfnamefont {Rastko}\ \bibnamefont
  {Sknepnek}}\ and\ \bibinfo {author} {\bibfnamefont {Silke}\ \bibnamefont
  {Henkes}},\ }\bibfield  {title} {\enquote {\bibinfo {title} {Active swarms on
  a sphere},}\ }\href@noop {} {\bibfield  {journal} {\bibinfo  {journal}
  {Physical Review E}\ }\textbf {\bibinfo {volume} {91}},\ \bibinfo {pages}
  {022306} (\bibinfo {year} {2015})}\BibitemShut {NoStop}%
\bibitem [{\citenamefont {Shankar}\ \emph {et~al.}(2017)\citenamefont
  {Shankar}, \citenamefont {Bowick},\ and\ \citenamefont
  {Marchetti}}]{shankar2017topological}%
  \BibitemOpen
  \bibfield  {author} {\bibinfo {author} {\bibfnamefont {Suraj}\ \bibnamefont
  {Shankar}}, \bibinfo {author} {\bibfnamefont {Mark~J}\ \bibnamefont
  {Bowick}}, \ and\ \bibinfo {author} {\bibfnamefont {M~Cristina}\ \bibnamefont
  {Marchetti}},\ }\bibfield  {title} {\enquote {\bibinfo {title} {Topological
  sound and flocking on curved surfaces},}\ }\href@noop {} {\bibfield
  {journal} {\bibinfo  {journal} {Physical Review X}\ }\textbf {\bibinfo
  {volume} {7}},\ \bibinfo {pages} {031039} (\bibinfo {year}
  {2017})}\BibitemShut {NoStop}%
\bibitem [{\citenamefont {Ellis}\ \emph {et~al.}(2018)\citenamefont {Ellis},
  \citenamefont {Pearce}, \citenamefont {Chang}, \citenamefont {Goldsztein},
  \citenamefont {Giomi},\ and\ \citenamefont
  {Fernandez-Nieves}}]{ellis2018curvature}%
  \BibitemOpen
  \bibfield  {author} {\bibinfo {author} {\bibfnamefont {Perry~W}\ \bibnamefont
  {Ellis}}, \bibinfo {author} {\bibfnamefont {Daniel~JG}\ \bibnamefont
  {Pearce}}, \bibinfo {author} {\bibfnamefont {Ya-Wen}\ \bibnamefont {Chang}},
  \bibinfo {author} {\bibfnamefont {Guillermo}\ \bibnamefont {Goldsztein}},
  \bibinfo {author} {\bibfnamefont {Luca}\ \bibnamefont {Giomi}}, \ and\
  \bibinfo {author} {\bibfnamefont {Alberto}\ \bibnamefont
  {Fernandez-Nieves}},\ }\bibfield  {title} {\enquote {\bibinfo {title}
  {Curvature-induced defect unbinding and dynamics in active nematic
  toroids},}\ }\href {\doibase 10.1038/nphys4276} {\bibfield  {journal}
  {\bibinfo  {journal} {Nature Physics}\ }\textbf {\bibinfo {volume} {14}},\
  \bibinfo {pages} {85} (\bibinfo {year} {2018})}\BibitemShut {NoStop}%
\bibitem [{\citenamefont {Henkes}\ \emph {et~al.}(2018)\citenamefont {Henkes},
  \citenamefont {Marchetti},\ and\ \citenamefont
  {Sknepnek}}]{henkes2018dynamical}%
  \BibitemOpen
  \bibfield  {author} {\bibinfo {author} {\bibfnamefont {Silke}\ \bibnamefont
  {Henkes}}, \bibinfo {author} {\bibfnamefont {M~Cristina}\ \bibnamefont
  {Marchetti}}, \ and\ \bibinfo {author} {\bibfnamefont {Rastko}\ \bibnamefont
  {Sknepnek}},\ }\bibfield  {title} {\enquote {\bibinfo {title} {Dynamical
  patterns in nematic active matter on a sphere},}\ }\href@noop {} {\bibfield
  {journal} {\bibinfo  {journal} {Physical Review E}\ }\textbf {\bibinfo
  {volume} {97}},\ \bibinfo {pages} {042605} (\bibinfo {year}
  {2018})}\BibitemShut {NoStop}%
\bibitem [{\citenamefont {Guillamat}\ \emph {et~al.}(2016)\citenamefont
  {Guillamat}, \citenamefont {Ign{\'e}s-Mullol},\ and\ \citenamefont
  {Sagu{\'e}s}}]{guillamat2016control}%
  \BibitemOpen
  \bibfield  {author} {\bibinfo {author} {\bibfnamefont {Pau}\ \bibnamefont
  {Guillamat}}, \bibinfo {author} {\bibfnamefont {Jordi}\ \bibnamefont
  {Ign{\'e}s-Mullol}}, \ and\ \bibinfo {author} {\bibfnamefont {Francesc}\
  \bibnamefont {Sagu{\'e}s}},\ }\bibfield  {title} {\enquote {\bibinfo {title}
  {Control of active liquid crystals with a magnetic field},}\ }\href@noop {}
  {\bibfield  {journal} {\bibinfo  {journal} {Proceedings of the National
  Academy of Sciences}\ }\textbf {\bibinfo {volume} {113}},\ \bibinfo {pages}
  {5498--5502} (\bibinfo {year} {2016})}\BibitemShut {NoStop}%
\bibitem [{\citenamefont {Ross}\ \emph {et~al.}(2019)\citenamefont {Ross},
  \citenamefont {Lee}, \citenamefont {Qu}, \citenamefont {Banks}, \citenamefont
  {Phillips},\ and\ \citenamefont {Thomson}}]{ross2019controlling}%
  \BibitemOpen
  \bibfield  {author} {\bibinfo {author} {\bibfnamefont {Tyler~D}\ \bibnamefont
  {Ross}}, \bibinfo {author} {\bibfnamefont {Heun~Jin}\ \bibnamefont {Lee}},
  \bibinfo {author} {\bibfnamefont {Zijie}\ \bibnamefont {Qu}}, \bibinfo
  {author} {\bibfnamefont {Rachel~A}\ \bibnamefont {Banks}}, \bibinfo {author}
  {\bibfnamefont {Rob}\ \bibnamefont {Phillips}}, \ and\ \bibinfo {author}
  {\bibfnamefont {Matt}\ \bibnamefont {Thomson}},\ }\bibfield  {title}
  {\enquote {\bibinfo {title} {Controlling organization and forces in active
  matter through optically defined boundaries},}\ }\href {\doibase
  10.1038/s41586-019-1447-1} {\bibfield  {journal} {\bibinfo  {journal}
  {Nature}\ }\textbf {\bibinfo {volume} {572}},\ \bibinfo {pages} {224--229}
  (\bibinfo {year} {2019})}\BibitemShut {NoStop}%
\bibitem [{\citenamefont {Gong}\ \emph {et~al.}(2020)\citenamefont {Gong},
  \citenamefont {Mathijssen}, \citenamefont {Bryant},\ and\ \citenamefont
  {Prakash}}]{gong2020engineering}%
  \BibitemOpen
  \bibfield  {author} {\bibinfo {author} {\bibfnamefont {Xingting}\
  \bibnamefont {Gong}}, \bibinfo {author} {\bibfnamefont {Arnold}\ \bibnamefont
  {Mathijssen}}, \bibinfo {author} {\bibfnamefont {Zev}\ \bibnamefont
  {Bryant}}, \ and\ \bibinfo {author} {\bibfnamefont {Manu}\ \bibnamefont
  {Prakash}},\ }\bibfield  {title} {\enquote {\bibinfo {title} {Engineering
  reconfigurable flow patterns via surface-driven light-controlled active
  matter},}\ }\href@noop {} {\bibfield  {journal} {\bibinfo  {journal} {arXiv
  preprint arXiv:2004.01368}\ } (\bibinfo {year} {2020})}\BibitemShut {NoStop}%
\bibitem [{\citenamefont {Zhang}\ \emph {et~al.}(2021)\citenamefont {Zhang},
  \citenamefont {Redford}, \citenamefont {Ruijgrok}, \citenamefont {Kumar},
  \citenamefont {Mozaffari}, \citenamefont {Zemsky}, \citenamefont {Dinner},
  \citenamefont {Vitelli}, \citenamefont {Bryant}, \citenamefont {Gardel} \emph
  {et~al.}}]{zhang2021spatiotemporal}%
  \BibitemOpen
  \bibfield  {author} {\bibinfo {author} {\bibfnamefont {Rui}\ \bibnamefont
  {Zhang}}, \bibinfo {author} {\bibfnamefont {Steven~A}\ \bibnamefont
  {Redford}}, \bibinfo {author} {\bibfnamefont {Paul~V}\ \bibnamefont
  {Ruijgrok}}, \bibinfo {author} {\bibfnamefont {Nitin}\ \bibnamefont {Kumar}},
  \bibinfo {author} {\bibfnamefont {Ali}\ \bibnamefont {Mozaffari}}, \bibinfo
  {author} {\bibfnamefont {Sasha}\ \bibnamefont {Zemsky}}, \bibinfo {author}
  {\bibfnamefont {Aaron~R}\ \bibnamefont {Dinner}}, \bibinfo {author}
  {\bibfnamefont {Vincenzo}\ \bibnamefont {Vitelli}}, \bibinfo {author}
  {\bibfnamefont {Zev}\ \bibnamefont {Bryant}}, \bibinfo {author}
  {\bibfnamefont {Margaret~L}\ \bibnamefont {Gardel}},  \emph {et~al.},\
  }\bibfield  {title} {\enquote {\bibinfo {title} {Spatiotemporal control of
  liquid crystal structure and dynamics through activity patterning},}\
  }\href@noop {} {\bibfield  {journal} {\bibinfo  {journal} {Nature Materials}\
  ,\ \bibinfo {pages} {1--8}} (\bibinfo {year} {2021})}\BibitemShut {NoStop}%
\bibitem [{\citenamefont {Saintillan}(2018)}]{saintillan2018rheology}%
  \BibitemOpen
  \bibfield  {author} {\bibinfo {author} {\bibfnamefont {David}\ \bibnamefont
  {Saintillan}},\ }\bibfield  {title} {\enquote {\bibinfo {title} {Rheology of
  active fluids},}\ }\href@noop {} {\bibfield  {journal} {\bibinfo  {journal}
  {Annual Review of Fluid Mechanics}\ }\textbf {\bibinfo {volume} {50}},\
  \bibinfo {pages} {563--592} (\bibinfo {year} {2018})}\BibitemShut {NoStop}%
\bibitem [{\citenamefont {Emmanuel}\ \emph {et~al.}(2020)\citenamefont
  {Emmanuel}, \citenamefont {Yeomans},\ and\ \citenamefont
  {Doostmohammadi}}]{emmanuel2020active}%
  \BibitemOpen
  \bibfield  {author} {\bibinfo {author} {\bibfnamefont {LC}~\bibnamefont
  {Emmanuel}}, \bibinfo {author} {\bibfnamefont {Julia~M}\ \bibnamefont
  {Yeomans}}, \ and\ \bibinfo {author} {\bibfnamefont {Amin}\ \bibnamefont
  {Doostmohammadi}},\ }\bibfield  {title} {\enquote {\bibinfo {title} {Active
  matter in a viscoelastic environment},}\ }\href@noop {} {\bibfield  {journal}
  {\bibinfo  {journal} {Physical Review Fluids}\ }\textbf {\bibinfo {volume}
  {5}},\ \bibinfo {pages} {023102} (\bibinfo {year} {2020})}\BibitemShut
  {NoStop}%
\bibitem [{\citenamefont {Bozorgi}\ and\ \citenamefont
  {Underhill}(2014)}]{bozorgi2014effects}%
  \BibitemOpen
  \bibfield  {author} {\bibinfo {author} {\bibfnamefont {Yaser}\ \bibnamefont
  {Bozorgi}}\ and\ \bibinfo {author} {\bibfnamefont {Patrick~T}\ \bibnamefont
  {Underhill}},\ }\bibfield  {title} {\enquote {\bibinfo {title} {Effects of
  elasticity on the nonlinear collective dynamics of self-propelled
  particles},}\ }\href {\doibase 10.1016/j.jnnfm.2014.09.016} {\bibfield
  {journal} {\bibinfo  {journal} {Journal of Non-Newtonian Fluid Mechanics}\
  }\textbf {\bibinfo {volume} {214}},\ \bibinfo {pages} {69--77} (\bibinfo
  {year} {2014})}\BibitemShut {NoStop}%
\bibitem [{\citenamefont {Hemingway}\ \emph {et~al.}(2015)\citenamefont
  {Hemingway}, \citenamefont {Maitra}, \citenamefont {Banerjee}, \citenamefont
  {Marchetti}, \citenamefont {Ramaswamy}, \citenamefont {Fielding},\ and\
  \citenamefont {Cates}}]{hemingway2015active}%
  \BibitemOpen
  \bibfield  {author} {\bibinfo {author} {\bibfnamefont {EJ}~\bibnamefont
  {Hemingway}}, \bibinfo {author} {\bibfnamefont {Ananyo}\ \bibnamefont
  {Maitra}}, \bibinfo {author} {\bibfnamefont {Shiladitya}\ \bibnamefont
  {Banerjee}}, \bibinfo {author} {\bibfnamefont {M~Cristina}\ \bibnamefont
  {Marchetti}}, \bibinfo {author} {\bibfnamefont {Sriram}\ \bibnamefont
  {Ramaswamy}}, \bibinfo {author} {\bibfnamefont {Suzanne~M}\ \bibnamefont
  {Fielding}}, \ and\ \bibinfo {author} {\bibfnamefont {Michael~E}\
  \bibnamefont {Cates}},\ }\bibfield  {title} {\enquote {\bibinfo {title}
  {Active viscoelastic matter: from bacterial drag reduction to turbulent
  solids},}\ }\href {\doibase 10.1103/PhysRevLett.114.098302} {\bibfield
  {journal} {\bibinfo  {journal} {Physical Review Letters}\ }\textbf {\bibinfo
  {volume} {114}},\ \bibinfo {pages} {098302} (\bibinfo {year}
  {2015})}\BibitemShut {NoStop}%
\bibitem [{\citenamefont {Hemingway}\ \emph {et~al.}(2016)\citenamefont
  {Hemingway}, \citenamefont {Cates},\ and\ \citenamefont
  {Fielding}}]{hemingway2016viscoelastic}%
  \BibitemOpen
  \bibfield  {author} {\bibinfo {author} {\bibfnamefont {Ewan~J}\ \bibnamefont
  {Hemingway}}, \bibinfo {author} {\bibfnamefont {ME}~\bibnamefont {Cates}}, \
  and\ \bibinfo {author} {\bibfnamefont {SM}~\bibnamefont {Fielding}},\
  }\bibfield  {title} {\enquote {\bibinfo {title} {Viscoelastic and elastomeric
  active matter: linear instability and nonlinear dynamics},}\ }\href {\doibase
  10.1103/PhysRevE.93.032702} {\bibfield  {journal} {\bibinfo  {journal}
  {Physical Review E}\ }\textbf {\bibinfo {volume} {93}},\ \bibinfo {pages}
  {032702} (\bibinfo {year} {2016})}\BibitemShut {NoStop}%
\bibitem [{\citenamefont {Li}\ and\ \citenamefont
  {Ardekani}(2016)}]{li2016collective}%
  \BibitemOpen
  \bibfield  {author} {\bibinfo {author} {\bibfnamefont {Gaojin}\ \bibnamefont
  {Li}}\ and\ \bibinfo {author} {\bibfnamefont {Arezoo~M}\ \bibnamefont
  {Ardekani}},\ }\bibfield  {title} {\enquote {\bibinfo {title} {Collective
  motion of microorganisms in a viscoelastic fluid},}\ }\href {\doibase
  10.1103/PhysRevLett.117.118001} {\bibfield  {journal} {\bibinfo  {journal}
  {Physical Review Letters}\ }\textbf {\bibinfo {volume} {117}},\ \bibinfo
  {pages} {118001} (\bibinfo {year} {2016})}\BibitemShut {NoStop}%
\bibitem [{\citenamefont {Liu}\ \emph {et~al.}(2021)\citenamefont {Liu},
  \citenamefont {Shankar}, \citenamefont {Marchetti},\ and\ \citenamefont
  {Wu}}]{liu2021viscoelastic}%
  \BibitemOpen
  \bibfield  {author} {\bibinfo {author} {\bibfnamefont {Song}\ \bibnamefont
  {Liu}}, \bibinfo {author} {\bibfnamefont {Suraj}\ \bibnamefont {Shankar}},
  \bibinfo {author} {\bibfnamefont {M~Cristina}\ \bibnamefont {Marchetti}}, \
  and\ \bibinfo {author} {\bibfnamefont {Yilin}\ \bibnamefont {Wu}},\
  }\bibfield  {title} {\enquote {\bibinfo {title} {Viscoelastic control of
  spatiotemporal order in bacterial active matter},}\ }\href@noop {} {\bibfield
   {journal} {\bibinfo  {journal} {Nature}\ }\textbf {\bibinfo {volume}
  {590}},\ \bibinfo {pages} {80--84} (\bibinfo {year} {2021})}\BibitemShut
  {NoStop}%
\bibitem [{\citenamefont {Kramar}\ and\ \citenamefont
  {Alim}(2021)}]{kramar2021encoding}%
  \BibitemOpen
  \bibfield  {author} {\bibinfo {author} {\bibfnamefont {Mirna}\ \bibnamefont
  {Kramar}}\ and\ \bibinfo {author} {\bibfnamefont {Karen}\ \bibnamefont
  {Alim}},\ }\bibfield  {title} {\enquote {\bibinfo {title} {Encoding memory in
  tube diameter hierarchy of living flow network},}\ }\href@noop {} {\bibfield
  {journal} {\bibinfo  {journal} {Proceedings of the National Academy of
  Sciences}\ }\textbf {\bibinfo {volume} {118}} (\bibinfo {year}
  {2021})}\BibitemShut {NoStop}%
\bibitem [{\citenamefont {Prakash}\ and\ \citenamefont
  {Gershenfeld}(2007)}]{prakash2007microfluidic}%
  \BibitemOpen
  \bibfield  {author} {\bibinfo {author} {\bibfnamefont {Manu}\ \bibnamefont
  {Prakash}}\ and\ \bibinfo {author} {\bibfnamefont {Neil}\ \bibnamefont
  {Gershenfeld}},\ }\bibfield  {title} {\enquote {\bibinfo {title}
  {Microfluidic bubble logic},}\ }\href@noop {} {\bibfield  {journal} {\bibinfo
   {journal} {Science}\ }\textbf {\bibinfo {volume} {315}},\ \bibinfo {pages}
  {832--835} (\bibinfo {year} {2007})}\BibitemShut {NoStop}%
\bibitem [{\citenamefont {Woodhouse}\ and\ \citenamefont
  {Dunkel}(2017)}]{woodhouse2017active}%
  \BibitemOpen
  \bibfield  {author} {\bibinfo {author} {\bibfnamefont {Francis~G}\
  \bibnamefont {Woodhouse}}\ and\ \bibinfo {author} {\bibfnamefont {J{\"o}rn}\
  \bibnamefont {Dunkel}},\ }\bibfield  {title} {\enquote {\bibinfo {title}
  {Active matter logic for autonomous microfluidics},}\ }\href@noop {}
  {\bibfield  {journal} {\bibinfo  {journal} {Nature communications}\ }\textbf
  {\bibinfo {volume} {8}},\ \bibinfo {pages} {1--7} (\bibinfo {year}
  {2017})}\BibitemShut {NoStop}%
\bibitem [{\citenamefont {Romeo}\ \emph {et~al.}(2021)\citenamefont {Romeo},
  \citenamefont {Hastewell}, \citenamefont {Mietke},\ and\ \citenamefont
  {Dunkel}}]{romeo2021learning}%
  \BibitemOpen
  \bibfield  {author} {\bibinfo {author} {\bibfnamefont {Nicolas}\ \bibnamefont
  {Romeo}}, \bibinfo {author} {\bibfnamefont {Alasdair}\ \bibnamefont
  {Hastewell}}, \bibinfo {author} {\bibfnamefont {Alexander}\ \bibnamefont
  {Mietke}}, \ and\ \bibinfo {author} {\bibfnamefont {J{\"o}rn}\ \bibnamefont
  {Dunkel}},\ }\bibfield  {title} {\enquote {\bibinfo {title} {Learning
  developmental mode dynamics from single-cell trajectories},}\ }\href@noop {}
  {\bibfield  {journal} {\bibinfo  {journal} {arXiv preprint arXiv:2103.08130}\
  } (\bibinfo {year} {2021})}\BibitemShut {NoStop}%
\bibitem [{\citenamefont {Br{\"u}ckner}\ \emph {et~al.}(2020)\citenamefont
  {Br{\"u}ckner}, \citenamefont {Ronceray},\ and\ \citenamefont
  {Broedersz}}]{bruckner2020inferring}%
  \BibitemOpen
  \bibfield  {author} {\bibinfo {author} {\bibfnamefont {David~B}\ \bibnamefont
  {Br{\"u}ckner}}, \bibinfo {author} {\bibfnamefont {Pierre}\ \bibnamefont
  {Ronceray}}, \ and\ \bibinfo {author} {\bibfnamefont {Chase~P}\ \bibnamefont
  {Broedersz}},\ }\bibfield  {title} {\enquote {\bibinfo {title} {Inferring the
  dynamics of underdamped stochastic systems},}\ }\href@noop {} {\bibfield
  {journal} {\bibinfo  {journal} {Physical review letters}\ }\textbf {\bibinfo
  {volume} {125}},\ \bibinfo {pages} {058103} (\bibinfo {year}
  {2020})}\BibitemShut {NoStop}%
\bibitem [{\citenamefont {Br{\"u}ckner}\ \emph {et~al.}(2021)\citenamefont
  {Br{\"u}ckner}, \citenamefont {Arlt}, \citenamefont {Fink}, \citenamefont
  {Ronceray}, \citenamefont {R{\"a}dler},\ and\ \citenamefont
  {Broedersz}}]{bruckner2021learning}%
  \BibitemOpen
  \bibfield  {author} {\bibinfo {author} {\bibfnamefont {David~B}\ \bibnamefont
  {Br{\"u}ckner}}, \bibinfo {author} {\bibfnamefont {Nicolas}\ \bibnamefont
  {Arlt}}, \bibinfo {author} {\bibfnamefont {Alexandra}\ \bibnamefont {Fink}},
  \bibinfo {author} {\bibfnamefont {Pierre}\ \bibnamefont {Ronceray}}, \bibinfo
  {author} {\bibfnamefont {Joachim~O}\ \bibnamefont {R{\"a}dler}}, \ and\
  \bibinfo {author} {\bibfnamefont {Chase~P}\ \bibnamefont {Broedersz}},\
  }\bibfield  {title} {\enquote {\bibinfo {title} {Learning the dynamics of
  cell--cell interactions in confined cell migration},}\ }\href@noop {}
  {\bibfield  {journal} {\bibinfo  {journal} {Proceedings of the National
  Academy of Sciences}\ }\textbf {\bibinfo {volume} {118}} (\bibinfo {year}
  {2021})}\BibitemShut {NoStop}%
\bibitem [{\citenamefont {Colen}\ \emph {et~al.}(2021)\citenamefont {Colen},
  \citenamefont {Han}, \citenamefont {Zhang}, \citenamefont {Redford},
  \citenamefont {Lemma}, \citenamefont {Morgan}, \citenamefont {Ruijgrok},
  \citenamefont {Adkins}, \citenamefont {Bryant}, \citenamefont {Dogic} \emph
  {et~al.}}]{colen2021machine}%
  \BibitemOpen
  \bibfield  {author} {\bibinfo {author} {\bibfnamefont {Jonathan}\
  \bibnamefont {Colen}}, \bibinfo {author} {\bibfnamefont {Ming}\ \bibnamefont
  {Han}}, \bibinfo {author} {\bibfnamefont {Rui}\ \bibnamefont {Zhang}},
  \bibinfo {author} {\bibfnamefont {Steven~A}\ \bibnamefont {Redford}},
  \bibinfo {author} {\bibfnamefont {Linnea~M}\ \bibnamefont {Lemma}}, \bibinfo
  {author} {\bibfnamefont {Link}\ \bibnamefont {Morgan}}, \bibinfo {author}
  {\bibfnamefont {Paul~V}\ \bibnamefont {Ruijgrok}}, \bibinfo {author}
  {\bibfnamefont {Raymond}\ \bibnamefont {Adkins}}, \bibinfo {author}
  {\bibfnamefont {Zev}\ \bibnamefont {Bryant}}, \bibinfo {author}
  {\bibfnamefont {Zvonimir}\ \bibnamefont {Dogic}},  \emph {et~al.},\
  }\bibfield  {title} {\enquote {\bibinfo {title} {Machine learning
  active-nematic hydrodynamics},}\ }\href@noop {} {\bibfield  {journal}
  {\bibinfo  {journal} {Proceedings of the National Academy of Sciences}\
  }\textbf {\bibinfo {volume} {118}} (\bibinfo {year} {2021})}\BibitemShut
  {NoStop}%
\bibitem [{\citenamefont {Jeckel}\ \emph {et~al.}(2019)\citenamefont {Jeckel},
  \citenamefont {Jelli}, \citenamefont {Hartmann}, \citenamefont {Singh},
  \citenamefont {Mok}, \citenamefont {Totz}, \citenamefont {Vidakovic},
  \citenamefont {Eckhardt}, \citenamefont {Dunkel},\ and\ \citenamefont
  {Drescher}}]{jeckel2019learning}%
  \BibitemOpen
  \bibfield  {author} {\bibinfo {author} {\bibfnamefont {Hannah}\ \bibnamefont
  {Jeckel}}, \bibinfo {author} {\bibfnamefont {Eric}\ \bibnamefont {Jelli}},
  \bibinfo {author} {\bibfnamefont {Raimo}\ \bibnamefont {Hartmann}}, \bibinfo
  {author} {\bibfnamefont {Praveen~K}\ \bibnamefont {Singh}}, \bibinfo {author}
  {\bibfnamefont {Rachel}\ \bibnamefont {Mok}}, \bibinfo {author}
  {\bibfnamefont {Jan~Frederik}\ \bibnamefont {Totz}}, \bibinfo {author}
  {\bibfnamefont {Lucia}\ \bibnamefont {Vidakovic}}, \bibinfo {author}
  {\bibfnamefont {Bruno}\ \bibnamefont {Eckhardt}}, \bibinfo {author}
  {\bibfnamefont {J{\"o}rn}\ \bibnamefont {Dunkel}}, \ and\ \bibinfo {author}
  {\bibfnamefont {Knut}\ \bibnamefont {Drescher}},\ }\bibfield  {title}
  {\enquote {\bibinfo {title} {Learning the space-time phase diagram of
  bacterial swarm expansion},}\ }\href@noop {} {\bibfield  {journal} {\bibinfo
  {journal} {Proceedings of the National Academy of Sciences}\ }\textbf
  {\bibinfo {volume} {116}},\ \bibinfo {pages} {1489--1494} (\bibinfo {year}
  {2019})}\BibitemShut {NoStop}%
\bibitem [{\citenamefont {Paxton}\ \emph {et~al.}(2004)\citenamefont {Paxton},
  \citenamefont {Kistler}, \citenamefont {Olmeda}, \citenamefont {Sen},
  \citenamefont {St.~Angelo}, \citenamefont {Cao}, \citenamefont {Mallouk},
  \citenamefont {Lammert},\ and\ \citenamefont {Crespi}}]{paxton2004catalytic}%
  \BibitemOpen
  \bibfield  {author} {\bibinfo {author} {\bibfnamefont {Walter~F}\
  \bibnamefont {Paxton}}, \bibinfo {author} {\bibfnamefont {Kevin~C}\
  \bibnamefont {Kistler}}, \bibinfo {author} {\bibfnamefont {Christine~C}\
  \bibnamefont {Olmeda}}, \bibinfo {author} {\bibfnamefont {Ayusman}\
  \bibnamefont {Sen}}, \bibinfo {author} {\bibfnamefont {Sarah~K}\ \bibnamefont
  {St.~Angelo}}, \bibinfo {author} {\bibfnamefont {Yanyan}\ \bibnamefont
  {Cao}}, \bibinfo {author} {\bibfnamefont {Thomas~E}\ \bibnamefont {Mallouk}},
  \bibinfo {author} {\bibfnamefont {Paul~E}\ \bibnamefont {Lammert}}, \ and\
  \bibinfo {author} {\bibfnamefont {Vincent~H}\ \bibnamefont {Crespi}},\
  }\bibfield  {title} {\enquote {\bibinfo {title} {Catalytic nanomotors:
  autonomous movement of striped nanorods},}\ }\href@noop {} {\bibfield
  {journal} {\bibinfo  {journal} {Journal of the American Chemical Society}\
  }\textbf {\bibinfo {volume} {126}},\ \bibinfo {pages} {13424--13431}
  (\bibinfo {year} {2004})}\BibitemShut {NoStop}%
\bibitem [{\citenamefont {Needleman}\ and\ \citenamefont
  {Dogic}(2017)}]{needleman2017active}%
  \BibitemOpen
  \bibfield  {author} {\bibinfo {author} {\bibfnamefont {Daniel}\ \bibnamefont
  {Needleman}}\ and\ \bibinfo {author} {\bibfnamefont {Zvonimir}\ \bibnamefont
  {Dogic}},\ }\bibfield  {title} {\enquote {\bibinfo {title} {Active matter at
  the interface between materials science and cell biology},}\ }\href@noop {}
  {\bibfield  {journal} {\bibinfo  {journal} {Nature Reviews Materials}\
  }\textbf {\bibinfo {volume} {2}},\ \bibinfo {pages} {1--14} (\bibinfo {year}
  {2017})}\BibitemShut {NoStop}%
\bibitem [{\citenamefont {Golestanian}\ \emph {et~al.}(2005)\citenamefont
  {Golestanian}, \citenamefont {Liverpool},\ and\ \citenamefont
  {Ajdari}}]{golestanian2005propulsion}%
  \BibitemOpen
  \bibfield  {author} {\bibinfo {author} {\bibfnamefont {Ramin}\ \bibnamefont
  {Golestanian}}, \bibinfo {author} {\bibfnamefont {Tanniemola~B}\ \bibnamefont
  {Liverpool}}, \ and\ \bibinfo {author} {\bibfnamefont {Armand}\ \bibnamefont
  {Ajdari}},\ }\bibfield  {title} {\enquote {\bibinfo {title} {Propulsion of a
  molecular machine by asymmetric distribution of reaction products},}\
  }\href@noop {} {\bibfield  {journal} {\bibinfo  {journal} {Physical review
  letters}\ }\textbf {\bibinfo {volume} {94}},\ \bibinfo {pages} {220801}
  (\bibinfo {year} {2005})}\BibitemShut {NoStop}%
\bibitem [{\citenamefont {Soto}\ and\ \citenamefont
  {Golestanian}(2014)}]{soto2014self}%
  \BibitemOpen
  \bibfield  {author} {\bibinfo {author} {\bibfnamefont {Rodrigo}\ \bibnamefont
  {Soto}}\ and\ \bibinfo {author} {\bibfnamefont {Ramin}\ \bibnamefont
  {Golestanian}},\ }\bibfield  {title} {\enquote {\bibinfo {title}
  {Self-assembly of catalytically active colloidal molecules: Tailoring
  activity through surface chemistry},}\ }\href@noop {} {\bibfield  {journal}
  {\bibinfo  {journal} {Physical review letters}\ }\textbf {\bibinfo {volume}
  {112}},\ \bibinfo {pages} {068301} (\bibinfo {year} {2014})}\BibitemShut
  {NoStop}%
\bibitem [{\citenamefont {Nasouri}\ and\ \citenamefont
  {Golestanian}(2020)}]{nasouri2020exact}%
  \BibitemOpen
  \bibfield  {author} {\bibinfo {author} {\bibfnamefont {Babak}\ \bibnamefont
  {Nasouri}}\ and\ \bibinfo {author} {\bibfnamefont {Ramin}\ \bibnamefont
  {Golestanian}},\ }\bibfield  {title} {\enquote {\bibinfo {title} {Exact
  phoretic interaction of two chemically active particles},}\ }\href@noop {}
  {\bibfield  {journal} {\bibinfo  {journal} {Physical review letters}\
  }\textbf {\bibinfo {volume} {124}},\ \bibinfo {pages} {168003} (\bibinfo
  {year} {2020})}\BibitemShut {NoStop}%
\bibitem [{\citenamefont {You}\ \emph {et~al.}(2020)\citenamefont {You},
  \citenamefont {Baskaran},\ and\ \citenamefont
  {Marchetti}}]{you2020nonreciprocity}%
  \BibitemOpen
  \bibfield  {author} {\bibinfo {author} {\bibfnamefont {Zhihong}\ \bibnamefont
  {You}}, \bibinfo {author} {\bibfnamefont {Aparna}\ \bibnamefont {Baskaran}},
  \ and\ \bibinfo {author} {\bibfnamefont {M~Cristina}\ \bibnamefont
  {Marchetti}},\ }\bibfield  {title} {\enquote {\bibinfo {title}
  {Nonreciprocity as a generic route to traveling states},}\ }\href@noop {}
  {\bibfield  {journal} {\bibinfo  {journal} {Proceedings of the National
  Academy of Sciences}\ }\textbf {\bibinfo {volume} {117}},\ \bibinfo {pages}
  {19767--19772} (\bibinfo {year} {2020})}\BibitemShut {NoStop}%
\bibitem [{\citenamefont {Saha}\ \emph {et~al.}(2020)\citenamefont {Saha},
  \citenamefont {Agudo-Canalejo},\ and\ \citenamefont
  {Golestanian}}]{saha2020scalar}%
  \BibitemOpen
  \bibfield  {author} {\bibinfo {author} {\bibfnamefont {Suropriya}\
  \bibnamefont {Saha}}, \bibinfo {author} {\bibfnamefont {Jaime}\ \bibnamefont
  {Agudo-Canalejo}}, \ and\ \bibinfo {author} {\bibfnamefont {Ramin}\
  \bibnamefont {Golestanian}},\ }\bibfield  {title} {\enquote {\bibinfo {title}
  {Scalar active mixtures: The nonreciprocal cahn-hilliard model},}\
  }\href@noop {} {\bibfield  {journal} {\bibinfo  {journal} {Physical Review
  X}\ }\textbf {\bibinfo {volume} {10}},\ \bibinfo {pages} {041009} (\bibinfo
  {year} {2020})}\BibitemShut {NoStop}%
\bibitem [{\citenamefont {Saha}\ \emph {et~al.}(2019)\citenamefont {Saha},
  \citenamefont {Ramaswamy},\ and\ \citenamefont {Golestanian}}]{Saha_2019}%
  \BibitemOpen
  \bibfield  {author} {\bibinfo {author} {\bibfnamefont {Suropriya}\
  \bibnamefont {Saha}}, \bibinfo {author} {\bibfnamefont {Sriram}\ \bibnamefont
  {Ramaswamy}}, \ and\ \bibinfo {author} {\bibfnamefont {Ramin}\ \bibnamefont
  {Golestanian}},\ }\bibfield  {title} {\enquote {\bibinfo {title} {Pairing,
  waltzing and scattering of chemotactic active colloids},}\ }\href {\doibase
  10.1088/1367-2630/ab20fd} {\bibfield  {journal} {\bibinfo  {journal} {New
  Journal of Physics}\ }\textbf {\bibinfo {volume} {21}},\ \bibinfo {pages}
  {063006} (\bibinfo {year} {2019})}\BibitemShut {NoStop}%
\bibitem [{\citenamefont {Saha}\ \emph {et~al.}(2014)\citenamefont {Saha},
  \citenamefont {Golestanian},\ and\ \citenamefont
  {Ramaswamy}}]{saha2014clusters}%
  \BibitemOpen
  \bibfield  {author} {\bibinfo {author} {\bibfnamefont {Suropriya}\
  \bibnamefont {Saha}}, \bibinfo {author} {\bibfnamefont {Ramin}\ \bibnamefont
  {Golestanian}}, \ and\ \bibinfo {author} {\bibfnamefont {Sriram}\
  \bibnamefont {Ramaswamy}},\ }\bibfield  {title} {\enquote {\bibinfo {title}
  {Clusters, asters, and collective oscillations in chemotactic colloids},}\
  }\href {\doibase 10.1103/PhysRevE.89.062316} {\bibfield  {journal} {\bibinfo
  {journal} {Phys. Rev. E}\ }\textbf {\bibinfo {volume} {89}},\ \bibinfo
  {pages} {062316} (\bibinfo {year} {2014})}\BibitemShut {NoStop}%
\bibitem [{\citenamefont {Matas-Navarro}\ \emph {et~al.}(2014)\citenamefont
  {Matas-Navarro}, \citenamefont {Golestanian}, \citenamefont {Liverpool},\
  and\ \citenamefont {Fielding}}]{matas2014hydrodynamic}%
  \BibitemOpen
  \bibfield  {author} {\bibinfo {author} {\bibfnamefont {Ricard}\ \bibnamefont
  {Matas-Navarro}}, \bibinfo {author} {\bibfnamefont {Ramin}\ \bibnamefont
  {Golestanian}}, \bibinfo {author} {\bibfnamefont {Tanniemola~B}\ \bibnamefont
  {Liverpool}}, \ and\ \bibinfo {author} {\bibfnamefont {Suzanne~M}\
  \bibnamefont {Fielding}},\ }\bibfield  {title} {\enquote {\bibinfo {title}
  {Hydrodynamic suppression of phase separation in active suspensions},}\
  }\href@noop {} {\bibfield  {journal} {\bibinfo  {journal} {Physical Review
  E}\ }\textbf {\bibinfo {volume} {90}},\ \bibinfo {pages} {032304} (\bibinfo
  {year} {2014})}\BibitemShut {NoStop}%
\bibitem [{\citenamefont {Zhou}\ \emph {et~al.}(2021)\citenamefont {Zhou},
  \citenamefont {Joshi}, \citenamefont {Liu}, \citenamefont {Norton},
  \citenamefont {Lemma}, \citenamefont {Dogic}, \citenamefont {Hagan},
  \citenamefont {Fraden},\ and\ \citenamefont {Hong}}]{zhou2021machine}%
  \BibitemOpen
  \bibfield  {author} {\bibinfo {author} {\bibfnamefont {Zhengyang}\
  \bibnamefont {Zhou}}, \bibinfo {author} {\bibfnamefont {Chaitanya}\
  \bibnamefont {Joshi}}, \bibinfo {author} {\bibfnamefont {Ruoshi}\
  \bibnamefont {Liu}}, \bibinfo {author} {\bibfnamefont {Michael~M}\
  \bibnamefont {Norton}}, \bibinfo {author} {\bibfnamefont {Linnea}\
  \bibnamefont {Lemma}}, \bibinfo {author} {\bibfnamefont {Zvonimir}\
  \bibnamefont {Dogic}}, \bibinfo {author} {\bibfnamefont {Michael~F}\
  \bibnamefont {Hagan}}, \bibinfo {author} {\bibfnamefont {Seth}\ \bibnamefont
  {Fraden}}, \ and\ \bibinfo {author} {\bibfnamefont {Pengyu}\ \bibnamefont
  {Hong}},\ }\bibfield  {title} {\enquote {\bibinfo {title} {Machine learning
  forecasting of active nematics},}\ }\href@noop {} {\bibfield  {journal}
  {\bibinfo  {journal} {Soft matter}\ }\textbf {\bibinfo {volume} {17}},\
  \bibinfo {pages} {738--747} (\bibinfo {year} {2021})}\BibitemShut {NoStop}%
\bibitem [{\citenamefont {Gelimson}\ \emph {et~al.}(2016)\citenamefont
  {Gelimson}, \citenamefont {Zhao}, \citenamefont {Lee}, \citenamefont {Kranz},
  \citenamefont {Wong},\ and\ \citenamefont
  {Golestanian}}]{gelimson2016multicellular}%
  \BibitemOpen
  \bibfield  {author} {\bibinfo {author} {\bibfnamefont {Anatolij}\
  \bibnamefont {Gelimson}}, \bibinfo {author} {\bibfnamefont {Kun}\
  \bibnamefont {Zhao}}, \bibinfo {author} {\bibfnamefont {Calvin~K}\
  \bibnamefont {Lee}}, \bibinfo {author} {\bibfnamefont {W~Till}\ \bibnamefont
  {Kranz}}, \bibinfo {author} {\bibfnamefont {Gerard~CL}\ \bibnamefont {Wong}},
  \ and\ \bibinfo {author} {\bibfnamefont {Ramin}\ \bibnamefont
  {Golestanian}},\ }\bibfield  {title} {\enquote {\bibinfo {title}
  {Multicellular self-organization of p. aeruginosa due to interactions with
  secreted trails},}\ }\href@noop {} {\bibfield  {journal} {\bibinfo  {journal}
  {Physical review letters}\ }\textbf {\bibinfo {volume} {117}},\ \bibinfo
  {pages} {178102} (\bibinfo {year} {2016})}\BibitemShut {NoStop}%
\bibitem [{\citenamefont {Husain}\ and\ \citenamefont
  {Rao}(2017)}]{husain2017emergent}%
  \BibitemOpen
  \bibfield  {author} {\bibinfo {author} {\bibfnamefont {Kabir}\ \bibnamefont
  {Husain}}\ and\ \bibinfo {author} {\bibfnamefont {Madan}\ \bibnamefont
  {Rao}},\ }\bibfield  {title} {\enquote {\bibinfo {title} {Emergent structures
  in an active polar fluid: Dynamics of shape, scattering, and merger},}\
  }\href {https://journals.aps.org/prl/abstract/10.1103/PhysRevLett.118.078104}
  {\bibfield  {journal} {\bibinfo  {journal} {Physical review letters}\
  }\textbf {\bibinfo {volume} {118}},\ \bibinfo {pages} {078104} (\bibinfo
  {year} {2017})}\BibitemShut {NoStop}%
\bibitem [{\citenamefont {Sohn}\ \emph {et~al.}(2019)\citenamefont {Sohn},
  \citenamefont {Liu},\ and\ \citenamefont {Smalyukh}}]{sohn2019schools}%
  \BibitemOpen
  \bibfield  {author} {\bibinfo {author} {\bibfnamefont {Hayley~RO}\
  \bibnamefont {Sohn}}, \bibinfo {author} {\bibfnamefont {Changda~D}\
  \bibnamefont {Liu}}, \ and\ \bibinfo {author} {\bibfnamefont {Ivan~I}\
  \bibnamefont {Smalyukh}},\ }\bibfield  {title} {\enquote {\bibinfo {title}
  {Schools of skyrmions with electrically tunable elastic interactions},}\
  }\href {https://www.nature.com/articles/s41467-019-12723-3} {\bibfield
  {journal} {\bibinfo  {journal} {Nature communications}\ }\textbf {\bibinfo
  {volume} {10}},\ \bibinfo {pages} {1--11} (\bibinfo {year}
  {2019})}\BibitemShut {NoStop}%
\bibitem [{\citenamefont {Collinet}\ and\ \citenamefont
  {Lecuit}(2021)}]{2021Lecuit}%
  \BibitemOpen
  \bibfield  {author} {\bibinfo {author} {\bibfnamefont {Claudio}\ \bibnamefont
  {Collinet}}\ and\ \bibinfo {author} {\bibfnamefont {Thomas}\ \bibnamefont
  {Lecuit}},\ }\bibfield  {title} {\enquote {\bibinfo {title} {Programmed and
  self-organized flow of information during morphogenesis},}\ }\href {\doibase
  10.1038/s41580-020-00318-6} {\bibfield  {journal} {\bibinfo  {journal}
  {Nature Reviews Molecular Cell Biology}\ }\textbf {\bibinfo {volume} {22}},\
  \bibinfo {pages} {245--265} (\bibinfo {year} {2021})}\BibitemShut {NoStop}%
\bibitem [{\citenamefont {Gross}\ \emph {et~al.}(2017)\citenamefont {Gross},
  \citenamefont {Kumar},\ and\ \citenamefont {Grill}}]{2017Grill}%
  \BibitemOpen
  \bibfield  {author} {\bibinfo {author} {\bibfnamefont {Peter}\ \bibnamefont
  {Gross}}, \bibinfo {author} {\bibfnamefont {K.~Vijay}\ \bibnamefont {Kumar}},
  \ and\ \bibinfo {author} {\bibfnamefont {Stephan~W.}\ \bibnamefont {Grill}},\
  }\bibfield  {title} {\enquote {\bibinfo {title} {How active mechanics and
  regulatory biochemistry combine to form patterns in development},}\
  }\href@noop {} {\bibfield  {journal} {\bibinfo  {journal} {Annual Review of
  Biophysics}\ }\textbf {\bibinfo {volume} {46}},\ \bibinfo {pages} {337--356}
  (\bibinfo {year} {2017})}\BibitemShut {NoStop}%
\bibitem [{\citenamefont {Frey}\ and\ \citenamefont
  {Brauns}(2020)}]{frey2020selforganisation}%
  \BibitemOpen
  \bibfield  {author} {\bibinfo {author} {\bibfnamefont {Erwin}\ \bibnamefont
  {Frey}}\ and\ \bibinfo {author} {\bibfnamefont {Fridtjof}\ \bibnamefont
  {Brauns}},\ }\href@noop {} {\enquote {\bibinfo {title} {Self-organisation of
  protein patterns},}\ } (\bibinfo {year} {2020}),\ \Eprint
  {http://arxiv.org/abs/2012.01797} {arXiv:2012.01797 [physics.bio-ph]}
  \BibitemShut {NoStop}%
\bibitem [{\citenamefont {Wigbers}\ \emph {et~al.}(2021)\citenamefont
  {Wigbers}, \citenamefont {Tan}, \citenamefont {Brauns}, \citenamefont {Liu},
  \citenamefont {Swartz}, \citenamefont {Frey},\ and\ \citenamefont
  {Fakhri}}]{Wigbers2021}%
  \BibitemOpen
  \bibfield  {author} {\bibinfo {author} {\bibfnamefont {Manon~C.}\
  \bibnamefont {Wigbers}}, \bibinfo {author} {\bibfnamefont {Tzer~Han}\
  \bibnamefont {Tan}}, \bibinfo {author} {\bibfnamefont {Fridtjof}\
  \bibnamefont {Brauns}}, \bibinfo {author} {\bibfnamefont {Jinghui}\
  \bibnamefont {Liu}}, \bibinfo {author} {\bibfnamefont {S.~Zachary}\
  \bibnamefont {Swartz}}, \bibinfo {author} {\bibfnamefont {Erwin}\
  \bibnamefont {Frey}}, \ and\ \bibinfo {author} {\bibfnamefont {Nikta}\
  \bibnamefont {Fakhri}},\ }\bibfield  {title} {\enquote {\bibinfo {title} {A
  hierarchy of protein patterns robustly decodes cell shape information},}\
  }\href {\doibase 10.1038/s41567-021-01164-9} {\bibfield  {journal} {\bibinfo
  {journal} {Nat. Phys.}\ }\textbf {\bibinfo {volume} {17}},\ \bibinfo {pages}
  {578--584} (\bibinfo {year} {2021})}\BibitemShut {NoStop}%
\bibitem [{\citenamefont {Khadka}\ \emph {et~al.}(2018)\citenamefont {Khadka},
  \citenamefont {Holubec}, \citenamefont {Yang},\ and\ \citenamefont
  {Cichos}}]{khadka2018active}%
  \BibitemOpen
  \bibfield  {author} {\bibinfo {author} {\bibfnamefont {Utsab}\ \bibnamefont
  {Khadka}}, \bibinfo {author} {\bibfnamefont {Viktor}\ \bibnamefont
  {Holubec}}, \bibinfo {author} {\bibfnamefont {Haw}\ \bibnamefont {Yang}}, \
  and\ \bibinfo {author} {\bibfnamefont {Frank}\ \bibnamefont {Cichos}},\
  }\bibfield  {title} {\enquote {\bibinfo {title} {Active particles bound by
  information flows},}\ }\href@noop {} {\bibfield  {journal} {\bibinfo
  {journal} {Nature communications}\ }\textbf {\bibinfo {volume} {9}},\
  \bibinfo {pages} {1--9} (\bibinfo {year} {2018})}\BibitemShut {NoStop}%
\bibitem [{\citenamefont {Geiss}\ \emph {et~al.}(2019)\citenamefont {Geiss},
  \citenamefont {Kroy},\ and\ \citenamefont {Holubec}}]{geiss2019brownian}%
  \BibitemOpen
  \bibfield  {author} {\bibinfo {author} {\bibfnamefont {Daniel}\ \bibnamefont
  {Geiss}}, \bibinfo {author} {\bibfnamefont {Klaus}\ \bibnamefont {Kroy}}, \
  and\ \bibinfo {author} {\bibfnamefont {Viktor}\ \bibnamefont {Holubec}},\
  }\bibfield  {title} {\enquote {\bibinfo {title} {Brownian molecules formed by
  delayed harmonic interactions},}\ }\href@noop {} {\bibfield  {journal}
  {\bibinfo  {journal} {New Journal of Physics}\ }\textbf {\bibinfo {volume}
  {21}},\ \bibinfo {pages} {093014} (\bibinfo {year} {2019})}\BibitemShut
  {NoStop}%
\bibitem [{\citenamefont {Lavergne}\ \emph {et~al.}(2019)\citenamefont
  {Lavergne}, \citenamefont {Wendehenne}, \citenamefont {B{\"a}uerle},\ and\
  \citenamefont {Bechinger}}]{lavergne2019group}%
  \BibitemOpen
  \bibfield  {author} {\bibinfo {author} {\bibfnamefont {Fran{\c{c}}ois~A}\
  \bibnamefont {Lavergne}}, \bibinfo {author} {\bibfnamefont {Hugo}\
  \bibnamefont {Wendehenne}}, \bibinfo {author} {\bibfnamefont {Tobias}\
  \bibnamefont {B{\"a}uerle}}, \ and\ \bibinfo {author} {\bibfnamefont
  {Clemens}\ \bibnamefont {Bechinger}},\ }\bibfield  {title} {\enquote
  {\bibinfo {title} {Group formation and cohesion of active particles with
  visual perception--dependent motility},}\ }\href@noop {} {\bibfield
  {journal} {\bibinfo  {journal} {Science}\ }\textbf {\bibinfo {volume}
  {364}},\ \bibinfo {pages} {70--74} (\bibinfo {year} {2019})}\BibitemShut
  {NoStop}%
\bibitem [{\citenamefont {Dasgupta}\ \emph {et~al.}(2020)\citenamefont
  {Dasgupta}, \citenamefont {Pally}, \citenamefont {Saini}, \citenamefont
  {Bhat},\ and\ \citenamefont {Ghosh}}]{dasgupta2020nanomotors}%
  \BibitemOpen
  \bibfield  {author} {\bibinfo {author} {\bibfnamefont {Debayan}\ \bibnamefont
  {Dasgupta}}, \bibinfo {author} {\bibfnamefont {Dharma}\ \bibnamefont
  {Pally}}, \bibinfo {author} {\bibfnamefont {Deepak~K}\ \bibnamefont {Saini}},
  \bibinfo {author} {\bibfnamefont {Ramray}\ \bibnamefont {Bhat}}, \ and\
  \bibinfo {author} {\bibfnamefont {Ambarish}\ \bibnamefont {Ghosh}},\
  }\bibfield  {title} {\enquote {\bibinfo {title} {Nanomotors sense local
  physicochemical heterogeneities in tumor microenvironments},}\ }\href
  {https://onlinelibrary.wiley.com/doi/full/10.1002/anie.202008681} {\bibfield
  {journal} {\bibinfo  {journal} {Angewandte Chemie International Edition}\
  }\textbf {\bibinfo {volume} {59}},\ \bibinfo {pages} {23690--23696} (\bibinfo
  {year} {2020})}\BibitemShut {NoStop}%
\bibitem [{\citenamefont {Loos}\ and\ \citenamefont
  {Klapp}(2020)}]{loos2020irreversibility}%
  \BibitemOpen
  \bibfield  {author} {\bibinfo {author} {\bibfnamefont {Sarah~AM}\
  \bibnamefont {Loos}}\ and\ \bibinfo {author} {\bibfnamefont {Sabine~HL}\
  \bibnamefont {Klapp}},\ }\bibfield  {title} {\enquote {\bibinfo {title}
  {Irreversibility, heat and information flows induced by non-reciprocal
  interactions},}\ }\href@noop {} {\bibfield  {journal} {\bibinfo  {journal}
  {New Journal of Physics}\ }\textbf {\bibinfo {volume} {22}},\ \bibinfo
  {pages} {123051} (\bibinfo {year} {2020})}\BibitemShut {NoStop}%
\bibitem [{\citenamefont {Wang}\ \emph {et~al.}(2021)\citenamefont {Wang},
  \citenamefont {Phan}, \citenamefont {Li}, \citenamefont {Wombacher},
  \citenamefont {Qu}, \citenamefont {Peng}, \citenamefont {Chen}, \citenamefont
  {Goldman}, \citenamefont {Levin}, \citenamefont {Austin},\ and\ \citenamefont
  {Liu}}]{2021Austin}%
  \BibitemOpen
  \bibfield  {author} {\bibinfo {author} {\bibfnamefont {Gao}\ \bibnamefont
  {Wang}}, \bibinfo {author} {\bibfnamefont {Trung~V.}\ \bibnamefont {Phan}},
  \bibinfo {author} {\bibfnamefont {Shengkai}\ \bibnamefont {Li}}, \bibinfo
  {author} {\bibfnamefont {Michael}\ \bibnamefont {Wombacher}}, \bibinfo
  {author} {\bibfnamefont {Junle}\ \bibnamefont {Qu}}, \bibinfo {author}
  {\bibfnamefont {Yan}\ \bibnamefont {Peng}}, \bibinfo {author} {\bibfnamefont
  {Guo}\ \bibnamefont {Chen}}, \bibinfo {author} {\bibfnamefont {Daniel~I.}\
  \bibnamefont {Goldman}}, \bibinfo {author} {\bibfnamefont {Simon~A.}\
  \bibnamefont {Levin}}, \bibinfo {author} {\bibfnamefont {Robert~H.}\
  \bibnamefont {Austin}}, \ and\ \bibinfo {author} {\bibfnamefont {Liyu}\
  \bibnamefont {Liu}},\ }\bibfield  {title} {\enquote {\bibinfo {title}
  {Emergent field-driven robot swarm states},}\ }\href {\doibase
  10.1103/PhysRevLett.126.108002} {\bibfield  {journal} {\bibinfo  {journal}
  {Phys. Rev. Lett.}\ }\textbf {\bibinfo {volume} {126}},\ \bibinfo {pages}
  {108002} (\bibinfo {year} {2021})}\BibitemShut {NoStop}%
\bibitem [{\citenamefont {Tang}\ and\ \citenamefont
  {Golestanian}(2020)}]{Tang_2020}%
  \BibitemOpen
  \bibfield  {author} {\bibinfo {author} {\bibfnamefont {Evelyn}\ \bibnamefont
  {Tang}}\ and\ \bibinfo {author} {\bibfnamefont {Ramin}\ \bibnamefont
  {Golestanian}},\ }\bibfield  {title} {\enquote {\bibinfo {title} {Quantifying
  configurational information for a stochastic particle in a flow-field},}\
  }\href {\doibase 10.1088/1367-2630/aba76b} {\bibfield  {journal} {\bibinfo
  {journal} {New Journal of Physics}\ }\textbf {\bibinfo {volume} {22}},\
  \bibinfo {pages} {083060} (\bibinfo {year} {2020})}\BibitemShut {NoStop}%
\bibitem [{\citenamefont {Ivlev}\ \emph {et~al.}(2015)\citenamefont {Ivlev},
  \citenamefont {Bartnick}, \citenamefont {Heinen}, \citenamefont {Du},
  \citenamefont {Nosenko},\ and\ \citenamefont
  {L{\"o}wen}}]{ivlev2015statistical}%
  \BibitemOpen
  \bibfield  {author} {\bibinfo {author} {\bibfnamefont {Alexei~V}\
  \bibnamefont {Ivlev}}, \bibinfo {author} {\bibfnamefont {J{\"o}rg}\
  \bibnamefont {Bartnick}}, \bibinfo {author} {\bibfnamefont {Marco}\
  \bibnamefont {Heinen}}, \bibinfo {author} {\bibfnamefont {C-R}\ \bibnamefont
  {Du}}, \bibinfo {author} {\bibfnamefont {V}~\bibnamefont {Nosenko}}, \ and\
  \bibinfo {author} {\bibfnamefont {Hartmut}\ \bibnamefont {L{\"o}wen}},\
  }\bibfield  {title} {\enquote {\bibinfo {title} {Statistical mechanics where
  newton’s third law is broken},}\ }\href@noop {} {\bibfield  {journal}
  {\bibinfo  {journal} {Physical Review X}\ }\textbf {\bibinfo {volume} {5}},\
  \bibinfo {pages} {011035} (\bibinfo {year} {2015})}\BibitemShut {NoStop}%
\bibitem [{\citenamefont {Kryuchkov}\ \emph {et~al.}(2018)\citenamefont
  {Kryuchkov}, \citenamefont {Ivlev},\ and\ \citenamefont
  {Yurchenko}}]{kryuchkov2018dissipative}%
  \BibitemOpen
  \bibfield  {author} {\bibinfo {author} {\bibfnamefont {Nikita~P}\
  \bibnamefont {Kryuchkov}}, \bibinfo {author} {\bibfnamefont {Alexei~V}\
  \bibnamefont {Ivlev}}, \ and\ \bibinfo {author} {\bibfnamefont {Stanislav~O}\
  \bibnamefont {Yurchenko}},\ }\bibfield  {title} {\enquote {\bibinfo {title}
  {Dissipative phase transitions in systems with nonreciprocal effective
  interactions},}\ }\href@noop {} {\bibfield  {journal} {\bibinfo  {journal}
  {Soft Matter}\ }\textbf {\bibinfo {volume} {14}},\ \bibinfo {pages}
  {9720--9729} (\bibinfo {year} {2018})}\BibitemShut {NoStop}%
\bibitem [{\citenamefont {Meredith}\ \emph {et~al.}(2020)\citenamefont
  {Meredith}, \citenamefont {Moerman}, \citenamefont {Groenewold},
  \citenamefont {Chiu}, \citenamefont {Kegel}, \citenamefont {van Blaaderen},\
  and\ \citenamefont {Zarzar}}]{meredith2020predator}%
  \BibitemOpen
  \bibfield  {author} {\bibinfo {author} {\bibfnamefont {Caleb~H}\ \bibnamefont
  {Meredith}}, \bibinfo {author} {\bibfnamefont {Pepijn~G}\ \bibnamefont
  {Moerman}}, \bibinfo {author} {\bibfnamefont {Jan}\ \bibnamefont
  {Groenewold}}, \bibinfo {author} {\bibfnamefont {Yu-Jen}\ \bibnamefont
  {Chiu}}, \bibinfo {author} {\bibfnamefont {Willem~K}\ \bibnamefont {Kegel}},
  \bibinfo {author} {\bibfnamefont {Alfons}\ \bibnamefont {van Blaaderen}}, \
  and\ \bibinfo {author} {\bibfnamefont {Lauren~D}\ \bibnamefont {Zarzar}},\
  }\bibfield  {title} {\enquote {\bibinfo {title} {Predator--prey interactions
  between droplets driven by non-reciprocal oil exchange},}\ }\href@noop {}
  {\bibfield  {journal} {\bibinfo  {journal} {Nature Chemistry}\ }\textbf
  {\bibinfo {volume} {12}},\ \bibinfo {pages} {1136--1142} (\bibinfo {year}
  {2020})}\BibitemShut {NoStop}%
\bibitem [{\citenamefont {Durve}\ \emph {et~al.}(2018)\citenamefont {Durve},
  \citenamefont {Saha},\ and\ \citenamefont {Sayeed}}]{durve2018active}%
  \BibitemOpen
  \bibfield  {author} {\bibinfo {author} {\bibfnamefont {Mihir}\ \bibnamefont
  {Durve}}, \bibinfo {author} {\bibfnamefont {Arnab}\ \bibnamefont {Saha}}, \
  and\ \bibinfo {author} {\bibfnamefont {Ahmed}\ \bibnamefont {Sayeed}},\
  }\bibfield  {title} {\enquote {\bibinfo {title} {Active particle condensation
  by non-reciprocal and time-delayed interactions},}\ }\href@noop {} {\bibfield
   {journal} {\bibinfo  {journal} {The European Physical Journal E}\ }\textbf
  {\bibinfo {volume} {41}},\ \bibinfo {pages} {49} (\bibinfo {year}
  {2018})}\BibitemShut {NoStop}%
\bibitem [{\citenamefont {Helbing}\ and\ \citenamefont
  {Molnar}(1995)}]{helbing1995social}%
  \BibitemOpen
  \bibfield  {author} {\bibinfo {author} {\bibfnamefont {Dirk}\ \bibnamefont
  {Helbing}}\ and\ \bibinfo {author} {\bibfnamefont {Peter}\ \bibnamefont
  {Molnar}},\ }\bibfield  {title} {\enquote {\bibinfo {title} {Social force
  model for pedestrian dynamics},}\ }\href@noop {} {\bibfield  {journal}
  {\bibinfo  {journal} {Physical review E}\ }\textbf {\bibinfo {volume} {51}},\
  \bibinfo {pages} {4282} (\bibinfo {year} {1995})}\BibitemShut {NoStop}%
\bibitem [{\citenamefont {Hong}\ and\ \citenamefont
  {Strogatz}(2011)}]{hong2011kuramoto}%
  \BibitemOpen
  \bibfield  {author} {\bibinfo {author} {\bibfnamefont {Hyunsuk}\ \bibnamefont
  {Hong}}\ and\ \bibinfo {author} {\bibfnamefont {Steven~H}\ \bibnamefont
  {Strogatz}},\ }\bibfield  {title} {\enquote {\bibinfo {title} {Kuramoto model
  of coupled oscillators with positive and negative coupling parameters: an
  example of conformist and contrarian oscillators},}\ }\href@noop {}
  {\bibfield  {journal} {\bibinfo  {journal} {Physical Review Letters}\
  }\textbf {\bibinfo {volume} {106}},\ \bibinfo {pages} {054102} (\bibinfo
  {year} {2011})}\BibitemShut {NoStop}%
\bibitem [{\citenamefont {Theveneau}\ \emph {et~al.}(2013)\citenamefont
  {Theveneau}, \citenamefont {Steventon}, \citenamefont {Scarpa}, \citenamefont
  {Garcia}, \citenamefont {Trepat}, \citenamefont {Streit},\ and\ \citenamefont
  {Mayor}}]{theveneau2013chase}%
  \BibitemOpen
  \bibfield  {author} {\bibinfo {author} {\bibfnamefont {Eric}\ \bibnamefont
  {Theveneau}}, \bibinfo {author} {\bibfnamefont {Benjamin}\ \bibnamefont
  {Steventon}}, \bibinfo {author} {\bibfnamefont {Elena}\ \bibnamefont
  {Scarpa}}, \bibinfo {author} {\bibfnamefont {Simon}\ \bibnamefont {Garcia}},
  \bibinfo {author} {\bibfnamefont {Xavier}\ \bibnamefont {Trepat}}, \bibinfo
  {author} {\bibfnamefont {Andrea}\ \bibnamefont {Streit}}, \ and\ \bibinfo
  {author} {\bibfnamefont {Roberto}\ \bibnamefont {Mayor}},\ }\bibfield
  {title} {\enquote {\bibinfo {title} {Chase-and-run between adjacent cell
  populations promotes directional collective migration},}\ }\href@noop {}
  {\bibfield  {journal} {\bibinfo  {journal} {Nature cell biology}\ }\textbf
  {\bibinfo {volume} {15}},\ \bibinfo {pages} {763--772} (\bibinfo {year}
  {2013})}\BibitemShut {NoStop}%
\bibitem [{\citenamefont {Xiong}\ \emph {et~al.}(2020)\citenamefont {Xiong},
  \citenamefont {Cao}, \citenamefont {Cooper}, \citenamefont {Rappel},
  \citenamefont {Hasty},\ and\ \citenamefont {Tsimring}}]{xiong2020flower}%
  \BibitemOpen
  \bibfield  {author} {\bibinfo {author} {\bibfnamefont {Liyang}\ \bibnamefont
  {Xiong}}, \bibinfo {author} {\bibfnamefont {Yuansheng}\ \bibnamefont {Cao}},
  \bibinfo {author} {\bibfnamefont {Robert}\ \bibnamefont {Cooper}}, \bibinfo
  {author} {\bibfnamefont {Wouter-Jan}\ \bibnamefont {Rappel}}, \bibinfo
  {author} {\bibfnamefont {Jeff}\ \bibnamefont {Hasty}}, \ and\ \bibinfo
  {author} {\bibfnamefont {Lev}\ \bibnamefont {Tsimring}},\ }\bibfield  {title}
  {\enquote {\bibinfo {title} {Flower-like patterns in multi-species bacterial
  colonies},}\ }\href@noop {} {\bibfield  {journal} {\bibinfo  {journal}
  {Elife}\ }\textbf {\bibinfo {volume} {9}},\ \bibinfo {pages} {e48885}
  (\bibinfo {year} {2020})}\BibitemShut {NoStop}%
\bibitem [{\citenamefont {Tsyganov}\ \emph {et~al.}(2003)\citenamefont
  {Tsyganov}, \citenamefont {Brindley}, \citenamefont {Holden},\ and\
  \citenamefont {Biktashev}}]{tsyganov2003quasisoliton}%
  \BibitemOpen
  \bibfield  {author} {\bibinfo {author} {\bibfnamefont {MA}~\bibnamefont
  {Tsyganov}}, \bibinfo {author} {\bibfnamefont {J}~\bibnamefont {Brindley}},
  \bibinfo {author} {\bibfnamefont {AV}~\bibnamefont {Holden}}, \ and\ \bibinfo
  {author} {\bibfnamefont {VN}~\bibnamefont {Biktashev}},\ }\bibfield  {title}
  {\enquote {\bibinfo {title} {Quasisoliton interaction of pursuit-evasion
  waves in a predator-prey system},}\ }\href@noop {} {\bibfield  {journal}
  {\bibinfo  {journal} {Physical review letters}\ }\textbf {\bibinfo {volume}
  {91}},\ \bibinfo {pages} {218102} (\bibinfo {year} {2003})}\BibitemShut
  {NoStop}%
\bibitem [{\citenamefont {Fruchart}\ \emph {et~al.}(2021)\citenamefont
  {Fruchart}, \citenamefont {Hanai}, \citenamefont {Littlewood},\ and\
  \citenamefont {Vitelli}}]{fruchart2021non}%
  \BibitemOpen
  \bibfield  {author} {\bibinfo {author} {\bibfnamefont {Michel}\ \bibnamefont
  {Fruchart}}, \bibinfo {author} {\bibfnamefont {Ryo}\ \bibnamefont {Hanai}},
  \bibinfo {author} {\bibfnamefont {Peter~B}\ \bibnamefont {Littlewood}}, \
  and\ \bibinfo {author} {\bibfnamefont {Vincenzo}\ \bibnamefont {Vitelli}},\
  }\bibfield  {title} {\enquote {\bibinfo {title} {Non-reciprocal phase
  transitions},}\ }\href {\doibase 10.1038/s41586-021-03375-9} {\bibfield
  {journal} {\bibinfo  {journal} {Nature}\ }\textbf {\bibinfo {volume} {592}},\
  \bibinfo {pages} {363--369} (\bibinfo {year} {2021})}\BibitemShut {NoStop}%
\bibitem [{\citenamefont {Frohoff-H{\"u}lsmann}\ \emph
  {et~al.}(2021)\citenamefont {Frohoff-H{\"u}lsmann}, \citenamefont {Wrembel},\
  and\ \citenamefont {Thiele}}]{frohoff2021suppression}%
  \BibitemOpen
  \bibfield  {author} {\bibinfo {author} {\bibfnamefont {Tobias}\ \bibnamefont
  {Frohoff-H{\"u}lsmann}}, \bibinfo {author} {\bibfnamefont {Jana}\
  \bibnamefont {Wrembel}}, \ and\ \bibinfo {author} {\bibfnamefont {Uwe}\
  \bibnamefont {Thiele}},\ }\bibfield  {title} {\enquote {\bibinfo {title}
  {Suppression of coarsening and emergence of oscillatory behavior in a
  cahn-hilliard model with nonvariational coupling},}\ }\href@noop {}
  {\bibfield  {journal} {\bibinfo  {journal} {Physical Review E}\ }\textbf
  {\bibinfo {volume} {103}},\ \bibinfo {pages} {042602} (\bibinfo {year}
  {2021})}\BibitemShut {NoStop}%
\bibitem [{\citenamefont {Yanni}\ \emph {et~al.}(2019)\citenamefont {Yanni},
  \citenamefont {M{\'a}rquez-Zacar{\'\i}as}, \citenamefont {Yunker},\ and\
  \citenamefont {Ratcliff}}]{yanni2019drivers}%
  \BibitemOpen
  \bibfield  {author} {\bibinfo {author} {\bibfnamefont {David}\ \bibnamefont
  {Yanni}}, \bibinfo {author} {\bibfnamefont {Pedro}\ \bibnamefont
  {M{\'a}rquez-Zacar{\'\i}as}}, \bibinfo {author} {\bibfnamefont {Peter~J}\
  \bibnamefont {Yunker}}, \ and\ \bibinfo {author} {\bibfnamefont {William~C}\
  \bibnamefont {Ratcliff}},\ }\bibfield  {title} {\enquote {\bibinfo {title}
  {Drivers of spatial structure in social microbial communities},}\ }\href@noop
  {} {\bibfield  {journal} {\bibinfo  {journal} {Current Biology}\ }\textbf
  {\bibinfo {volume} {29}},\ \bibinfo {pages} {R545--R550} (\bibinfo {year}
  {2019})}\BibitemShut {NoStop}%
\bibitem [{\citenamefont {Curatolo}\ \emph {et~al.}(2019)\citenamefont
  {Curatolo}, \citenamefont {Zhou}, \citenamefont {Zhao}, \citenamefont {Liu},
  \citenamefont {Daerr}, \citenamefont {Tailleur},\ and\ \citenamefont
  {Huang}}]{curatolo2019engineering}%
  \BibitemOpen
  \bibfield  {author} {\bibinfo {author} {\bibfnamefont {AI}~\bibnamefont
  {Curatolo}}, \bibinfo {author} {\bibfnamefont {N}~\bibnamefont {Zhou}},
  \bibinfo {author} {\bibfnamefont {Y}~\bibnamefont {Zhao}}, \bibinfo {author}
  {\bibfnamefont {C}~\bibnamefont {Liu}}, \bibinfo {author} {\bibfnamefont
  {A}~\bibnamefont {Daerr}}, \bibinfo {author} {\bibfnamefont {J}~\bibnamefont
  {Tailleur}}, \ and\ \bibinfo {author} {\bibfnamefont {J}~\bibnamefont
  {Huang}},\ }\bibfield  {title} {\enquote {\bibinfo {title} {Engineering
  cooperative patterns in multi-species bacterial colonies},}\ }\href@noop {}
  {\bibfield  {journal} {\bibinfo  {journal} {bioRxiv}\ ,\ \bibinfo {pages}
  {798827}} (\bibinfo {year} {2019})}\BibitemShut {NoStop}%
\bibitem [{\citenamefont {Bender}(2007)}]{bender2007making}%
  \BibitemOpen
  \bibfield  {author} {\bibinfo {author} {\bibfnamefont {Carl~M}\ \bibnamefont
  {Bender}},\ }\bibfield  {title} {\enquote {\bibinfo {title} {Making sense of
  non-hermitian hamiltonians},}\ }\href@noop {} {\bibfield  {journal} {\bibinfo
   {journal} {Reports on Progress in Physics}\ }\textbf {\bibinfo {volume}
  {70}},\ \bibinfo {pages} {947} (\bibinfo {year} {2007})}\BibitemShut
  {NoStop}%
\bibitem [{\citenamefont {Khemani}\ \emph {et~al.}(2019)\citenamefont
  {Khemani}, \citenamefont {Moessner},\ and\ \citenamefont
  {Sondhi}}]{khemani2019brief}%
  \BibitemOpen
  \bibfield  {author} {\bibinfo {author} {\bibfnamefont {Vedika}\ \bibnamefont
  {Khemani}}, \bibinfo {author} {\bibfnamefont {Roderich}\ \bibnamefont
  {Moessner}}, \ and\ \bibinfo {author} {\bibfnamefont {SL}~\bibnamefont
  {Sondhi}},\ }\bibfield  {title} {\enquote {\bibinfo {title} {A brief history
  of time crystals},}\ }\href@noop {} {\bibfield  {journal} {\bibinfo
  {journal} {arXiv preprint arXiv:1910.10745}\ } (\bibinfo {year}
  {2019})}\BibitemShut {NoStop}%
\bibitem [{\citenamefont {McDonald}\ \emph {et~al.}(2021)\citenamefont
  {McDonald}, \citenamefont {Hanai},\ and\ \citenamefont
  {Clerk}}]{mcdonald2021non}%
  \BibitemOpen
  \bibfield  {author} {\bibinfo {author} {\bibfnamefont {Alexander}\
  \bibnamefont {McDonald}}, \bibinfo {author} {\bibfnamefont {Ryo}\
  \bibnamefont {Hanai}}, \ and\ \bibinfo {author} {\bibfnamefont {Aashish~A}\
  \bibnamefont {Clerk}},\ }\bibfield  {title} {\enquote {\bibinfo {title}
  {Non-equilibrium stationary states of quantum non-hermitian lattice
  models},}\ }\href@noop {} {\bibfield  {journal} {\bibinfo  {journal} {arXiv
  preprint arXiv:2103.01941}\ } (\bibinfo {year} {2021})}\BibitemShut {NoStop}%
\bibitem [{\citenamefont {Scheibner}\ \emph {et~al.}(2020)\citenamefont
  {Scheibner}, \citenamefont {Souslov}, \citenamefont {Banerjee}, \citenamefont
  {Surowka}, \citenamefont {Irvine},\ and\ \citenamefont
  {Vitelli}}]{scheibner2020odd}%
  \BibitemOpen
  \bibfield  {author} {\bibinfo {author} {\bibfnamefont {Colin}\ \bibnamefont
  {Scheibner}}, \bibinfo {author} {\bibfnamefont {Anton}\ \bibnamefont
  {Souslov}}, \bibinfo {author} {\bibfnamefont {Debarghya}\ \bibnamefont
  {Banerjee}}, \bibinfo {author} {\bibfnamefont {Piotr}\ \bibnamefont
  {Surowka}}, \bibinfo {author} {\bibfnamefont {William~TM}\ \bibnamefont
  {Irvine}}, \ and\ \bibinfo {author} {\bibfnamefont {Vincenzo}\ \bibnamefont
  {Vitelli}},\ }\bibfield  {title} {\enquote {\bibinfo {title} {Odd
  elasticity},}\ }\href@noop {} {\bibfield  {journal} {\bibinfo  {journal}
  {Nature Physics}\ }\textbf {\bibinfo {volume} {16}},\ \bibinfo {pages}
  {475--480} (\bibinfo {year} {2020})}\BibitemShut {NoStop}%
\bibitem [{\citenamefont {O’Byrne}\ and\ \citenamefont
  {Tailleur}(2020)}]{o2020lamellar}%
  \BibitemOpen
  \bibfield  {author} {\bibinfo {author} {\bibfnamefont {J{\'e}r{\'e}my}\
  \bibnamefont {O’Byrne}}\ and\ \bibinfo {author} {\bibfnamefont {Julien}\
  \bibnamefont {Tailleur}},\ }\bibfield  {title} {\enquote {\bibinfo {title}
  {Lamellar to micellar phases and beyond: When tactic active systems admit
  free energy functionals},}\ }\href
  {https://doi.org/10.1103/PhysRevLett.125.208003} {\bibfield  {journal}
  {\bibinfo  {journal} {Physical Review Letters}\ }\textbf {\bibinfo {volume}
  {125}},\ \bibinfo {pages} {208003} (\bibinfo {year} {2020})}\BibitemShut
  {NoStop}%
\bibitem [{\citenamefont {Gupta}\ \emph {et~al.}(2020)\citenamefont {Gupta},
  \citenamefont {Kant}, \citenamefont {Soni}, \citenamefont {Sood},\ and\
  \citenamefont {Ramaswamy}}]{gupta2020active}%
  \BibitemOpen
  \bibfield  {author} {\bibinfo {author} {\bibfnamefont {Rahul~Kumar}\
  \bibnamefont {Gupta}}, \bibinfo {author} {\bibfnamefont {Raushan}\
  \bibnamefont {Kant}}, \bibinfo {author} {\bibfnamefont {Harsh}\ \bibnamefont
  {Soni}}, \bibinfo {author} {\bibfnamefont {AK}~\bibnamefont {Sood}}, \ and\
  \bibinfo {author} {\bibfnamefont {Sriram}\ \bibnamefont {Ramaswamy}},\
  }\bibfield  {title} {\enquote {\bibinfo {title} {Active nonreciprocal
  attraction between motile particles in an elastic medium},}\ }\href
  {https://arxiv.org/abs/2007.04860} {\bibfield  {journal} {\bibinfo  {journal}
  {arXiv preprint arXiv:2007.04860}\ } (\bibinfo {year} {2020})}\BibitemShut
  {NoStop}%
\bibitem [{\citenamefont {Kruse}\ \emph {et~al.}(2005)\citenamefont {Kruse},
  \citenamefont {Joanny}, \citenamefont {J{\"u}licher}, \citenamefont {Prost},\
  and\ \citenamefont {Sekimoto}}]{kruse2005generic}%
  \BibitemOpen
  \bibfield  {author} {\bibinfo {author} {\bibfnamefont {Karsten}\ \bibnamefont
  {Kruse}}, \bibinfo {author} {\bibfnamefont {Jean-Francois}\ \bibnamefont
  {Joanny}}, \bibinfo {author} {\bibfnamefont {Frank}\ \bibnamefont
  {J{\"u}licher}}, \bibinfo {author} {\bibfnamefont {Jacques}\ \bibnamefont
  {Prost}}, \ and\ \bibinfo {author} {\bibfnamefont {Ken}\ \bibnamefont
  {Sekimoto}},\ }\bibfield  {title} {\enquote {\bibinfo {title} {Generic theory
  of active polar gels: a paradigm for cytoskeletal dynamics},}\ }\href@noop {}
  {\bibfield  {journal} {\bibinfo  {journal} {The European Physical Journal E}\
  }\textbf {\bibinfo {volume} {16}},\ \bibinfo {pages} {5--16} (\bibinfo {year}
  {2005})}\BibitemShut {NoStop}%
\bibitem [{\citenamefont {Ramaswamy}(2017)}]{Ramaswamy_2017}%
  \BibitemOpen
  \bibfield  {author} {\bibinfo {author} {\bibfnamefont {Sriram}\ \bibnamefont
  {Ramaswamy}},\ }\bibfield  {title} {\enquote {\bibinfo {title} {Active
  matter},}\ }\href {\doibase 10.1088/1742-5468/aa6bc5} {\bibfield  {journal}
  {\bibinfo  {journal} {Journal of Statistical Mechanics: Theory and
  Experiment}\ }\textbf {\bibinfo {volume} {2017}},\ \bibinfo {pages} {054002}
  (\bibinfo {year} {2017})}\BibitemShut {NoStop}%
\bibitem [{\citenamefont {Dadhichi}\ \emph {et~al.}(2018)\citenamefont
  {Dadhichi}, \citenamefont {Maitra},\ and\ \citenamefont
  {Ramaswamy}}]{dadhichi2018origins}%
  \BibitemOpen
  \bibfield  {author} {\bibinfo {author} {\bibfnamefont {Lokrshi~Prawar}\
  \bibnamefont {Dadhichi}}, \bibinfo {author} {\bibfnamefont {Ananyo}\
  \bibnamefont {Maitra}}, \ and\ \bibinfo {author} {\bibfnamefont {Sriram}\
  \bibnamefont {Ramaswamy}},\ }\bibfield  {title} {\enquote {\bibinfo {title}
  {Origins and diagnostics of the nonequilibrium character of active
  systems},}\ }\href {\doibase 10.1088/1742-5468/aae852/meta} {\bibfield
  {journal} {\bibinfo  {journal} {Journal of Statistical Mechanics: Theory and
  Experiment}\ }\textbf {\bibinfo {volume} {2018}},\ \bibinfo {pages} {123201}
  (\bibinfo {year} {2018})}\BibitemShut {NoStop}%
\bibitem [{\citenamefont {Dadhichi}\ \emph {et~al.}(2020)\citenamefont
  {Dadhichi}, \citenamefont {Kethapelli}, \citenamefont {Chajwa}, \citenamefont
  {Ramaswamy},\ and\ \citenamefont {Maitra}}]{dadhichi2020nonmutual}%
  \BibitemOpen
  \bibfield  {author} {\bibinfo {author} {\bibfnamefont {Lokrshi~Prawar}\
  \bibnamefont {Dadhichi}}, \bibinfo {author} {\bibfnamefont {Jitendra}\
  \bibnamefont {Kethapelli}}, \bibinfo {author} {\bibfnamefont {Rahul}\
  \bibnamefont {Chajwa}}, \bibinfo {author} {\bibfnamefont {Sriram}\
  \bibnamefont {Ramaswamy}}, \ and\ \bibinfo {author} {\bibfnamefont {Ananyo}\
  \bibnamefont {Maitra}},\ }\bibfield  {title} {\enquote {\bibinfo {title}
  {Nonmutual torques and the unimportance of motility for long-range order in
  two-dimensional flocks},}\ }\href {\doibase 10.1103/PhysRevE.101.052601}
  {\bibfield  {journal} {\bibinfo  {journal} {Physical Review E}\ }\textbf
  {\bibinfo {volume} {101}},\ \bibinfo {pages} {052601} (\bibinfo {year}
  {2020})}\BibitemShut {NoStop}%
\bibitem [{\citenamefont {Kelvin}(1894)}]{kelvin1894molecular}%
  \BibitemOpen
  \bibfield  {author} {\bibinfo {author} {\bibfnamefont {Baron}\ \bibnamefont
  {Kelvin}, \bibfnamefont {William~Thomson}},\ }\href
  {https://archive.org/details/moleculartactic00kelvgoog} {\emph {\bibinfo
  {title} {The molecular tactics of a crystal}}}\ (\bibinfo  {publisher}
  {Clarendon Press, Oxford},\ \bibinfo {year} {1894})\BibitemShut {NoStop}%
\bibitem [{\citenamefont {Pasteur}(1848)}]{pasteur1848relations}%
  \BibitemOpen
  \bibfield  {author} {\bibinfo {author} {\bibfnamefont {Louis}\ \bibnamefont
  {Pasteur}},\ }\bibfield  {title} {\enquote {\bibinfo {title} {Sur les
  relations qui peuvent exister entre la forme crystalline, la composition
  chimique et le sens de la polarization rotatoire},}\ }\href
  {https://wellcomecollection.org/works/rdepqfzw/download?sierraId=b30379696}
  {\bibfield  {journal} {\bibinfo  {journal} {Annales Chimie Phys.}\ }\textbf
  {\bibinfo {volume} {24}},\ \bibinfo {pages} {442--459} (\bibinfo {year}
  {1848})}\BibitemShut {NoStop}%
\bibitem [{\citenamefont {Lubensky}\ \emph {et~al.}(1998)\citenamefont
  {Lubensky}, \citenamefont {Harris}, \citenamefont {Kamien},\ and\
  \citenamefont {Yan}}]{lubensky1998chirality}%
  \BibitemOpen
  \bibfield  {author} {\bibinfo {author} {\bibfnamefont {Tom~C}\ \bibnamefont
  {Lubensky}}, \bibinfo {author} {\bibfnamefont {A~Brooks}\ \bibnamefont
  {Harris}}, \bibinfo {author} {\bibfnamefont {Randall~D}\ \bibnamefont
  {Kamien}}, \ and\ \bibinfo {author} {\bibfnamefont {Gu}~\bibnamefont {Yan}},\
  }\bibfield  {title} {\enquote {\bibinfo {title} {Chirality in liquid
  crystals: From microscopic origins to macroscopic structure},}\ }\href
  {https://www.tandfonline.com/doi/abs/10.1080/00150199808217346} {\bibfield
  {journal} {\bibinfo  {journal} {Ferroelectrics}\ }\textbf {\bibinfo {volume}
  {212}},\ \bibinfo {pages} {1--20} (\bibinfo {year} {1998})}\BibitemShut
  {NoStop}%
\bibitem [{\citenamefont {Harris}\ \emph {et~al.}(1999)\citenamefont {Harris},
  \citenamefont {Kamien},\ and\ \citenamefont
  {Lubensky}}]{harris1999molecular}%
  \BibitemOpen
  \bibfield  {author} {\bibinfo {author} {\bibfnamefont {A~Brooks}\
  \bibnamefont {Harris}}, \bibinfo {author} {\bibfnamefont {Randall~D}\
  \bibnamefont {Kamien}}, \ and\ \bibinfo {author} {\bibfnamefont {Tom~C}\
  \bibnamefont {Lubensky}},\ }\bibfield  {title} {\enquote {\bibinfo {title}
  {Molecular chirality and chiral parameters},}\ }\href
  {https://journals.aps.org/rmp/abstract/10.1103/RevModPhys.71.1745} {\bibfield
   {journal} {\bibinfo  {journal} {Reviews of Modern Physics}\ }\textbf
  {\bibinfo {volume} {71}},\ \bibinfo {pages} {1745} (\bibinfo {year}
  {1999})}\BibitemShut {NoStop}%
\bibitem [{\citenamefont {Efrati}\ and\ \citenamefont
  {Irvine}(2014)}]{efrati2014orientation}%
  \BibitemOpen
  \bibfield  {author} {\bibinfo {author} {\bibfnamefont {Efi}\ \bibnamefont
  {Efrati}}\ and\ \bibinfo {author} {\bibfnamefont {William~TM}\ \bibnamefont
  {Irvine}},\ }\bibfield  {title} {\enquote {\bibinfo {title}
  {Orientation-dependent handedness and chiral design},}\ }\href {\doibase
  10.1103/PhysRevX.4.011003} {\bibfield  {journal} {\bibinfo  {journal}
  {Physical Review X}\ }\textbf {\bibinfo {volume} {4}},\ \bibinfo {pages}
  {011003} (\bibinfo {year} {2014})}\BibitemShut {NoStop}%
\bibitem [{\citenamefont {Dietler}\ \emph {et~al.}(2020)\citenamefont
  {Dietler}, \citenamefont {Kusner}, \citenamefont {Kusner}, \citenamefont
  {Rawdon},\ and\ \citenamefont {Szymczak}}]{dietler2020chirality}%
  \BibitemOpen
  \bibfield  {author} {\bibinfo {author} {\bibfnamefont {Giovanni}\
  \bibnamefont {Dietler}}, \bibinfo {author} {\bibfnamefont {Robert}\
  \bibnamefont {Kusner}}, \bibinfo {author} {\bibfnamefont {W{\"{o}}den}\
  \bibnamefont {Kusner}}, \bibinfo {author} {\bibfnamefont {Eric}\ \bibnamefont
  {Rawdon}}, \ and\ \bibinfo {author} {\bibfnamefont {Piotr}\ \bibnamefont
  {Szymczak}},\ }\bibfield  {title} {\enquote {\bibinfo {title} {Chirality for
  crooked curves},}\ }\href {https://arxiv.org/abs/2004.10338} {\bibfield
  {journal} {\bibinfo  {journal} {arXiv preprint arXiv:2004.10338}\ } (\bibinfo
  {year} {2020})}\BibitemShut {NoStop}%
\bibitem [{\citenamefont {Lubensky}(1972)}]{lubensky1972hydrodynamics}%
  \BibitemOpen
  \bibfield  {author} {\bibinfo {author} {\bibfnamefont {Tom~C}\ \bibnamefont
  {Lubensky}},\ }\bibfield  {title} {\enquote {\bibinfo {title} {Hydrodynamics
  of cholesteric liquid crystals},}\ }\href
  {https://journals.aps.org/pra/abstract/10.1103/PhysRevA.6.452} {\bibfield
  {journal} {\bibinfo  {journal} {Physical Review A}\ }\textbf {\bibinfo
  {volume} {6}},\ \bibinfo {pages} {452} (\bibinfo {year} {1972})}\BibitemShut
  {NoStop}%
\bibitem [{\citenamefont {Radzihovsky}\ and\ \citenamefont
  {Lubensky}(2011)}]{radzihovsky2011nonlinear}%
  \BibitemOpen
  \bibfield  {author} {\bibinfo {author} {\bibfnamefont {Leo}\ \bibnamefont
  {Radzihovsky}}\ and\ \bibinfo {author} {\bibfnamefont {TC}~\bibnamefont
  {Lubensky}},\ }\bibfield  {title} {\enquote {\bibinfo {title} {Nonlinear
  smectic elasticity of helical state in cholesteric liquid crystals and
  helimagnets},}\ }\href
  {https://journals.aps.org/pre/abstract/10.1103/PhysRevE.83.051701} {\bibfield
   {journal} {\bibinfo  {journal} {Physical Review E}\ }\textbf {\bibinfo
  {volume} {83}},\ \bibinfo {pages} {051701} (\bibinfo {year}
  {2011})}\BibitemShut {NoStop}%
\bibitem [{\citenamefont {Maitra}\ and\ \citenamefont
  {Ramaswamy}(2019)}]{maitra2019oriented}%
  \BibitemOpen
  \bibfield  {author} {\bibinfo {author} {\bibfnamefont {Ananyo}\ \bibnamefont
  {Maitra}}\ and\ \bibinfo {author} {\bibfnamefont {Sriram}\ \bibnamefont
  {Ramaswamy}},\ }\bibfield  {title} {\enquote {\bibinfo {title} {Oriented
  active solids},}\ }\href
  {https://journals.aps.org/prl/abstract/10.1103/PhysRevLett.123.238001}
  {\bibfield  {journal} {\bibinfo  {journal} {Physical review letters}\
  }\textbf {\bibinfo {volume} {123}},\ \bibinfo {pages} {238001} (\bibinfo
  {year} {2019})}\BibitemShut {NoStop}%
\bibitem [{\citenamefont {Adhyapak}\ \emph {et~al.}(2013)\citenamefont
  {Adhyapak}, \citenamefont {Ramaswamy},\ and\ \citenamefont
  {Toner}}]{adhyapak2013live}%
  \BibitemOpen
  \bibfield  {author} {\bibinfo {author} {\bibfnamefont {Tapan~Chandra}\
  \bibnamefont {Adhyapak}}, \bibinfo {author} {\bibfnamefont {Sriram}\
  \bibnamefont {Ramaswamy}}, \ and\ \bibinfo {author} {\bibfnamefont {John}\
  \bibnamefont {Toner}},\ }\bibfield  {title} {\enquote {\bibinfo {title} {Live
  soap: stability, order, and fluctuations in apolar active smectics},}\ }\href
  {https://journals.aps.org/prl/abstract/10.1103/PhysRevLett.110.118102}
  {\bibfield  {journal} {\bibinfo  {journal} {Physical review letters}\
  }\textbf {\bibinfo {volume} {110}},\ \bibinfo {pages} {118102} (\bibinfo
  {year} {2013})}\BibitemShut {NoStop}%
\bibitem [{\citenamefont {Kole}\ \emph {et~al.}(2020)\citenamefont {Kole},
  \citenamefont {Alexander}, \citenamefont {Ramaswamy},\ and\ \citenamefont
  {Maitra}}]{kole2020active}%
  \BibitemOpen
  \bibfield  {author} {\bibinfo {author} {\bibfnamefont {SJ}~\bibnamefont
  {Kole}}, \bibinfo {author} {\bibfnamefont {Gareth~P}\ \bibnamefont
  {Alexander}}, \bibinfo {author} {\bibfnamefont {Sriram}\ \bibnamefont
  {Ramaswamy}}, \ and\ \bibinfo {author} {\bibfnamefont {Ananyo}\ \bibnamefont
  {Maitra}},\ }\bibfield  {title} {\enquote {\bibinfo {title} {Layered chiral
  active matter: Beyond odd elasticity},}\ }\href
  {https://journals.aps.org/prl/abstract/10.1103/PhysRevLett.126.248001}
  {\bibfield  {journal} {\bibinfo  {journal} {Physical Review Letters}\
  }\textbf {\bibinfo {volume} {126}},\ \bibinfo {pages} {248001} (\bibinfo
  {year} {2020})}\BibitemShut {NoStop}%
\bibitem [{\citenamefont {L{\"o}wen}(2016)}]{lowen2016chirality}%
  \BibitemOpen
  \bibfield  {author} {\bibinfo {author} {\bibfnamefont {Hartmut}\ \bibnamefont
  {L{\"o}wen}},\ }\bibfield  {title} {\enquote {\bibinfo {title} {Chirality in
  microswimmer motion: From circle swimmers to active turbulence},}\ }\href
  {https://link.springer.com/article/10.1140/epjst/e2016-60054-6} {\bibfield
  {journal} {\bibinfo  {journal} {The European Physical Journal Special
  Topics}\ }\textbf {\bibinfo {volume} {225}},\ \bibinfo {pages} {2319--2331}
  (\bibinfo {year} {2016})}\BibitemShut {NoStop}%
\bibitem [{\citenamefont {Liebchen}\ and\ \citenamefont
  {Levis}(2017)}]{liebchen2017collective}%
  \BibitemOpen
  \bibfield  {author} {\bibinfo {author} {\bibfnamefont {Benno}\ \bibnamefont
  {Liebchen}}\ and\ \bibinfo {author} {\bibfnamefont {Demian}\ \bibnamefont
  {Levis}},\ }\bibfield  {title} {\enquote {\bibinfo {title} {Collective
  behavior of chiral active matter: pattern formation and enhanced flocking},}\
  }\href {https://journals.aps.org/prl/abstract/10.1103/PhysRevLett.119.058002}
  {\bibfield  {journal} {\bibinfo  {journal} {Physical review letters}\
  }\textbf {\bibinfo {volume} {119}},\ \bibinfo {pages} {058002} (\bibinfo
  {year} {2017})}\BibitemShut {NoStop}%
\bibitem [{\citenamefont {van Zuiden}\ \emph
  {et~al.}(2016{\natexlab{a}})\citenamefont {van Zuiden}, \citenamefont
  {Paulose}, \citenamefont {Irvine}, \citenamefont {Bartolo},\ and\
  \citenamefont {Vitelli}}]{vanZuiden12919}%
  \BibitemOpen
  \bibfield  {author} {\bibinfo {author} {\bibfnamefont {Benjamin~C.}\
  \bibnamefont {van Zuiden}}, \bibinfo {author} {\bibfnamefont {Jayson}\
  \bibnamefont {Paulose}}, \bibinfo {author} {\bibfnamefont {William T.~M.}\
  \bibnamefont {Irvine}}, \bibinfo {author} {\bibfnamefont {Denis}\
  \bibnamefont {Bartolo}}, \ and\ \bibinfo {author} {\bibfnamefont {Vincenzo}\
  \bibnamefont {Vitelli}},\ }\bibfield  {title} {\enquote {\bibinfo {title}
  {Spatiotemporal order and emergent edge currents in active spinner
  materials},}\ }\href {\doibase 10.1073/pnas.1609572113} {\bibfield  {journal}
  {\bibinfo  {journal} {Proceedings of the National Academy of Sciences}\
  }\textbf {\bibinfo {volume} {113}},\ \bibinfo {pages} {12919--12924}
  (\bibinfo {year} {2016}{\natexlab{a}})}\BibitemShut {NoStop}%
\bibitem [{\citenamefont {Aragones~Gomez}\ \emph {et~al.}(2016)\citenamefont
  {Aragones~Gomez}, \citenamefont {Steimel},\ and\ \citenamefont
  {Alexander-Katz}}]{aragones2016elasticity}%
  \BibitemOpen
  \bibfield  {author} {\bibinfo {author} {\bibfnamefont {Juan~Luis}\
  \bibnamefont {Aragones~Gomez}}, \bibinfo {author} {\bibfnamefont {Joshua~P}\
  \bibnamefont {Steimel}}, \ and\ \bibinfo {author} {\bibfnamefont {Alfredo}\
  \bibnamefont {Alexander-Katz}},\ }\bibfield  {title} {\enquote {\bibinfo
  {title} {Elasticity-induced force reversal between active spinning particles
  in dense passive media},}\ }\href
  {https://www.nature.com/articles/ncomms11325} {\ \textbf {\bibinfo {volume}
  {7}},\ \bibinfo {pages} {11325} (\bibinfo {year} {2016})}\BibitemShut
  {NoStop}%
\bibitem [{\citenamefont {Tsai}\ \emph {et~al.}(2005)\citenamefont {Tsai},
  \citenamefont {Ye}, \citenamefont {Rodriguez}, \citenamefont {Gollub},\ and\
  \citenamefont {Lubensky}}]{Tsai2005chiral}%
  \BibitemOpen
  \bibfield  {author} {\bibinfo {author} {\bibfnamefont {J.-C.}\ \bibnamefont
  {Tsai}}, \bibinfo {author} {\bibfnamefont {Fangfu}\ \bibnamefont {Ye}},
  \bibinfo {author} {\bibfnamefont {Juan}\ \bibnamefont {Rodriguez}}, \bibinfo
  {author} {\bibfnamefont {J.~P.}\ \bibnamefont {Gollub}}, \ and\ \bibinfo
  {author} {\bibfnamefont {T.~C.}\ \bibnamefont {Lubensky}},\ }\bibfield
  {title} {\enquote {\bibinfo {title} {A chiral granular gas},}\ }\href
  {\doibase 10.1103/PhysRevLett.94.214301} {\bibfield  {journal} {\bibinfo
  {journal} {Phys. Rev. Lett.}\ }\textbf {\bibinfo {volume} {94}},\ \bibinfo
  {pages} {214301} (\bibinfo {year} {2005})}\BibitemShut {NoStop}%
\bibitem [{\citenamefont {Ahmad}\ \emph {et~al.}(2013)\citenamefont {Ahmad},
  \citenamefont {Routh}, \citenamefont {Kamarthapu}, \citenamefont
  {Chalissery}, \citenamefont {Muthukumar}, \citenamefont {Hussain},
  \citenamefont {Kruparani}, \citenamefont {Deshmukh},\ and\ \citenamefont
  {Sankaranarayanan}}]{ahmad2013mechanism}%
  \BibitemOpen
  \bibfield  {author} {\bibinfo {author} {\bibfnamefont {Sadeem}\ \bibnamefont
  {Ahmad}}, \bibinfo {author} {\bibfnamefont {Satya~Brata}\ \bibnamefont
  {Routh}}, \bibinfo {author} {\bibfnamefont {Venu}\ \bibnamefont
  {Kamarthapu}}, \bibinfo {author} {\bibfnamefont {Jisha}\ \bibnamefont
  {Chalissery}}, \bibinfo {author} {\bibfnamefont {Sowndarya}\ \bibnamefont
  {Muthukumar}}, \bibinfo {author} {\bibfnamefont {Tanweer}\ \bibnamefont
  {Hussain}}, \bibinfo {author} {\bibfnamefont {Shobha~P}\ \bibnamefont
  {Kruparani}}, \bibinfo {author} {\bibfnamefont {Mandar~V}\ \bibnamefont
  {Deshmukh}}, \ and\ \bibinfo {author} {\bibfnamefont {Rajan}\ \bibnamefont
  {Sankaranarayanan}},\ }\bibfield  {title} {\enquote {\bibinfo {title}
  {Mechanism of chiral proofreading during translation of the genetic code},}\
  }\href {https://elifesciences.org/articles/01519} {\bibfield  {journal}
  {\bibinfo  {journal} {Elife}\ }\textbf {\bibinfo {volume} {2}},\ \bibinfo
  {pages} {e01519} (\bibinfo {year} {2013})}\BibitemShut {NoStop}%
\bibitem [{\citenamefont {Arora}\ \emph {et~al.}(2021)\citenamefont {Arora},
  \citenamefont {Sood},\ and\ \citenamefont {Ganapathy}}]{arora2021emergent}%
  \BibitemOpen
  \bibfield  {author} {\bibinfo {author} {\bibfnamefont {Pragya}\ \bibnamefont
  {Arora}}, \bibinfo {author} {\bibfnamefont {AK}~\bibnamefont {Sood}}, \ and\
  \bibinfo {author} {\bibfnamefont {Rajesh}\ \bibnamefont {Ganapathy}},\
  }\bibfield  {title} {\enquote {\bibinfo {title} {Emergent stereoselective
  interactions and self-recognition in polar chiral active ellipsoids},}\
  }\href {https://advances.sciencemag.org/content/7/9/eabd0331} {\bibfield
  {journal} {\bibinfo  {journal} {Science Advances}\ }\textbf {\bibinfo
  {volume} {7}},\ \bibinfo {pages} {eabd0331} (\bibinfo {year}
  {2021})}\BibitemShut {NoStop}%
\bibitem [{\citenamefont {F{\"u}rthauer}\ \emph {et~al.}(2012)\citenamefont
  {F{\"u}rthauer}, \citenamefont {Strempel}, \citenamefont {Grill},\ and\
  \citenamefont {J{\"u}licher}}]{furthauer2012active}%
  \BibitemOpen
  \bibfield  {author} {\bibinfo {author} {\bibfnamefont {S}~\bibnamefont
  {F{\"u}rthauer}}, \bibinfo {author} {\bibfnamefont {M}~\bibnamefont
  {Strempel}}, \bibinfo {author} {\bibfnamefont {Stephan~W}\ \bibnamefont
  {Grill}}, \ and\ \bibinfo {author} {\bibfnamefont {Frank}\ \bibnamefont
  {J{\"u}licher}},\ }\bibfield  {title} {\enquote {\bibinfo {title} {Active
  chiral fluids},}\ }\href
  {https://link.springer.com/article/10.1140/epje/i2012-12089-6} {\bibfield
  {journal} {\bibinfo  {journal} {The European Physical Journal E}\ }\textbf
  {\bibinfo {volume} {35}},\ \bibinfo {pages} {1--13} (\bibinfo {year}
  {2012})}\BibitemShut {NoStop}%
\bibitem [{\citenamefont {F{\"u}rthauer}\ \emph {et~al.}(2013)\citenamefont
  {F{\"u}rthauer}, \citenamefont {Strempel}, \citenamefont {Grill},\ and\
  \citenamefont {J{\"u}licher}}]{furthauer2013active}%
  \BibitemOpen
  \bibfield  {author} {\bibinfo {author} {\bibfnamefont {Sebastian}\
  \bibnamefont {F{\"u}rthauer}}, \bibinfo {author} {\bibfnamefont
  {M}~\bibnamefont {Strempel}}, \bibinfo {author} {\bibfnamefont {Stephan~W}\
  \bibnamefont {Grill}}, \ and\ \bibinfo {author} {\bibfnamefont {Frank}\
  \bibnamefont {J{\"u}licher}},\ }\bibfield  {title} {\enquote {\bibinfo
  {title} {Active chiral processes in thin films},}\ }\href
  {https://journals.aps.org/prl/abstract/10.1103/PhysRevLett.110.048103}
  {\bibfield  {journal} {\bibinfo  {journal} {Physical Review Letters}\
  }\textbf {\bibinfo {volume} {110}},\ \bibinfo {pages} {048103} (\bibinfo
  {year} {2013})}\BibitemShut {NoStop}%
\bibitem [{\citenamefont {Markovich}\ \emph {et~al.}(2019)\citenamefont
  {Markovich}, \citenamefont {Tjhung},\ and\ \citenamefont
  {Cates}}]{markovich2019chiral}%
  \BibitemOpen
  \bibfield  {author} {\bibinfo {author} {\bibfnamefont {Tomer}\ \bibnamefont
  {Markovich}}, \bibinfo {author} {\bibfnamefont {Elsen}\ \bibnamefont
  {Tjhung}}, \ and\ \bibinfo {author} {\bibfnamefont {Michael~E}\ \bibnamefont
  {Cates}},\ }\bibfield  {title} {\enquote {\bibinfo {title} {Chiral active
  matter: microscopic ‘torque dipoles’ have more than one hydrodynamic
  description},}\ }\href {https://doi.org/10.1088/1367-2630/ab54af} {\bibfield
  {journal} {\bibinfo  {journal} {New Journal of Physics}\ }\textbf {\bibinfo
  {volume} {21}},\ \bibinfo {pages} {112001} (\bibinfo {year}
  {2019})}\BibitemShut {NoStop}%
\bibitem [{\citenamefont {Maitra}\ and\ \citenamefont
  {Lenz}(2019)}]{maitra2019spontaneous}%
  \BibitemOpen
  \bibfield  {author} {\bibinfo {author} {\bibfnamefont {Ananyo}\ \bibnamefont
  {Maitra}}\ and\ \bibinfo {author} {\bibfnamefont {Martin}\ \bibnamefont
  {Lenz}},\ }\bibfield  {title} {\enquote {\bibinfo {title} {Spontaneous
  rotation can stabilise ordered chiral active fluids},}\ }\href
  {https://www.nature.com/articles/s41467-019-08914-7} {\bibfield  {journal}
  {\bibinfo  {journal} {Nature communications}\ }\textbf {\bibinfo {volume}
  {10}},\ \bibinfo {pages} {1--6} (\bibinfo {year} {2019})}\BibitemShut
  {NoStop}%
\bibitem [{\citenamefont {Simha}\ and\ \citenamefont
  {Ramaswamy}(2002{\natexlab{b}})}]{simha2002hydrodynamic}%
  \BibitemOpen
  \bibfield  {author} {\bibinfo {author} {\bibfnamefont {R~Aditi}\ \bibnamefont
  {Simha}}\ and\ \bibinfo {author} {\bibfnamefont {Sriram}\ \bibnamefont
  {Ramaswamy}},\ }\bibfield  {title} {\enquote {\bibinfo {title} {Hydrodynamic
  fluctuations and instabilities in ordered suspensions of self-propelled
  particles},}\ }\href
  {https://journals.aps.org/prl/abstract/10.1103/PhysRevLett.89.058101}
  {\bibfield  {journal} {\bibinfo  {journal} {Physical review letters}\
  }\textbf {\bibinfo {volume} {89}},\ \bibinfo {pages} {058101} (\bibinfo
  {year} {2002}{\natexlab{b}})}\BibitemShut {NoStop}%
\bibitem [{\citenamefont {Naganathan}\ \emph {et~al.}(2014)\citenamefont
  {Naganathan}, \citenamefont {F{\"u}rthauer}, \citenamefont {Nishikawa},
  \citenamefont {J{\"u}licher},\ and\ \citenamefont
  {Grill}}]{naganathan2014active}%
  \BibitemOpen
  \bibfield  {author} {\bibinfo {author} {\bibfnamefont {Sundar~Ram}\
  \bibnamefont {Naganathan}}, \bibinfo {author} {\bibfnamefont {Sebastian}\
  \bibnamefont {F{\"u}rthauer}}, \bibinfo {author} {\bibfnamefont {Masatoshi}\
  \bibnamefont {Nishikawa}}, \bibinfo {author} {\bibfnamefont {Frank}\
  \bibnamefont {J{\"u}licher}}, \ and\ \bibinfo {author} {\bibfnamefont
  {Stephan~W}\ \bibnamefont {Grill}},\ }\bibfield  {title} {\enquote {\bibinfo
  {title} {Active torque generation by the actomyosin cell cortex drives
  left--right symmetry breaking},}\ }\href
  {https://elifesciences.org/articles/04165} {\bibfield  {journal} {\bibinfo
  {journal} {elife}\ }\textbf {\bibinfo {volume} {3}},\ \bibinfo {pages}
  {e04165} (\bibinfo {year} {2014})}\BibitemShut {NoStop}%
\bibitem [{\citenamefont {Pimpale}\ \emph {et~al.}(2020)\citenamefont
  {Pimpale}, \citenamefont {Middelkoop}, \citenamefont {Mietke},\ and\
  \citenamefont {Grill}}]{pimpale2020cell}%
  \BibitemOpen
  \bibfield  {author} {\bibinfo {author} {\bibfnamefont {Lokesh~G}\
  \bibnamefont {Pimpale}}, \bibinfo {author} {\bibfnamefont {Teije~C}\
  \bibnamefont {Middelkoop}}, \bibinfo {author} {\bibfnamefont {Alexander}\
  \bibnamefont {Mietke}}, \ and\ \bibinfo {author} {\bibfnamefont {Stephan~W}\
  \bibnamefont {Grill}},\ }\bibfield  {title} {\enquote {\bibinfo {title} {Cell
  lineage-dependent chiral actomyosin flows drive cellular rearrangements in
  early caenorhabditis elegans development},}\ }\href
  {https://elifesciences.org/articles/54930} {\bibfield  {journal} {\bibinfo
  {journal} {Elife}\ }\textbf {\bibinfo {volume} {9}},\ \bibinfo {pages}
  {e54930} (\bibinfo {year} {2020})}\BibitemShut {NoStop}%
\bibitem [{\citenamefont {Tan}\ \emph {et~al.}(2021{\natexlab{a}})\citenamefont
  {Tan}, \citenamefont {Mietke}, \citenamefont {Higinbotham}, \citenamefont
  {Li}, \citenamefont {Chen}, \citenamefont {Foster}, \citenamefont {Gokhale},
  \citenamefont {Dunkel},\ and\ \citenamefont {Fakhri}}]{tan2021development}%
  \BibitemOpen
  \bibfield  {author} {\bibinfo {author} {\bibfnamefont {Tzer~Han}\
  \bibnamefont {Tan}}, \bibinfo {author} {\bibfnamefont {Alexander}\
  \bibnamefont {Mietke}}, \bibinfo {author} {\bibfnamefont {Hugh}\ \bibnamefont
  {Higinbotham}}, \bibinfo {author} {\bibfnamefont {Junang}\ \bibnamefont
  {Li}}, \bibinfo {author} {\bibfnamefont {Yuchao}\ \bibnamefont {Chen}},
  \bibinfo {author} {\bibfnamefont {Peter~J.}\ \bibnamefont {Foster}}, \bibinfo
  {author} {\bibfnamefont {Shreyas}\ \bibnamefont {Gokhale}}, \bibinfo {author}
  {\bibfnamefont {Jörn}\ \bibnamefont {Dunkel}}, \ and\ \bibinfo {author}
  {\bibfnamefont {Nikta}\ \bibnamefont {Fakhri}},\ }\bibfield  {title}
  {\enquote {\bibinfo {title} {Development drives dynamics of living chiral
  crystals},}\ }\href@noop {} {\  (\bibinfo {year} {2021}{\natexlab{a}})},\
  \Eprint {http://arxiv.org/abs/2105.07507} {arXiv:2105.07507 [cond-mat.soft]}
  \BibitemShut {NoStop}%
\bibitem [{\citenamefont {Helfrich}(1970)}]{helfrich1970deformation}%
  \BibitemOpen
  \bibfield  {author} {\bibinfo {author} {\bibfnamefont {W}~\bibnamefont
  {Helfrich}},\ }\bibfield  {title} {\enquote {\bibinfo {title} {Deformation of
  cholesteric liquid crystals with low threshold voltage},}\ }\href
  {https://doi.org/10.1063/1.1653297} {\bibfield  {journal} {\bibinfo
  {journal} {Applied Physics Letters}\ }\textbf {\bibinfo {volume} {17}},\
  \bibinfo {pages} {531--532} (\bibinfo {year} {1970})}\BibitemShut {NoStop}%
\bibitem [{\citenamefont {Hurault}(1973)}]{hurault1973static}%
  \BibitemOpen
  \bibfield  {author} {\bibinfo {author} {\bibfnamefont {JP}~\bibnamefont
  {Hurault}},\ }\bibfield  {title} {\enquote {\bibinfo {title} {Static
  distortions of a cholesteric planar structure induced by magnetic or ac
  electric fields},}\ }\href {https://doi.org/10.1063/1.1680293} {\bibfield
  {journal} {\bibinfo  {journal} {The Journal of Chemical Physics}\ }\textbf
  {\bibinfo {volume} {59}},\ \bibinfo {pages} {2068--2075} (\bibinfo {year}
  {1973})}\BibitemShut {NoStop}%
\bibitem [{\citenamefont {Whitfield}\ \emph {et~al.}(2017)\citenamefont
  {Whitfield}, \citenamefont {Adhyapak}, \citenamefont {Tiribocchi},
  \citenamefont {Alexander}, \citenamefont {Marenduzzo},\ and\ \citenamefont
  {Ramaswamy}}]{whitfield2017hydrodynamic}%
  \BibitemOpen
  \bibfield  {author} {\bibinfo {author} {\bibfnamefont {Carl~A}\ \bibnamefont
  {Whitfield}}, \bibinfo {author} {\bibfnamefont {Tapan~Chandra}\ \bibnamefont
  {Adhyapak}}, \bibinfo {author} {\bibfnamefont {Adriano}\ \bibnamefont
  {Tiribocchi}}, \bibinfo {author} {\bibfnamefont {Gareth~P}\ \bibnamefont
  {Alexander}}, \bibinfo {author} {\bibfnamefont {Davide}\ \bibnamefont
  {Marenduzzo}}, \ and\ \bibinfo {author} {\bibfnamefont {Sriram}\ \bibnamefont
  {Ramaswamy}},\ }\bibfield  {title} {\enquote {\bibinfo {title} {Hydrodynamic
  instabilities in active cholesteric liquid crystals},}\ }\href
  {https://link.springer.com/article/10.1140/epje/i2017-11536-2} {\bibfield
  {journal} {\bibinfo  {journal} {The European Physical Journal E}\ }\textbf
  {\bibinfo {volume} {40}},\ \bibinfo {pages} {1--16} (\bibinfo {year}
  {2017})}\BibitemShut {NoStop}%
\bibitem [{\citenamefont {van Zuiden}\ \emph
  {et~al.}(2016{\natexlab{b}})\citenamefont {van Zuiden}, \citenamefont
  {Paulose}, \citenamefont {Irvine}, \citenamefont {Bartolo},\ and\
  \citenamefont {Vitelli}}]{van2016spatiotemporal}%
  \BibitemOpen
  \bibfield  {author} {\bibinfo {author} {\bibfnamefont {Benjamin~C}\
  \bibnamefont {van Zuiden}}, \bibinfo {author} {\bibfnamefont {Jayson}\
  \bibnamefont {Paulose}}, \bibinfo {author} {\bibfnamefont {William~TM}\
  \bibnamefont {Irvine}}, \bibinfo {author} {\bibfnamefont {Denis}\
  \bibnamefont {Bartolo}}, \ and\ \bibinfo {author} {\bibfnamefont {Vincenzo}\
  \bibnamefont {Vitelli}},\ }\bibfield  {title} {\enquote {\bibinfo {title}
  {Spatiotemporal order and emergent edge currents in active spinner
  materials},}\ }\href {https://www.pnas.org/content/113/46/12919.short}
  {\bibfield  {journal} {\bibinfo  {journal} {Proceedings of the national
  academy of sciences}\ }\textbf {\bibinfo {volume} {113}},\ \bibinfo {pages}
  {12919--12924} (\bibinfo {year} {2016}{\natexlab{b}})}\BibitemShut {NoStop}%
\bibitem [{\citenamefont {Gibaud}\ \emph {et~al.}(2017)\citenamefont {Gibaud},
  \citenamefont {Kaplan}, \citenamefont {Sharma}, \citenamefont {Zakhary},
  \citenamefont {Ward}, \citenamefont {Oldenbourg}, \citenamefont {Meyer},
  \citenamefont {Kamien}, \citenamefont {Powers},\ and\ \citenamefont
  {Dogic}}]{gibaud2017achiral}%
  \BibitemOpen
  \bibfield  {author} {\bibinfo {author} {\bibfnamefont {Thomas}\ \bibnamefont
  {Gibaud}}, \bibinfo {author} {\bibfnamefont {C~Nadir}\ \bibnamefont
  {Kaplan}}, \bibinfo {author} {\bibfnamefont {Prerna}\ \bibnamefont {Sharma}},
  \bibinfo {author} {\bibfnamefont {Mark~J}\ \bibnamefont {Zakhary}}, \bibinfo
  {author} {\bibfnamefont {Andrew}\ \bibnamefont {Ward}}, \bibinfo {author}
  {\bibfnamefont {Rudolf}\ \bibnamefont {Oldenbourg}}, \bibinfo {author}
  {\bibfnamefont {Robert~B}\ \bibnamefont {Meyer}}, \bibinfo {author}
  {\bibfnamefont {Randall~D}\ \bibnamefont {Kamien}}, \bibinfo {author}
  {\bibfnamefont {Thomas~R}\ \bibnamefont {Powers}}, \ and\ \bibinfo {author}
  {\bibfnamefont {Zvonimir}\ \bibnamefont {Dogic}},\ }\bibfield  {title}
  {\enquote {\bibinfo {title} {Achiral symmetry breaking and positive gaussian
  modulus lead to scalloped colloidal membranes},}\ }\href
  {https://www.pnas.org/content/114/17/E3376.short} {\bibfield  {journal}
  {\bibinfo  {journal} {Proceedings of the National Academy of Sciences}\
  }\textbf {\bibinfo {volume} {114}},\ \bibinfo {pages} {E3376--E3384}
  (\bibinfo {year} {2017})}\BibitemShut {NoStop}%
\bibitem [{\citenamefont {Chugh}\ and\ \citenamefont
  {Paluch}(2018)}]{chugh2018actin}%
  \BibitemOpen
  \bibfield  {author} {\bibinfo {author} {\bibfnamefont {Priyamvada}\
  \bibnamefont {Chugh}}\ and\ \bibinfo {author} {\bibfnamefont {Ewa~K}\
  \bibnamefont {Paluch}},\ }\bibfield  {title} {\enquote {\bibinfo {title} {The
  actin cortex at a glance},}\ }\href@noop {} {\bibfield  {journal} {\bibinfo
  {journal} {J Cell Sci}\ }\textbf {\bibinfo {volume} {131}},\ \bibinfo {pages}
  {jcs186254} (\bibinfo {year} {2018})}\BibitemShut {NoStop}%
\bibitem [{\citenamefont {Tan}\ \emph {et~al.}(2018)\citenamefont {Tan},
  \citenamefont {Malik-Garbi}, \citenamefont {Abu-Shah}, \citenamefont {Li},
  \citenamefont {Sharma}, \citenamefont {MacKintosh}, \citenamefont {Keren},
  \citenamefont {Schmidt},\ and\ \citenamefont {Fakhri}}]{tan2018self}%
  \BibitemOpen
  \bibfield  {author} {\bibinfo {author} {\bibfnamefont {Tzer~Han}\
  \bibnamefont {Tan}}, \bibinfo {author} {\bibfnamefont {Maya}\ \bibnamefont
  {Malik-Garbi}}, \bibinfo {author} {\bibfnamefont {Enas}\ \bibnamefont
  {Abu-Shah}}, \bibinfo {author} {\bibfnamefont {Junang}\ \bibnamefont {Li}},
  \bibinfo {author} {\bibfnamefont {Abhinav}\ \bibnamefont {Sharma}}, \bibinfo
  {author} {\bibfnamefont {Fred~C}\ \bibnamefont {MacKintosh}}, \bibinfo
  {author} {\bibfnamefont {Kinneret}\ \bibnamefont {Keren}}, \bibinfo {author}
  {\bibfnamefont {Christoph~F}\ \bibnamefont {Schmidt}}, \ and\ \bibinfo
  {author} {\bibfnamefont {Nikta}\ \bibnamefont {Fakhri}},\ }\bibfield  {title}
  {\enquote {\bibinfo {title} {Self-organized stress patterns drive state
  transitions in actin cortices},}\ }\href@noop {} {\bibfield  {journal}
  {\bibinfo  {journal} {Science advances}\ }\textbf {\bibinfo {volume} {4}},\
  \bibinfo {pages} {eaar2847} (\bibinfo {year} {2018})}\BibitemShut {NoStop}%
\bibitem [{\citenamefont {Sawin}\ \emph {et~al.}(1992)\citenamefont {Sawin},
  \citenamefont {LeGuellec}, \citenamefont {Philippe},\ and\ \citenamefont
  {Mitchison}}]{sawin1992mitotic}%
  \BibitemOpen
  \bibfield  {author} {\bibinfo {author} {\bibfnamefont {Kenneth~E}\
  \bibnamefont {Sawin}}, \bibinfo {author} {\bibfnamefont {Katherine}\
  \bibnamefont {LeGuellec}}, \bibinfo {author} {\bibfnamefont {Michel}\
  \bibnamefont {Philippe}}, \ and\ \bibinfo {author} {\bibfnamefont
  {Timothy~J}\ \bibnamefont {Mitchison}},\ }\bibfield  {title} {\enquote
  {\bibinfo {title} {Mitotic spindle organization by a plus-end-directed
  microtubule motor.}}\ }\href@noop {} {\bibfield  {journal} {\bibinfo
  {journal} {Nature}\ }\textbf {\bibinfo {volume} {359}},\ \bibinfo {pages}
  {540--543} (\bibinfo {year} {1992})}\BibitemShut {NoStop}%
\bibitem [{\citenamefont {Brugu{\'e}s}\ and\ \citenamefont
  {Needleman}(2014)}]{brugues2014physical}%
  \BibitemOpen
  \bibfield  {author} {\bibinfo {author} {\bibfnamefont {Jan}\ \bibnamefont
  {Brugu{\'e}s}}\ and\ \bibinfo {author} {\bibfnamefont {Daniel}\ \bibnamefont
  {Needleman}},\ }\bibfield  {title} {\enquote {\bibinfo {title} {Physical
  basis of spindle self-organization},}\ }\href@noop {} {\bibfield  {journal}
  {\bibinfo  {journal} {Proceedings of the National Academy of Sciences}\
  }\textbf {\bibinfo {volume} {111}},\ \bibinfo {pages} {18496--18500}
  (\bibinfo {year} {2014})}\BibitemShut {NoStop}%
\bibitem [{\citenamefont {Seifert}(2012)}]{seifert2012stochastic}%
  \BibitemOpen
  \bibfield  {author} {\bibinfo {author} {\bibfnamefont {Udo}\ \bibnamefont
  {Seifert}},\ }\bibfield  {title} {\enquote {\bibinfo {title} {Stochastic
  thermodynamics, fluctuation theorems and molecular machines},}\ }\href@noop
  {} {\bibfield  {journal} {\bibinfo  {journal} {Reports on progress in
  physics}\ }\textbf {\bibinfo {volume} {75}},\ \bibinfo {pages} {126001}
  (\bibinfo {year} {2012})}\BibitemShut {NoStop}%
\bibitem [{\citenamefont {Chun}\ \emph {et~al.}(2021)\citenamefont {Chun},
  \citenamefont {Gao},\ and\ \citenamefont
  {Horowitz}}]{chun2021nonequilibrium}%
  \BibitemOpen
  \bibfield  {author} {\bibinfo {author} {\bibfnamefont {Hyun-Myung}\
  \bibnamefont {Chun}}, \bibinfo {author} {\bibfnamefont {Qi}~\bibnamefont
  {Gao}}, \ and\ \bibinfo {author} {\bibfnamefont {Jordan~M}\ \bibnamefont
  {Horowitz}},\ }\bibfield  {title} {\enquote {\bibinfo {title} {Nonequilibrium
  green-kubo relations for hydrodynamic transport from an equilibrium-like
  fluctuation-response equality},}\ }\href@noop {} {\bibfield  {journal}
  {\bibinfo  {journal} {arXiv preprint arXiv:2103.09288}\ } (\bibinfo {year}
  {2021})}\BibitemShut {NoStop}%
\bibitem [{\citenamefont {Han}\ \emph {et~al.}(2020)\citenamefont {Han},
  \citenamefont {Fruchart}, \citenamefont {Scheibner}, \citenamefont
  {Vaikuntanathan}, \citenamefont {Irvine}, \citenamefont {de~Pablo},\ and\
  \citenamefont {Vitelli}}]{han2020statistical}%
  \BibitemOpen
  \bibfield  {author} {\bibinfo {author} {\bibfnamefont {Ming}\ \bibnamefont
  {Han}}, \bibinfo {author} {\bibfnamefont {Michel}\ \bibnamefont {Fruchart}},
  \bibinfo {author} {\bibfnamefont {Colin}\ \bibnamefont {Scheibner}}, \bibinfo
  {author} {\bibfnamefont {Suriyanarayanan}\ \bibnamefont {Vaikuntanathan}},
  \bibinfo {author} {\bibfnamefont {William}\ \bibnamefont {Irvine}}, \bibinfo
  {author} {\bibfnamefont {Juan}\ \bibnamefont {de~Pablo}}, \ and\ \bibinfo
  {author} {\bibfnamefont {Vincenzo}\ \bibnamefont {Vitelli}},\ }\href@noop {}
  {\enquote {\bibinfo {title} {Statistical mechanics of a chiral active
  fluid},}\ } (\bibinfo {year} {2020}),\ \Eprint
  {http://arxiv.org/abs/2002.07679} {arXiv:2002.07679 [cond-mat.soft]}
  \BibitemShut {NoStop}%
\bibitem [{\citenamefont {Horowitz}\ and\ \citenamefont
  {Gingrich}(2020)}]{2020Horowitz}%
  \BibitemOpen
  \bibfield  {author} {\bibinfo {author} {\bibfnamefont {Jordan~M.}\
  \bibnamefont {Horowitz}}\ and\ \bibinfo {author} {\bibfnamefont {Todd~R.}\
  \bibnamefont {Gingrich}},\ }\bibfield  {title} {\enquote {\bibinfo {title}
  {Thermodynamic uncertainty relations constrain non-equilibrium
  fluctuations},}\ }\href@noop {} {\bibfield  {journal} {\bibinfo  {journal}
  {Nature Physics}\ }\textbf {\bibinfo {volume} {16}},\ \bibinfo {pages}
  {15--20} (\bibinfo {year} {2020})}\BibitemShut {NoStop}%
\bibitem [{\citenamefont {Li}\ \emph {et~al.}(2019)\citenamefont {Li},
  \citenamefont {Horowitz}, \citenamefont {Gingrich},\ and\ \citenamefont
  {Fakhri}}]{2019Li}%
  \BibitemOpen
  \bibfield  {author} {\bibinfo {author} {\bibfnamefont {Junang}\ \bibnamefont
  {Li}}, \bibinfo {author} {\bibfnamefont {Jordan~M.}\ \bibnamefont
  {Horowitz}}, \bibinfo {author} {\bibfnamefont {Todd~R.}\ \bibnamefont
  {Gingrich}}, \ and\ \bibinfo {author} {\bibfnamefont {Nikta}\ \bibnamefont
  {Fakhri}},\ }\bibfield  {title} {\enquote {\bibinfo {title} {Quantifying
  dissipation using fluctuating currents},}\ }\href@noop {} {\bibfield
  {journal} {\bibinfo  {journal} {Nature Communications}\ }\textbf {\bibinfo
  {volume} {10}},\ \bibinfo {pages} {1666} (\bibinfo {year}
  {2019})}\BibitemShut {NoStop}%
\bibitem [{\citenamefont {Mizuno}\ \emph {et~al.}(2007)\citenamefont {Mizuno},
  \citenamefont {Tardin}, \citenamefont {Schmidt},\ and\ \citenamefont
  {MacKintosh}}]{mizuno2007nonequilibrium}%
  \BibitemOpen
  \bibfield  {author} {\bibinfo {author} {\bibfnamefont {Daisuke}\ \bibnamefont
  {Mizuno}}, \bibinfo {author} {\bibfnamefont {Catherine}\ \bibnamefont
  {Tardin}}, \bibinfo {author} {\bibfnamefont {Christoph~F}\ \bibnamefont
  {Schmidt}}, \ and\ \bibinfo {author} {\bibfnamefont {Frederik~C}\
  \bibnamefont {MacKintosh}},\ }\bibfield  {title} {\enquote {\bibinfo {title}
  {Nonequilibrium mechanics of active cytoskeletal networks},}\ }\href@noop {}
  {\bibfield  {journal} {\bibinfo  {journal} {Science}\ }\textbf {\bibinfo
  {volume} {315}},\ \bibinfo {pages} {370--373} (\bibinfo {year}
  {2007})}\BibitemShut {NoStop}%
\bibitem [{\citenamefont {Wang}\ \emph {et~al.}(2016)\citenamefont {Wang},
  \citenamefont {Kawaguchi}, \citenamefont {Sasa},\ and\ \citenamefont
  {Tang}}]{Wang2016entropy}%
  \BibitemOpen
  \bibfield  {author} {\bibinfo {author} {\bibfnamefont {Shou-Wen}\
  \bibnamefont {Wang}}, \bibinfo {author} {\bibfnamefont {Kyogo}\ \bibnamefont
  {Kawaguchi}}, \bibinfo {author} {\bibfnamefont {Shin-ichi}\ \bibnamefont
  {Sasa}}, \ and\ \bibinfo {author} {\bibfnamefont {Lei-Han}\ \bibnamefont
  {Tang}},\ }\bibfield  {title} {\enquote {\bibinfo {title} {Entropy production
  of nanosystems with time scale separation},}\ }\href {\doibase
  10.1103/PhysRevLett.117.070601} {\bibfield  {journal} {\bibinfo  {journal}
  {Phys. Rev. Lett.}\ }\textbf {\bibinfo {volume} {117}},\ \bibinfo {pages}
  {070601} (\bibinfo {year} {2016})}\BibitemShut {NoStop}%
\bibitem [{\citenamefont {Levine}\ and\ \citenamefont
  {MacKintosh}(2009)}]{Levine2009}%
  \BibitemOpen
  \bibfield  {author} {\bibinfo {author} {\bibfnamefont {A.~J.}\ \bibnamefont
  {Levine}}\ and\ \bibinfo {author} {\bibfnamefont {F.~C.}\ \bibnamefont
  {MacKintosh}},\ }\bibfield  {title} {\enquote {\bibinfo {title} {The
  mechanics and fluctuation spectrum of active gels},}\ }\href@noop {}
  {\bibfield  {journal} {\bibinfo  {journal} {J. Phys. Chem. B}\ }\textbf
  {\bibinfo {volume} {113}},\ \bibinfo {pages} {3820--3830} (\bibinfo {year}
  {2009})}\BibitemShut {NoStop}%
\bibitem [{\citenamefont {Prost}\ \emph {et~al.}(2009)\citenamefont {Prost},
  \citenamefont {Joanny},\ and\ \citenamefont
  {Parrondo}}]{prost2009generalized}%
  \BibitemOpen
  \bibfield  {author} {\bibinfo {author} {\bibfnamefont {J}~\bibnamefont
  {Prost}}, \bibinfo {author} {\bibfnamefont {J-F}\ \bibnamefont {Joanny}}, \
  and\ \bibinfo {author} {\bibfnamefont {JMR}\ \bibnamefont {Parrondo}},\
  }\bibfield  {title} {\enquote {\bibinfo {title} {Generalized
  fluctuation-dissipation theorem for steady-state systems},}\ }\href@noop {}
  {\bibfield  {journal} {\bibinfo  {journal} {Physical review letters}\
  }\textbf {\bibinfo {volume} {103}},\ \bibinfo {pages} {090601} (\bibinfo
  {year} {2009})}\BibitemShut {NoStop}%
\bibitem [{\citenamefont {Fakhri}\ \emph {et~al.}(2014)\citenamefont {Fakhri},
  \citenamefont {Wessel}, \citenamefont {Willms}, \citenamefont {Pasquali},
  \citenamefont {Klopfenstein}, \citenamefont {MacKintosh},\ and\ \citenamefont
  {Schmidt}}]{fakhri2014high}%
  \BibitemOpen
  \bibfield  {author} {\bibinfo {author} {\bibfnamefont {Nikta}\ \bibnamefont
  {Fakhri}}, \bibinfo {author} {\bibfnamefont {Alok~D}\ \bibnamefont {Wessel}},
  \bibinfo {author} {\bibfnamefont {Charlotte}\ \bibnamefont {Willms}},
  \bibinfo {author} {\bibfnamefont {Matteo}\ \bibnamefont {Pasquali}}, \bibinfo
  {author} {\bibfnamefont {Dieter~R}\ \bibnamefont {Klopfenstein}}, \bibinfo
  {author} {\bibfnamefont {Frederick~C}\ \bibnamefont {MacKintosh}}, \ and\
  \bibinfo {author} {\bibfnamefont {Christoph~F}\ \bibnamefont {Schmidt}},\
  }\bibfield  {title} {\enquote {\bibinfo {title} {High-resolution mapping of
  intracellular fluctuations using carbon nanotubes},}\ }\href@noop {}
  {\bibfield  {journal} {\bibinfo  {journal} {Science}\ }\textbf {\bibinfo
  {volume} {344}},\ \bibinfo {pages} {1031--1035} (\bibinfo {year}
  {2014})}\BibitemShut {NoStop}%
\bibitem [{\citenamefont {Dinis}\ \emph {et~al.}(2012)\citenamefont {Dinis},
  \citenamefont {Martin}, \citenamefont {Barral}, \citenamefont {Prost},\ and\
  \citenamefont {Joanny}}]{dinis2012fluctuation}%
  \BibitemOpen
  \bibfield  {author} {\bibinfo {author} {\bibfnamefont {L}~\bibnamefont
  {Dinis}}, \bibinfo {author} {\bibfnamefont {P}~\bibnamefont {Martin}},
  \bibinfo {author} {\bibfnamefont {J}~\bibnamefont {Barral}}, \bibinfo
  {author} {\bibfnamefont {J}~\bibnamefont {Prost}}, \ and\ \bibinfo {author}
  {\bibfnamefont {JF}~\bibnamefont {Joanny}},\ }\bibfield  {title} {\enquote
  {\bibinfo {title} {Fluctuation-response theorem for the active noisy
  oscillator of the hair-cell bundle},}\ }\href
  {https://journals.aps.org/prl/abstract/10.1103/PhysRevLett.109.160602}
  {\bibfield  {journal} {\bibinfo  {journal} {Physical review letters}\
  }\textbf {\bibinfo {volume} {109}},\ \bibinfo {pages} {160602} (\bibinfo
  {year} {2012})}\BibitemShut {NoStop}%
\bibitem [{\citenamefont {Liphardt}\ \emph {et~al.}(2002)\citenamefont
  {Liphardt}, \citenamefont {Dumont}, \citenamefont {Smith}, \citenamefont
  {Tinoco},\ and\ \citenamefont {Bustamante}}]{Liphardt1832}%
  \BibitemOpen
  \bibfield  {author} {\bibinfo {author} {\bibfnamefont {Jan}\ \bibnamefont
  {Liphardt}}, \bibinfo {author} {\bibfnamefont {Sophie}\ \bibnamefont
  {Dumont}}, \bibinfo {author} {\bibfnamefont {Steven~B.}\ \bibnamefont
  {Smith}}, \bibinfo {author} {\bibfnamefont {Ignacio}\ \bibnamefont {Tinoco}},
  \ and\ \bibinfo {author} {\bibfnamefont {Carlos}\ \bibnamefont
  {Bustamante}},\ }\bibfield  {title} {\enquote {\bibinfo {title} {Equilibrium
  information from nonequilibrium measurements in an experimental test of
  jarzynski{\textquoteright}s equality},}\ }\href {\doibase
  10.1126/science.1071152} {\bibfield  {journal} {\bibinfo  {journal}
  {Science}\ }\textbf {\bibinfo {volume} {296}},\ \bibinfo {pages} {1832--1835}
  (\bibinfo {year} {2002})},\ \Eprint
  {http://arxiv.org/abs/https://science.sciencemag.org/content/296/5574/1832.full.pdf}
  {https://science.sciencemag.org/content/296/5574/1832.full.pdf} \BibitemShut
  {NoStop}%
\bibitem [{\citenamefont {Pietzonka}\ \emph {et~al.}(2019)\citenamefont
  {Pietzonka}, \citenamefont {Fodor}, \citenamefont {Lohrmann}, \citenamefont
  {Cates},\ and\ \citenamefont {Seifert}}]{2019Pietzonka}%
  \BibitemOpen
  \bibfield  {author} {\bibinfo {author} {\bibfnamefont {Patrick}\ \bibnamefont
  {Pietzonka}}, \bibinfo {author} {\bibfnamefont {\'Etienne}\ \bibnamefont
  {Fodor}}, \bibinfo {author} {\bibfnamefont {Christoph}\ \bibnamefont
  {Lohrmann}}, \bibinfo {author} {\bibfnamefont {Michael~E.}\ \bibnamefont
  {Cates}}, \ and\ \bibinfo {author} {\bibfnamefont {Udo}\ \bibnamefont
  {Seifert}},\ }\bibfield  {title} {\enquote {\bibinfo {title} {Autonomous
  engines driven by active matter: Energetics and design principles},}\ }\href
  {\doibase 10.1103/PhysRevX.9.041032} {\bibfield  {journal} {\bibinfo
  {journal} {Phys. Rev. X}\ }\textbf {\bibinfo {volume} {9}},\ \bibinfo {pages}
  {041032} (\bibinfo {year} {2019})}\BibitemShut {NoStop}%
\bibitem [{\citenamefont {Tan}\ \emph {et~al.}(2021{\natexlab{b}})\citenamefont
  {Tan}, \citenamefont {Watson}, \citenamefont {Chao}, \citenamefont {Li},
  \citenamefont {Gingrich}, \citenamefont {Horowitz},\ and\ \citenamefont
  {Fakhri}}]{Tan2021KLD}%
  \BibitemOpen
  \bibfield  {author} {\bibinfo {author} {\bibfnamefont {Tzer~Han}\
  \bibnamefont {Tan}}, \bibinfo {author} {\bibfnamefont {Garrett~A.}\
  \bibnamefont {Watson}}, \bibinfo {author} {\bibfnamefont {Yu-Chen}\
  \bibnamefont {Chao}}, \bibinfo {author} {\bibfnamefont {Junang}\ \bibnamefont
  {Li}}, \bibinfo {author} {\bibfnamefont {Todd~R.}\ \bibnamefont {Gingrich}},
  \bibinfo {author} {\bibfnamefont {Jordan~M.}\ \bibnamefont {Horowitz}}, \
  and\ \bibinfo {author} {\bibfnamefont {Nikta}\ \bibnamefont {Fakhri}},\
  }\href@noop {} {\enquote {\bibinfo {title} {Scale-dependent irreversibility
  in living matter},}\ } (\bibinfo {year} {2021}{\natexlab{b}}),\ \Eprint
  {http://arxiv.org/abs/2107.05701} {arXiv:2107.05701 [physics.bio-ph]}
  \BibitemShut {NoStop}%
\bibitem [{\citenamefont {Kawai}\ \emph {et~al.}(2007)\citenamefont {Kawai},
  \citenamefont {Parrondo},\ and\ \citenamefont {Van~den Broeck}}]{Kawai2007}%
  \BibitemOpen
  \bibfield  {author} {\bibinfo {author} {\bibfnamefont {R.}~\bibnamefont
  {Kawai}}, \bibinfo {author} {\bibfnamefont {J.~M.~R.}\ \bibnamefont
  {Parrondo}}, \ and\ \bibinfo {author} {\bibfnamefont {C.}~\bibnamefont
  {Van~den Broeck}},\ }\bibfield  {title} {\enquote {\bibinfo {title}
  {Dissipation: The phase-space perspective},}\ }\href@noop {} {\bibfield
  {journal} {\bibinfo  {journal} {Phys. Rev. Lett.}\ }\textbf {\bibinfo
  {volume} {98}},\ \bibinfo {pages} {080602} (\bibinfo {year}
  {2007})}\BibitemShut {NoStop}%
\bibitem [{\citenamefont {Rold\'an}\ and\ \citenamefont
  {Parrondo}(2012)}]{Roldan2012}%
  \BibitemOpen
  \bibfield  {author} {\bibinfo {author} {\bibfnamefont {\'E}\ \bibnamefont
  {Rold\'an}}\ and\ \bibinfo {author} {\bibfnamefont {J.~M.~R.}\ \bibnamefont
  {Parrondo}},\ }\bibfield  {title} {\enquote {\bibinfo {title} {Entropy
  production and kullback-leibler divergence between stationary trajectories of
  discrete systems},}\ }\href@noop {} {\bibfield  {journal} {\bibinfo
  {journal} {Phys. Rev. E}\ }\textbf {\bibinfo {volume} {85}},\ \bibinfo
  {pages} {031129} (\bibinfo {year} {2012})}\BibitemShut {NoStop}%
\bibitem [{\citenamefont {Parrondo}\ \emph {et~al.}(2009)\citenamefont
  {Parrondo}, \citenamefont {Van~den Broeck},\ and\ \citenamefont
  {Kawai}}]{Parrondo2009}%
  \BibitemOpen
  \bibfield  {author} {\bibinfo {author} {\bibfnamefont {J.~M.~R.}\
  \bibnamefont {Parrondo}}, \bibinfo {author} {\bibfnamefont {C.}~\bibnamefont
  {Van~den Broeck}}, \ and\ \bibinfo {author} {\bibfnamefont {R.}~\bibnamefont
  {Kawai}},\ }\bibfield  {title} {\enquote {\bibinfo {title} {Entropy
  production and the arrow of time},}\ }\href@noop {} {\bibfield  {journal}
  {\bibinfo  {journal} {New J. Phys.}\ }\textbf {\bibinfo {volume} {11}},\
  \bibinfo {pages} {073008} (\bibinfo {year} {2009})}\BibitemShut {NoStop}%
\bibitem [{\citenamefont {Gomez-Marin}\ \emph {et~al.}(2008)\citenamefont
  {Gomez-Marin}, \citenamefont {Parrondo},\ and\ \citenamefont {Van~den
  Broeck}}]{Gomez-Marin2008a}%
  \BibitemOpen
  \bibfield  {author} {\bibinfo {author} {\bibfnamefont {A.}~\bibnamefont
  {Gomez-Marin}}, \bibinfo {author} {\bibfnamefont {J.~M.~R.}\ \bibnamefont
  {Parrondo}}, \ and\ \bibinfo {author} {\bibfnamefont {C.}~\bibnamefont
  {Van~den Broeck}},\ }\bibfield  {title} {\enquote {\bibinfo {title} {The
  ``footprints'' of irreversibility},}\ }\href@noop {} {\bibfield  {journal}
  {\bibinfo  {journal} {Europhys. Lett.}\ }\textbf {\bibinfo {volume} {82}},\
  \bibinfo {pages} {50002} (\bibinfo {year} {2008})}\BibitemShut {NoStop}%
\bibitem [{\citenamefont {Horowitz}\ and\ \citenamefont
  {Jarzynski}(2009)}]{Horowitz2009b}%
  \BibitemOpen
  \bibfield  {author} {\bibinfo {author} {\bibfnamefont {J.~M.}\ \bibnamefont
  {Horowitz}}\ and\ \bibinfo {author} {\bibfnamefont {C.}~\bibnamefont
  {Jarzynski}},\ }\bibfield  {title} {\enquote {\bibinfo {title} {Illustrative
  example of the relationship between dissipation and relative entropy},}\
  }\href@noop {} {\bibfield  {journal} {\bibinfo  {journal} {Phys. Rev. E}\
  }\textbf {\bibinfo {volume} {79}},\ \bibinfo {pages} {021106} (\bibinfo
  {year} {2009})}\BibitemShut {NoStop}%
\bibitem [{\citenamefont {Rold{\'a}n}\ \emph {et~al.}(2018)\citenamefont
  {Rold{\'a}n}, \citenamefont {Barral}, \citenamefont {Martin}, \citenamefont
  {Parrondo},\ and\ \citenamefont {J{\"u}licher}}]{roldan2018arrow}%
  \BibitemOpen
  \bibfield  {author} {\bibinfo {author} {\bibfnamefont {{\'E}dgar}\
  \bibnamefont {Rold{\'a}n}}, \bibinfo {author} {\bibfnamefont
  {J{\'e}r{\'e}mie}\ \bibnamefont {Barral}}, \bibinfo {author} {\bibfnamefont
  {Pascal}\ \bibnamefont {Martin}}, \bibinfo {author} {\bibfnamefont {Juan~MR}\
  \bibnamefont {Parrondo}}, \ and\ \bibinfo {author} {\bibfnamefont {Frank}\
  \bibnamefont {J{\"u}licher}},\ }\bibfield  {title} {\enquote {\bibinfo
  {title} {Arrow of time in active fluctuations},}\ }\href@noop {} {\bibfield
  {journal} {\bibinfo  {journal} {arXiv preprint arXiv:1803.04743}\ } (\bibinfo
  {year} {2018})}\BibitemShut {NoStop}%
\bibitem [{\citenamefont {Mart\'inez}\ \emph {et~al.}(2019)\citenamefont
  {Mart\'inez}, \citenamefont {Bisker}, \citenamefont {Horowitz},\ and\
  \citenamefont {Parrondo}}]{Martinez2018}%
  \BibitemOpen
  \bibfield  {author} {\bibinfo {author} {\bibfnamefont {I.~A.}\ \bibnamefont
  {Mart\'inez}}, \bibinfo {author} {\bibfnamefont {G.}~\bibnamefont {Bisker}},
  \bibinfo {author} {\bibfnamefont {J.~M.}\ \bibnamefont {Horowitz}}, \ and\
  \bibinfo {author} {\bibfnamefont {J.~M.~R.}\ \bibnamefont {Parrondo}},\
  }\bibfield  {title} {\enquote {\bibinfo {title} {Inferring broken detailed
  balance in the absence of observable currents},}\ }\href@noop {} {\bibfield
  {journal} {\bibinfo  {journal} {Nat. Comm.}\ }\textbf {\bibinfo {volume}
  {10}},\ \bibinfo {pages} {1--10} (\bibinfo {year} {2019})}\BibitemShut
  {NoStop}%
\bibitem [{\citenamefont {Fodor}\ \emph {et~al.}(2016)\citenamefont {Fodor},
  \citenamefont {Nardini}, \citenamefont {Cates}, \citenamefont {Tailleur},
  \citenamefont {Visco},\ and\ \citenamefont {van Wijland}}]{2016Fodor}%
  \BibitemOpen
  \bibfield  {author} {\bibinfo {author} {\bibfnamefont {\'Etienne}\
  \bibnamefont {Fodor}}, \bibinfo {author} {\bibfnamefont {Cesare}\
  \bibnamefont {Nardini}}, \bibinfo {author} {\bibfnamefont {Michael~E.}\
  \bibnamefont {Cates}}, \bibinfo {author} {\bibfnamefont {Julien}\
  \bibnamefont {Tailleur}}, \bibinfo {author} {\bibfnamefont {Paolo}\
  \bibnamefont {Visco}}, \ and\ \bibinfo {author} {\bibfnamefont
  {Fr\'ed\'eric}\ \bibnamefont {van Wijland}},\ }\bibfield  {title} {\enquote
  {\bibinfo {title} {How far from equilibrium is active matter?}}\ }\href
  {\doibase 10.1103/PhysRevLett.117.038103} {\bibfield  {journal} {\bibinfo
  {journal} {Phys. Rev. Lett.}\ }\textbf {\bibinfo {volume} {117}},\ \bibinfo
  {pages} {038103} (\bibinfo {year} {2016})}\BibitemShut {NoStop}%
\bibitem [{\citenamefont {Nardini}\ \emph {et~al.}(2017)\citenamefont
  {Nardini}, \citenamefont {Fodor}, \citenamefont {Tjhung}, \citenamefont {van
  Wijland}, \citenamefont {Taileur},\ and\ \citenamefont
  {Cates}}]{Nardini2017}%
  \BibitemOpen
  \bibfield  {author} {\bibinfo {author} {\bibfnamefont {C.}~\bibnamefont
  {Nardini}}, \bibinfo {author} {\bibfnamefont {\'E}\ \bibnamefont {Fodor}},
  \bibinfo {author} {\bibfnamefont {E.}~\bibnamefont {Tjhung}}, \bibinfo
  {author} {\bibfnamefont {F.}~\bibnamefont {van Wijland}}, \bibinfo {author}
  {\bibfnamefont {J.}~\bibnamefont {Taileur}}, \ and\ \bibinfo {author}
  {\bibfnamefont {M.~E.}\ \bibnamefont {Cates}},\ }\bibfield  {title} {\enquote
  {\bibinfo {title} {Entropy production in field theoreis without time-reversal
  symmetry: quantifying the non-equilibrium character of active matter},}\
  }\href@noop {} {\bibfield  {journal} {\bibinfo  {journal} {Phys. Rev. X}\
  }\textbf {\bibinfo {volume} {7}},\ \bibinfo {pages} {021007} (\bibinfo {year}
  {2017})}\BibitemShut {NoStop}%
\bibitem [{\citenamefont {Tociu}\ \emph {et~al.}(2019)\citenamefont {Tociu},
  \citenamefont {Fodor}, \citenamefont {Nemoto},\ and\ \citenamefont
  {Vaikuntanathan}}]{2019Tociu}%
  \BibitemOpen
  \bibfield  {author} {\bibinfo {author} {\bibfnamefont {Laura}\ \bibnamefont
  {Tociu}}, \bibinfo {author} {\bibfnamefont {\'Etienne}\ \bibnamefont
  {Fodor}}, \bibinfo {author} {\bibfnamefont {Takahiro}\ \bibnamefont
  {Nemoto}}, \ and\ \bibinfo {author} {\bibfnamefont {Suriyanarayanan}\
  \bibnamefont {Vaikuntanathan}},\ }\bibfield  {title} {\enquote {\bibinfo
  {title} {How dissipation constrains fluctuations in nonequilibrium liquids:
  Diffusion, structure, and biased interactions},}\ }\href {\doibase
  10.1103/PhysRevX.9.041026} {\bibfield  {journal} {\bibinfo  {journal} {Phys.
  Rev. X}\ }\textbf {\bibinfo {volume} {9}},\ \bibinfo {pages} {041026}
  (\bibinfo {year} {2019})}\BibitemShut {NoStop}%
\bibitem [{\citenamefont {Fodor}\ \emph {et~al.}(2020)\citenamefont {Fodor},
  \citenamefont {Nemoto},\ and\ \citenamefont {Vaikuntanathan}}]{Fodor_2020}%
  \BibitemOpen
  \bibfield  {author} {\bibinfo {author} {\bibfnamefont {{\'{E}}tienne}\
  \bibnamefont {Fodor}}, \bibinfo {author} {\bibfnamefont {Takahiro}\
  \bibnamefont {Nemoto}}, \ and\ \bibinfo {author} {\bibfnamefont
  {Suriyanarayanan}\ \bibnamefont {Vaikuntanathan}},\ }\bibfield  {title}
  {\enquote {\bibinfo {title} {Dissipation controls transport and phase
  transitions in active fluids: mobility, diffusion and biased ensembles},}\
  }\href {\doibase 10.1088/1367-2630/ab6353} {\bibfield  {journal} {\bibinfo
  {journal} {New Journal of Physics}\ }\textbf {\bibinfo {volume} {22}},\
  \bibinfo {pages} {013052} (\bibinfo {year} {2020})}\BibitemShut {NoStop}%
\bibitem [{\citenamefont {Falasco}\ \emph {et~al.}(2018)\citenamefont
  {Falasco}, \citenamefont {Rao},\ and\ \citenamefont
  {Esposito}}]{2018Falasco}%
  \BibitemOpen
  \bibfield  {author} {\bibinfo {author} {\bibfnamefont {Gianmaria}\
  \bibnamefont {Falasco}}, \bibinfo {author} {\bibfnamefont {Riccardo}\
  \bibnamefont {Rao}}, \ and\ \bibinfo {author} {\bibfnamefont {Massimiliano}\
  \bibnamefont {Esposito}},\ }\bibfield  {title} {\enquote {\bibinfo {title}
  {Information thermodynamics of turing patterns},}\ }\href {\doibase
  10.1103/PhysRevLett.121.108301} {\bibfield  {journal} {\bibinfo  {journal}
  {Phys. Rev. Lett.}\ }\textbf {\bibinfo {volume} {121}},\ \bibinfo {pages}
  {108301} (\bibinfo {year} {2018})}\BibitemShut {NoStop}%
\bibitem [{\citenamefont {Rogers}\ and\ \citenamefont
  {Schier}(2011)}]{rogers2011morphogen}%
  \BibitemOpen
  \bibfield  {author} {\bibinfo {author} {\bibfnamefont {Katherine~W}\
  \bibnamefont {Rogers}}\ and\ \bibinfo {author} {\bibfnamefont {Alexander~F}\
  \bibnamefont {Schier}},\ }\bibfield  {title} {\enquote {\bibinfo {title}
  {Morphogen gradients: from generation to interpretation},}\ }\href@noop {}
  {\bibfield  {journal} {\bibinfo  {journal} {Annual review of cell and
  developmental biology}\ }\textbf {\bibinfo {volume} {27}},\ \bibinfo {pages}
  {377--407} (\bibinfo {year} {2011})}\BibitemShut {NoStop}%
\bibitem [{\citenamefont {Prost}\ \emph {et~al.}(2015)\citenamefont {Prost},
  \citenamefont {J{\"u}licher},\ and\ \citenamefont
  {Joanny}}]{prost2015active}%
  \BibitemOpen
  \bibfield  {author} {\bibinfo {author} {\bibfnamefont {Jacques}\ \bibnamefont
  {Prost}}, \bibinfo {author} {\bibfnamefont {Frank}\ \bibnamefont
  {J{\"u}licher}}, \ and\ \bibinfo {author} {\bibfnamefont
  {Jean-Fran{\c{c}}ois}\ \bibnamefont {Joanny}},\ }\bibfield  {title} {\enquote
  {\bibinfo {title} {Active gel physics},}\ }\href@noop {} {\bibfield
  {journal} {\bibinfo  {journal} {Nature physics}\ }\textbf {\bibinfo {volume}
  {11}},\ \bibinfo {pages} {111--117} (\bibinfo {year} {2015})}\BibitemShut
  {NoStop}%
\bibitem [{\citenamefont {Streichan}\ \emph {et~al.}(2018)\citenamefont
  {Streichan}, \citenamefont {Lefebvre}, \citenamefont {Noll}, \citenamefont
  {Wieschaus},\ and\ \citenamefont {Shraiman}}]{streichan2018global}%
  \BibitemOpen
  \bibfield  {author} {\bibinfo {author} {\bibfnamefont {Sebastian~J}\
  \bibnamefont {Streichan}}, \bibinfo {author} {\bibfnamefont {Matthew~F}\
  \bibnamefont {Lefebvre}}, \bibinfo {author} {\bibfnamefont {Nicholas}\
  \bibnamefont {Noll}}, \bibinfo {author} {\bibfnamefont {Eric~F}\ \bibnamefont
  {Wieschaus}}, \ and\ \bibinfo {author} {\bibfnamefont {Boris~I}\ \bibnamefont
  {Shraiman}},\ }\bibfield  {title} {\enquote {\bibinfo {title} {Global
  morphogenetic flow is accurately predicted by the spatial distribution of
  myosin motors},}\ }\href@noop {} {\bibfield  {journal} {\bibinfo  {journal}
  {Elife}\ }\textbf {\bibinfo {volume} {7}},\ \bibinfo {pages} {e27454}
  (\bibinfo {year} {2018})}\BibitemShut {NoStop}%
\bibitem [{\citenamefont {Morris}\ and\ \citenamefont
  {Rao}(2019)}]{morris2019active}%
  \BibitemOpen
  \bibfield  {author} {\bibinfo {author} {\bibfnamefont {Richard~G}\
  \bibnamefont {Morris}}\ and\ \bibinfo {author} {\bibfnamefont {Madan}\
  \bibnamefont {Rao}},\ }\bibfield  {title} {\enquote {\bibinfo {title} {Active
  morphogenesis of epithelial monolayers},}\ }\href
  {https://journals.aps.org/pre/abstract/10.1103/PhysRevE.100.022413}
  {\bibfield  {journal} {\bibinfo  {journal} {Physical Review E}\ }\textbf
  {\bibinfo {volume} {100}},\ \bibinfo {pages} {022413} (\bibinfo {year}
  {2019})}\BibitemShut {NoStop}%
\bibitem [{\citenamefont {Struhl}\ \emph {et~al.}(2012)\citenamefont {Struhl},
  \citenamefont {Casal},\ and\ \citenamefont
  {Lawrence}}]{struhl2012dissecting}%
  \BibitemOpen
  \bibfield  {author} {\bibinfo {author} {\bibfnamefont {Gary}\ \bibnamefont
  {Struhl}}, \bibinfo {author} {\bibfnamefont {Jos{\'e}}\ \bibnamefont
  {Casal}}, \ and\ \bibinfo {author} {\bibfnamefont {Peter~A}\ \bibnamefont
  {Lawrence}},\ }\bibfield  {title} {\enquote {\bibinfo {title} {Dissecting the
  molecular bridges that mediate the function of frizzled in planar cell
  polarity},}\ }\href@noop {} {\bibfield  {journal} {\bibinfo  {journal}
  {Development}\ }\textbf {\bibinfo {volume} {139}},\ \bibinfo {pages}
  {3665--3674} (\bibinfo {year} {2012})}\BibitemShut {NoStop}%
\bibitem [{\citenamefont {Etournay}\ \emph {et~al.}(2016)\citenamefont
  {Etournay}, \citenamefont {Merkel}, \citenamefont {Popović}, \citenamefont
  {Brandl}, \citenamefont {Dye}, \citenamefont {Aigouy}, \citenamefont
  {Salbreux}, \citenamefont {Eaton},\ and\ \citenamefont
  {Jülicher}}]{2016Eaton}%
  \BibitemOpen
  \bibfield  {author} {\bibinfo {author} {\bibfnamefont {Raphaël}\
  \bibnamefont {Etournay}}, \bibinfo {author} {\bibfnamefont {Matthias}\
  \bibnamefont {Merkel}}, \bibinfo {author} {\bibfnamefont {Marko}\
  \bibnamefont {Popović}}, \bibinfo {author} {\bibfnamefont {Holger}\
  \bibnamefont {Brandl}}, \bibinfo {author} {\bibfnamefont {Natalie~A}\
  \bibnamefont {Dye}}, \bibinfo {author} {\bibfnamefont {Benoît}\ \bibnamefont
  {Aigouy}}, \bibinfo {author} {\bibfnamefont {Guillaume}\ \bibnamefont
  {Salbreux}}, \bibinfo {author} {\bibfnamefont {Suzanne}\ \bibnamefont
  {Eaton}}, \ and\ \bibinfo {author} {\bibfnamefont {Frank}\ \bibnamefont
  {Jülicher}},\ }\bibfield  {title} {\enquote {\bibinfo {title} {Tissueminer:
  A multiscale analysis toolkit to quantify how cellular processes create
  tissue dynamics},}\ }\href {\doibase 10.7554/eLife.14334} {\bibfield
  {journal} {\bibinfo  {journal} {eLife}\ }\textbf {\bibinfo {volume} {5}},\
  \bibinfo {pages} {e14334} (\bibinfo {year} {2016})}\BibitemShut {NoStop}%
\bibitem [{\citenamefont {Balasubramaniam}\ \emph {et~al.}(2021)\citenamefont
  {Balasubramaniam}, \citenamefont {Doostmohammadi}, \citenamefont {Saw},
  \citenamefont {Narayana}, \citenamefont {Mueller}, \citenamefont {Dang},
  \citenamefont {Thomas}, \citenamefont {Gupta}, \citenamefont {Sonam},
  \citenamefont {Yap} \emph {et~al.}}]{balasubramaniam2021investigating}%
  \BibitemOpen
  \bibfield  {author} {\bibinfo {author} {\bibfnamefont {Lakshmi}\ \bibnamefont
  {Balasubramaniam}}, \bibinfo {author} {\bibfnamefont {Amin}\ \bibnamefont
  {Doostmohammadi}}, \bibinfo {author} {\bibfnamefont {Thuan~Beng}\
  \bibnamefont {Saw}}, \bibinfo {author} {\bibfnamefont {Gautham Hari
  Narayana~Sankara}\ \bibnamefont {Narayana}}, \bibinfo {author} {\bibfnamefont
  {Romain}\ \bibnamefont {Mueller}}, \bibinfo {author} {\bibfnamefont {Tien}\
  \bibnamefont {Dang}}, \bibinfo {author} {\bibfnamefont {Minnah}\ \bibnamefont
  {Thomas}}, \bibinfo {author} {\bibfnamefont {Shafali}\ \bibnamefont {Gupta}},
  \bibinfo {author} {\bibfnamefont {Surabhi}\ \bibnamefont {Sonam}}, \bibinfo
  {author} {\bibfnamefont {Alpha~S}\ \bibnamefont {Yap}},  \emph {et~al.},\
  }\bibfield  {title} {\enquote {\bibinfo {title} {Investigating the nature of
  active forces in tissues reveals how contractile cells can form extensile
  monolayers},}\ }\href@noop {} {\bibfield  {journal} {\bibinfo  {journal}
  {Nature materials}\ ,\ \bibinfo {pages} {1--11}} (\bibinfo {year}
  {2021})}\BibitemShut {NoStop}%
\bibitem [{\citenamefont {Serra-Picamal}\ \emph {et~al.}(2012)\citenamefont
  {Serra-Picamal}, \citenamefont {Conte}, \citenamefont {Vincent},
  \citenamefont {Anon}, \citenamefont {Tambe}, \citenamefont {Bazellieres},
  \citenamefont {Butler}, \citenamefont {Fredberg},\ and\ \citenamefont
  {Trepat}}]{serra2012mechanical}%
  \BibitemOpen
  \bibfield  {author} {\bibinfo {author} {\bibfnamefont {Xavier}\ \bibnamefont
  {Serra-Picamal}}, \bibinfo {author} {\bibfnamefont {Vito}\ \bibnamefont
  {Conte}}, \bibinfo {author} {\bibfnamefont {Romaric}\ \bibnamefont
  {Vincent}}, \bibinfo {author} {\bibfnamefont {Ester}\ \bibnamefont {Anon}},
  \bibinfo {author} {\bibfnamefont {Dhananjay~T}\ \bibnamefont {Tambe}},
  \bibinfo {author} {\bibfnamefont {Elsa}\ \bibnamefont {Bazellieres}},
  \bibinfo {author} {\bibfnamefont {James~P}\ \bibnamefont {Butler}}, \bibinfo
  {author} {\bibfnamefont {Jeffrey~J}\ \bibnamefont {Fredberg}}, \ and\
  \bibinfo {author} {\bibfnamefont {Xavier}\ \bibnamefont {Trepat}},\
  }\bibfield  {title} {\enquote {\bibinfo {title} {Mechanical waves during
  tissue expansion},}\ }\href@noop {} {\bibfield  {journal} {\bibinfo
  {journal} {Nature Physics}\ }\textbf {\bibinfo {volume} {8}},\ \bibinfo
  {pages} {628--634} (\bibinfo {year} {2012})}\BibitemShut {NoStop}%
\bibitem [{\citenamefont {Alert}\ and\ \citenamefont
  {Trepat}(2020)}]{alert2020physical}%
  \BibitemOpen
  \bibfield  {author} {\bibinfo {author} {\bibfnamefont {Ricard}\ \bibnamefont
  {Alert}}\ and\ \bibinfo {author} {\bibfnamefont {Xavier}\ \bibnamefont
  {Trepat}},\ }\bibfield  {title} {\enquote {\bibinfo {title} {Physical models
  of collective cell migration},}\ }\href@noop {} {\bibfield  {journal}
  {\bibinfo  {journal} {Annual Review of Condensed Matter Physics}\ }\textbf
  {\bibinfo {volume} {11}},\ \bibinfo {pages} {77--101} (\bibinfo {year}
  {2020})}\BibitemShut {NoStop}%
\bibitem [{\citenamefont {Duclut}\ \emph {et~al.}(2021)\citenamefont {Duclut},
  \citenamefont {Paijmans}, \citenamefont {Inamdar}, \citenamefont {Modes},\
  and\ \citenamefont {J{\"u}licher}}]{duclut2021nonlinear}%
  \BibitemOpen
  \bibfield  {author} {\bibinfo {author} {\bibfnamefont {Charlie}\ \bibnamefont
  {Duclut}}, \bibinfo {author} {\bibfnamefont {Joris}\ \bibnamefont
  {Paijmans}}, \bibinfo {author} {\bibfnamefont {Mandar~M}\ \bibnamefont
  {Inamdar}}, \bibinfo {author} {\bibfnamefont {Carl~D}\ \bibnamefont {Modes}},
  \ and\ \bibinfo {author} {\bibfnamefont {Frank}\ \bibnamefont
  {J{\"u}licher}},\ }\bibfield  {title} {\enquote {\bibinfo {title} {Nonlinear
  rheology of cellular networks},}\ }\href@noop {} {\bibfield  {journal}
  {\bibinfo  {journal} {arXiv preprint arXiv:2103.16462}\ } (\bibinfo {year}
  {2021})}\BibitemShut {NoStop}%
\bibitem [{\citenamefont {Pujic}\ and\ \citenamefont
  {Malicki}(2004)}]{pujic2004retinal}%
  \BibitemOpen
  \bibfield  {author} {\bibinfo {author} {\bibfnamefont {Zac}\ \bibnamefont
  {Pujic}}\ and\ \bibinfo {author} {\bibfnamefont {Jarema}\ \bibnamefont
  {Malicki}},\ }\bibfield  {title} {\enquote {\bibinfo {title} {Retinal pattern
  and the genetic basis of its formation in zebrafish},}\ }in\ \href@noop {}
  {\emph {\bibinfo {booktitle} {Seminars in cell \& developmental biology}}},\
  Vol.~\bibinfo {volume} {15}\ (\bibinfo {organization} {Elsevier},\ \bibinfo
  {year} {2004})\ pp.\ \bibinfo {pages} {105--114}\BibitemShut {NoStop}%
\bibitem [{\citenamefont {Classen}\ \emph {et~al.}(2005)\citenamefont
  {Classen}, \citenamefont {Anderson}, \citenamefont {Marois},\ and\
  \citenamefont {Eaton}}]{classen2005hexagonal}%
  \BibitemOpen
  \bibfield  {author} {\bibinfo {author} {\bibfnamefont {Anne-Kathrin}\
  \bibnamefont {Classen}}, \bibinfo {author} {\bibfnamefont {Kurt~I}\
  \bibnamefont {Anderson}}, \bibinfo {author} {\bibfnamefont {Eric}\
  \bibnamefont {Marois}}, \ and\ \bibinfo {author} {\bibfnamefont {Suzanne}\
  \bibnamefont {Eaton}},\ }\bibfield  {title} {\enquote {\bibinfo {title}
  {Hexagonal packing of drosophila wing epithelial cells by the planar cell
  polarity pathway},}\ }\href@noop {} {\bibfield  {journal} {\bibinfo
  {journal} {Developmental cell}\ }\textbf {\bibinfo {volume} {9}},\ \bibinfo
  {pages} {805--817} (\bibinfo {year} {2005})}\BibitemShut {NoStop}%
\bibitem [{\citenamefont {Sun}\ and\ \citenamefont
  {Patel}(2019)}]{sun2019amphipod}%
  \BibitemOpen
  \bibfield  {author} {\bibinfo {author} {\bibfnamefont {Dennis~A}\
  \bibnamefont {Sun}}\ and\ \bibinfo {author} {\bibfnamefont {Nipam~H}\
  \bibnamefont {Patel}},\ }\bibfield  {title} {\enquote {\bibinfo {title} {The
  amphipod crustacean parhyale hawaiensis: an emerging comparative model of
  arthropod development, evolution, and regeneration},}\ }\href@noop {}
  {\bibfield  {journal} {\bibinfo  {journal} {Wiley Interdisciplinary Reviews:
  Developmental Biology}\ }\textbf {\bibinfo {volume} {8}},\ \bibinfo {pages}
  {e355} (\bibinfo {year} {2019})}\BibitemShut {NoStop}%
\bibitem [{\citenamefont {Salbreux}\ \emph {et~al.}(2012)\citenamefont
  {Salbreux}, \citenamefont {Barthel}, \citenamefont {Raymond},\ and\
  \citenamefont {Lubensky}}]{salbreux2012coupling}%
  \BibitemOpen
  \bibfield  {author} {\bibinfo {author} {\bibfnamefont {Guillaume}\
  \bibnamefont {Salbreux}}, \bibinfo {author} {\bibfnamefont {Linda~K}\
  \bibnamefont {Barthel}}, \bibinfo {author} {\bibfnamefont {Pamela~A}\
  \bibnamefont {Raymond}}, \ and\ \bibinfo {author} {\bibfnamefont {David~K}\
  \bibnamefont {Lubensky}},\ }\bibfield  {title} {\enquote {\bibinfo {title}
  {Coupling mechanical deformations and planar cell polarity to create regular
  patterns in the zebrafish retina},}\ }\href@noop {} {\bibfield  {journal}
  {\bibinfo  {journal} {PLoS Comput Biol}\ }\textbf {\bibinfo {volume} {8}},\
  \bibinfo {pages} {e1002618} (\bibinfo {year} {2012})}\BibitemShut {NoStop}%
\bibitem [{\citenamefont {Ajeti}\ \emph {et~al.}(2019)\citenamefont {Ajeti},
  \citenamefont {Tabatabai}, \citenamefont {Fleszar}, \citenamefont {Staddon},
  \citenamefont {Seara}, \citenamefont {Suarez}, \citenamefont {Yousafzai},
  \citenamefont {Bi}, \citenamefont {Kovar}, \citenamefont {Banerjee} \emph
  {et~al.}}]{ajeti2019wound}%
  \BibitemOpen
  \bibfield  {author} {\bibinfo {author} {\bibfnamefont {Visar}\ \bibnamefont
  {Ajeti}}, \bibinfo {author} {\bibfnamefont {A~Pasha}\ \bibnamefont
  {Tabatabai}}, \bibinfo {author} {\bibfnamefont {Andrew~J}\ \bibnamefont
  {Fleszar}}, \bibinfo {author} {\bibfnamefont {Michael~F}\ \bibnamefont
  {Staddon}}, \bibinfo {author} {\bibfnamefont {Daniel~S}\ \bibnamefont
  {Seara}}, \bibinfo {author} {\bibfnamefont {Cristian}\ \bibnamefont
  {Suarez}}, \bibinfo {author} {\bibfnamefont {M~Sulaiman}\ \bibnamefont
  {Yousafzai}}, \bibinfo {author} {\bibfnamefont {Dapeng}\ \bibnamefont {Bi}},
  \bibinfo {author} {\bibfnamefont {David~R}\ \bibnamefont {Kovar}}, \bibinfo
  {author} {\bibfnamefont {Shiladitya}\ \bibnamefont {Banerjee}},  \emph
  {et~al.},\ }\bibfield  {title} {\enquote {\bibinfo {title} {Wound healing
  coordinates actin architectures to regulate mechanical work},}\ }\href@noop
  {} {\bibfield  {journal} {\bibinfo  {journal} {Nature physics}\ }\textbf
  {\bibinfo {volume} {15}},\ \bibinfo {pages} {696--705} (\bibinfo {year}
  {2019})}\BibitemShut {NoStop}%
\bibitem [{\citenamefont {Camalet}\ \emph {et~al.}(2000)\citenamefont
  {Camalet}, \citenamefont {Duke}, \citenamefont {J{\"u}licher},\ and\
  \citenamefont {Prost}}]{camalet2000auditory}%
  \BibitemOpen
  \bibfield  {author} {\bibinfo {author} {\bibfnamefont {S{\'e}bastien}\
  \bibnamefont {Camalet}}, \bibinfo {author} {\bibfnamefont {Thomas}\
  \bibnamefont {Duke}}, \bibinfo {author} {\bibfnamefont {Frank}\ \bibnamefont
  {J{\"u}licher}}, \ and\ \bibinfo {author} {\bibfnamefont {Jacques}\
  \bibnamefont {Prost}},\ }\bibfield  {title} {\enquote {\bibinfo {title}
  {Auditory sensitivity provided by self-tuned critical oscillations of hair
  cells},}\ }\href {https://www.pnas.org/content/97/7/3183.short} {\bibfield
  {journal} {\bibinfo  {journal} {Proceedings of the National Academy of
  Sciences}\ }\textbf {\bibinfo {volume} {97}},\ \bibinfo {pages} {3183--3188}
  (\bibinfo {year} {2000})}\BibitemShut {NoStop}%
\bibitem [{\citenamefont {Bialek}\ \emph {et~al.}(2012)\citenamefont {Bialek},
  \citenamefont {Cavagna}, \citenamefont {Giardina}, \citenamefont {Mora},
  \citenamefont {Silvestri}, \citenamefont {Viale},\ and\ \citenamefont
  {Walczak}}]{bialek2012statistical}%
  \BibitemOpen
  \bibfield  {author} {\bibinfo {author} {\bibfnamefont {William}\ \bibnamefont
  {Bialek}}, \bibinfo {author} {\bibfnamefont {Andrea}\ \bibnamefont
  {Cavagna}}, \bibinfo {author} {\bibfnamefont {Irene}\ \bibnamefont
  {Giardina}}, \bibinfo {author} {\bibfnamefont {Thierry}\ \bibnamefont
  {Mora}}, \bibinfo {author} {\bibfnamefont {Edmondo}\ \bibnamefont
  {Silvestri}}, \bibinfo {author} {\bibfnamefont {Massimiliano}\ \bibnamefont
  {Viale}}, \ and\ \bibinfo {author} {\bibfnamefont {Aleksandra~M}\
  \bibnamefont {Walczak}},\ }\bibfield  {title} {\enquote {\bibinfo {title}
  {Statistical mechanics for natural flocks of birds},}\ }\href
  {https://www.pnas.org/content/109/13/4786.short} {\bibfield  {journal}
  {\bibinfo  {journal} {Proceedings of the National Academy of Sciences}\
  }\textbf {\bibinfo {volume} {109}},\ \bibinfo {pages} {4786--4791} (\bibinfo
  {year} {2012})}\BibitemShut {NoStop}%
\bibitem [{\citenamefont {Krotov}\ \emph {et~al.}(2014)\citenamefont {Krotov},
  \citenamefont {Dubuis}, \citenamefont {Gregor},\ and\ \citenamefont
  {Bialek}}]{krotov2014morphogenesis}%
  \BibitemOpen
  \bibfield  {author} {\bibinfo {author} {\bibfnamefont {Dmitry}\ \bibnamefont
  {Krotov}}, \bibinfo {author} {\bibfnamefont {Julien~O}\ \bibnamefont
  {Dubuis}}, \bibinfo {author} {\bibfnamefont {Thomas}\ \bibnamefont {Gregor}},
  \ and\ \bibinfo {author} {\bibfnamefont {William}\ \bibnamefont {Bialek}},\
  }\bibfield  {title} {\enquote {\bibinfo {title} {Morphogenesis at
  criticality},}\ }\href {https://www.pnas.org/content/111/10/3683.short}
  {\bibfield  {journal} {\bibinfo  {journal} {Proceedings of the National
  Academy of Sciences}\ }\textbf {\bibinfo {volume} {111}},\ \bibinfo {pages}
  {3683--3688} (\bibinfo {year} {2014})}\BibitemShut {NoStop}%
\bibitem [{\citenamefont {Milewski}\ \emph {et~al.}(2017)\citenamefont
  {Milewski}, \citenamefont {Maoil{\'e}idigh}, \citenamefont {Salvi},\ and\
  \citenamefont {Hudspeth}}]{milewski2017homeostatic}%
  \BibitemOpen
  \bibfield  {author} {\bibinfo {author} {\bibfnamefont {Andrew~R}\
  \bibnamefont {Milewski}}, \bibinfo {author} {\bibfnamefont
  {D{\'a}ibhid~{\'O}}\ \bibnamefont {Maoil{\'e}idigh}}, \bibinfo {author}
  {\bibfnamefont {Joshua~D}\ \bibnamefont {Salvi}}, \ and\ \bibinfo {author}
  {\bibfnamefont {AJ}~\bibnamefont {Hudspeth}},\ }\bibfield  {title} {\enquote
  {\bibinfo {title} {Homeostatic enhancement of sensory transduction},}\ }\href
  {https://www.pnas.org/content/114/33/E6794.short} {\bibfield  {journal}
  {\bibinfo  {journal} {Proceedings of the National Academy of Sciences}\
  }\textbf {\bibinfo {volume} {114}},\ \bibinfo {pages} {E6794--E6803}
  (\bibinfo {year} {2017})}\BibitemShut {NoStop}%
\bibitem [{\citenamefont {Paoluzzi}\ \emph {et~al.}(2020)\citenamefont
  {Paoluzzi}, \citenamefont {Leoni},\ and\ \citenamefont
  {Marchetti}}]{paoluzzi2020information}%
  \BibitemOpen
  \bibfield  {author} {\bibinfo {author} {\bibfnamefont {Matteo}\ \bibnamefont
  {Paoluzzi}}, \bibinfo {author} {\bibfnamefont {Marco}\ \bibnamefont {Leoni}},
  \ and\ \bibinfo {author} {\bibfnamefont {M~Cristina}\ \bibnamefont
  {Marchetti}},\ }\bibfield  {title} {\enquote {\bibinfo {title} {Information
  and motility exchange in collectives of active particles},}\ }\href@noop {}
  {\bibfield  {journal} {\bibinfo  {journal} {Soft Matter}\ }\textbf {\bibinfo
  {volume} {16}},\ \bibinfo {pages} {6317--6327} (\bibinfo {year}
  {2020})}\BibitemShut {NoStop}%
\bibitem [{\citenamefont {Norambuena}\ \emph {et~al.}(2020)\citenamefont
  {Norambuena}, \citenamefont {Valencia},\ and\ \citenamefont
  {Guzm{\'a}n-Lastra}}]{norambuena2020understanding}%
  \BibitemOpen
  \bibfield  {author} {\bibinfo {author} {\bibfnamefont {Ariel}\ \bibnamefont
  {Norambuena}}, \bibinfo {author} {\bibfnamefont {Felipe~J}\ \bibnamefont
  {Valencia}}, \ and\ \bibinfo {author} {\bibfnamefont {Francisca}\
  \bibnamefont {Guzm{\'a}n-Lastra}},\ }\bibfield  {title} {\enquote {\bibinfo
  {title} {Understanding contagion dynamics through microscopic processes in
  active brownian particles},}\ }\href@noop {} {\bibfield  {journal} {\bibinfo
  {journal} {Scientific Reports}\ }\textbf {\bibinfo {volume} {10}},\ \bibinfo
  {pages} {1--7} (\bibinfo {year} {2020})}\BibitemShut {NoStop}%
\bibitem [{\citenamefont {Zhao}\ \emph {et~al.}(2021)\citenamefont {Zhao},
  \citenamefont {Huepe},\ and\ \citenamefont {Romanczuk}}]{zhao2021contagion}%
  \BibitemOpen
  \bibfield  {author} {\bibinfo {author} {\bibfnamefont {Yinong}\ \bibnamefont
  {Zhao}}, \bibinfo {author} {\bibfnamefont {Cristi{\'a}n}\ \bibnamefont
  {Huepe}}, \ and\ \bibinfo {author} {\bibfnamefont {Pawel}\ \bibnamefont
  {Romanczuk}},\ }\bibfield  {title} {\enquote {\bibinfo {title} {Contagion
  dynamics in self-organized systems of self-propelled agents},}\ }\href@noop
  {} {\bibfield  {journal} {\bibinfo  {journal} {arXiv preprint
  arXiv:2103.12618}\ } (\bibinfo {year} {2021})}\BibitemShut {NoStop}%
\bibitem [{\citenamefont {Ganai}\ \emph {et~al.}(2014)\citenamefont {Ganai},
  \citenamefont {Sengupta},\ and\ \citenamefont {Menon}}]{ganai2014chromosome}%
  \BibitemOpen
  \bibfield  {author} {\bibinfo {author} {\bibfnamefont {Nirmalendu}\
  \bibnamefont {Ganai}}, \bibinfo {author} {\bibfnamefont {Surajit}\
  \bibnamefont {Sengupta}}, \ and\ \bibinfo {author} {\bibfnamefont {Gautam~I}\
  \bibnamefont {Menon}},\ }\bibfield  {title} {\enquote {\bibinfo {title}
  {Chromosome positioning from activity-based segregation},}\ }\href
  {https://doi.org/10.1093/nar/gkt1417} {\bibfield  {journal} {\bibinfo
  {journal} {Nucleic acids research}\ }\textbf {\bibinfo {volume} {42}},\
  \bibinfo {pages} {4145--4159} (\bibinfo {year} {2014})}\BibitemShut {NoStop}%
\bibitem [{\citenamefont {Grosberg}\ and\ \citenamefont
  {Joanny}(2015)}]{joanny-2015}%
  \BibitemOpen
  \bibfield  {author} {\bibinfo {author} {\bibfnamefont {A.~Y.}\ \bibnamefont
  {Grosberg}}\ and\ \bibinfo {author} {\bibfnamefont {J.-F.}\ \bibnamefont
  {Joanny}},\ }\bibfield  {title} {\enquote {\bibinfo {title} {Nonequilibrium
  statistical mechanics of mixtures of particles in contact with different
  thermostats},}\ }\href {\doibase 10.1103/PhysRevE.92.032118} {\bibfield
  {journal} {\bibinfo  {journal} {Phys. Rev. E}\ }\textbf {\bibinfo {volume}
  {92}},\ \bibinfo {pages} {032118} (\bibinfo {year} {2015})}\BibitemShut
  {NoStop}%
\bibitem [{\citenamefont {Shankar}\ \emph {et~al.}(2020)\citenamefont
  {Shankar}, \citenamefont {Souslov}, \citenamefont {Bowick}, \citenamefont
  {Marchetti},\ and\ \citenamefont {Vitelli}}]{shankar2020topological}%
  \BibitemOpen
  \bibfield  {author} {\bibinfo {author} {\bibfnamefont {Suraj}\ \bibnamefont
  {Shankar}}, \bibinfo {author} {\bibfnamefont {Anton}\ \bibnamefont
  {Souslov}}, \bibinfo {author} {\bibfnamefont {Mark~J}\ \bibnamefont
  {Bowick}}, \bibinfo {author} {\bibfnamefont {M~Cristina}\ \bibnamefont
  {Marchetti}}, \ and\ \bibinfo {author} {\bibfnamefont {Vincenzo}\
  \bibnamefont {Vitelli}},\ }\bibfield  {title} {\enquote {\bibinfo {title}
  {Topological active matter},}\ }\href@noop {} {\bibfield  {journal} {\bibinfo
   {journal} {arXiv preprint arXiv:2010.00364}\ } (\bibinfo {year}
  {2020})}\BibitemShut {NoStop}%
\end{thebibliography}

\end{document}